\documentclass[english]{article}
\usepackage{mathpazo}
\usepackage[T1]{fontenc}
\usepackage{textcomp}
\usepackage[utf8]{inputenc}
\usepackage{babel}
\usepackage{calc}
\usepackage{tikz}
\usepackage{amsmath}
\usepackage{graphicx}
\usepackage[a4paper]{geometry}
\usepackage{setspace}
\usepackage[authoryear]{natbib}
\usepackage[pdfusetitle,
 bookmarks=true,bookmarksnumbered=false,bookmarksopen=false,
 breaklinks=false,pdfborder={0 0 0},pdfborderstyle={},backref=false,colorlinks=false]
 {hyperref}

\makeatletter

\newcommand*\LyXZeroWidthSpace{\hspace{0pt}}
\providecommand{\tabularnewline}{\\}
\providecolor{lyxadded}{rgb}{0,0,1}
\providecolor{lyxdeleted}{rgb}{1,0,0}
\usetikzlibrary{calc}
\newcommand{\lyxobjectsout}[1]{%
  \bgroup%
  \color{lyxdeleted}%
  \tikz{
    \node[inner sep=0pt,outer sep=0pt](lyxdelobj){#1};
    \draw($(lyxdelobj.south west)+(2em,.5em)$)--($(lyxdelobj.north east)-(2em,.5em)$);
  }
  \egroup%
}

\DeclareRobustCommand{\lyxdisplayobjdeleted}[4][]{%
  \ifx#4\empty\else%
     \texorpdfstring{\leavevmode\\\lyxobjectsout{\parbox{\linewidth}{#4}}}{}%
  \fi%
}
\DeclareRobustCommand{\lyxudisplayobjdeleted}[4][]{%
  \ifx#4\empty\else%
     \texorpdfstring{\leavevmode\\\raisebox{-\belowdisplayshortskip}{%
                \lyxobjectsout{\parbox[b]{\linewidth}{#4}}}}{}%
     \leavevmode\\%
  \fi%
}

\newcommand{\lyxaddress}[1]{
	\par {\raggedright #1
	\vspace{1.4em}
	\noindent\par}
}

\usepackage{textcomp} 
\usepackage{cuted}

\makeatother

\begin{document}
\title{Detection, attribution, and modeling of climate change: \\
key open issues}
\author{Nicola Scafetta\textsuperscript{1{*}}}
\maketitle

\lyxaddress{\textsuperscript{\textsuperscript{}1}Department of Earth Sciences,
Environment and Georesources, University of Naples Federico II, Complesso
Universitario di Monte S. Angelo, Naples, Italy.}

\lyxaddress{\textsuperscript{\textsuperscript{}{*}}Corresponding author: nicola.scafetta@unina.it}
\begin{abstract}
The Coupled Model Intercomparison Project (CMIP) global climate models
(GCMs) assess that nearly 100\% of global surface warming observed
between 1850--1900 and 2011--2020 is attributable to anthropogenic
drivers like greenhouse gas emissions. These models also generate
future climate projections based on shared socioeconomic pathways
(SSPs), aiding in risk assessment and the development of costly “\emph{Net-Zero}”
climate mitigation strategies. Yet, as this study discusses, the CMIP
GCMs face significant scientific challenges in attributing and modeling
climate change, particularly in capturing natural climate variability
over multiple timescales throughout the Holocene. Other key concerns
include the reliability of global surface temperature records, the
accuracy of solar irradiance models, and the robustness of climate
sensitivity estimates. Global warming estimates may be overstated
due to uncorrected non-climatic biases, and the GCMs may significantly
underestimate solar and astronomical influences on climate variations.
The equilibrium climate sensitivity (ECS) to radiative forcing could
be lower than commonly assumed; empirical findings suggest ECS values
lower than 3°C and possibly even closer to $1.1\pm0.4$ °C. Empirical
models incorporating natural variability suggest that the 21\textsuperscript{st}-century
global warming may remain moderate, even under SSP scenarios that
do not necessitate Net-Zero emission policies. These findings raise
important questions regarding the necessity and urgency of implementing
aggressive climate mitigation strategies. While GCMs remain essential
tools for climate research and policymaking, their scientific limitations
underscore the need for more refined modeling approaches to ensure
accurate future climate assessments. Addressing uncertainties related
to climate change detection, natural variability, solar influences,
and climate sensitivity to radiative forcing will enhance predictions
and better inform sustainable climate strategies. \\

\textbf{Keywords:} Climate change detection, attribution and hazard
assessment; Climate model validation; Anthropogenic vs. natural climate
drivers; Equilibrium climate sensitivity; Solar cycles and influence
on climate.\\

\textbf{Cite as:} Scafetta, N.: 2025. Detection, attribution, and
modeling of climate change: Key open issues. Gondwana Research. \href{https://doi.org/10.1016/j.gr.2025.05.001}{https://doi.org/10.1016/j.gr.2025.05.001}
\end{abstract}

\subsection*{Highlights:}

This paper argues that climate science still faces unresolved key
issues. Global Climate models struggle in reproducing natural variability
at all time scales and may exaggerate warming. Global Climate models
likely exgerate the ECS and significantly underestimate solar influences.
Empirical models projects moderate warming challenging strict net-zero
climate policies. New climate models that better reflect natural climate
drivers and variations are needed.

\begin{figure*}[!t]
\includegraphics[width=1\textwidth]{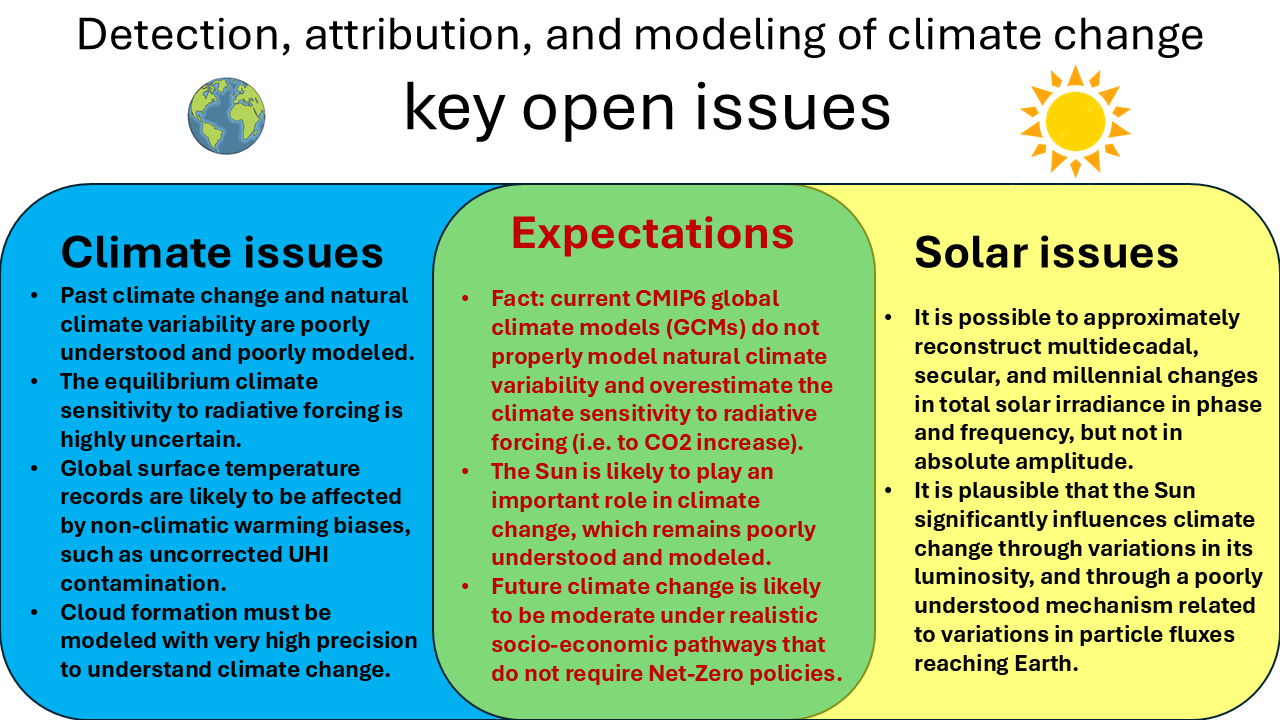}
\end{figure*}

\section{Introduction}

Throughout human history, climate change has profoundly influenced
both the natural environment and societal structures, leaving a lasting
imprint through its wide-ranging positive and negative effects. The
negative impacts encompass a range of potential threats, including
diminished food production, increased transmission of diseases, and
the disruption of ecosystems and infrastructure caused by extreme
weather events such as heatwaves, floods, hurricanes, sea level rise,
and droughts \citep{IPCC2023}. Importantly, the effects of climate
change are unevenly distributed across the globe, with socio-economic
disparities accentuating differences in exposure and vulnerability.
Marginalized communities, particularly those facing economic constraints,
are disproportionately affected by climate-related hazards. Such disparities
underscore the enduring concern humanity has had regarding climate
change. Nevertheless, the science underpinning climate change, its
causative factors, and the reliability of future climate projections
remain subjects of significant scientific uncertainty \citep{Curry2023}.
This contrasts with prevalent public assertions claiming that climate
change is ``\emph{settled science}''.

The United Nations Intergovernmental Panel on Climate Change (IPCC)
has published a series of Assessment Reports (ARs) since 1990, which
address the detection and attribution of climate change \citep{IPCC1990,IPCC1995,IPCC2001,IPCC2007,IPCC2013,IPCC2021}.
These reports are regarded by many scientists, policymakers, and members
of the public as providing robust evidence of a significant anthropogenic
contribution to unprecedented global warming since the pre-industrial
period of 1850--1900. Fossil fuel combustion --- resulting in the
release of greenhouse gases, particularly carbon dioxide (CO\textsubscript{2})
--- is identified as the primary driver of increased radiative forcing
leading to rising global surface temperatures. Global climate models
(GCMs) process a number of radiative forcing inputs and simulate global
temperature increases, attributed in gran part to the activation of
allegedly-positive climatic feedback mechanisms. The outputs of these
GCMs are subsequently employed to assess future climate scenarios
based on hypothetical shared socioeconomic pathways (SSPs) and associated
hazards. This information forms the basis for the IPCC recommendations
aimed at guiding policymakers in crafting comprehensive and costly
climate mitigation strategies to avert projected adverse climate change
outcomes \citep{IPCC2023}.

For instance, the European Union has adopted the “\emph{Green Deal}”,
a policy initiative targeting at least a 55\% reduction in net greenhouse
gas emissions by 2030 (compared to 1990 levels) and achieving carbon
neutrality, or “\emph{Net-Zero}”, by 2050 \citep{EUGreenDeal}. This
policy framework seeks to balance greenhouse gas emissions with their
removal via natural absorption processes and emerging carbon sequestration
technologies. The overarching goal of the Net-Zero policies is to
keep the ``\emph{increase in the global average temperature to well
below 2}°C\emph{ above pre-industrial levels}'' with efforts ``\emph{to
limit the temperature increase to 1.5}°C\emph{ above pre-industrial
levels}'' \citep{ParisAgreement}. Strategies include transitioning
from fossil fuels to renewable energy sources (geothermal, hydroelectric,
solar, and wind power) and replacing internal combustion vehicles
with electric alternatives.

However, the implementation of the Green Deal policies entails significant
economic investments, potentially jeopardizing industrial development
and economic growth. Therefore, a thorough evaluation of the feasibility
of these policies and the underlying scientific justifications is
imperative.

Serious economic challenges render Net-Zero targets difficult, if
not impossible to achieve. As of 2023, the European Union's 27 member
states collectively contributed only 6.08\% of global greenhouse gas
emission, which continues to grow globally by approximately 2\% annually,
despite reductions in Europe \citep{EU2024}. The continued global
reliance on fossil fuels --- evident in the approximately 1,000 coal-fired
power plants across Asia currently under construction or planned,
particularly in China and India \citep{GCPT2024} --- suggests that
achieving Net-Zero emissions may remain impractical throughout the
21\textsuperscript{st} century \citep{CAT2024,Hausfather2020,IEA2024}.
Also the depletion of critical metals essential for low-carbon technologies
and electric vehicles (e.g., Co, Ni, Cu, Se, Ag, Cd, In, Te, and Pt)
by 2060 could hinder further technological advancements \citep{Groves2023}.
Furthermore, economic meta-analyses indicate that the costs associated
with implementing Net-Zero policies to meet the targets of the Paris
Agreement could outweigh the anticipated benefits even in the most
favorable scenario \citep{Tol2023}. The above findings suggest the
need of a meticulous evaluation of energy and climate policies by
carefully weighing their potential advantages and disadvantages \citep{OhAiseadha2020}.

To accurately assess the risks associated with projected climate changes
and the effectiveness of mitigation policies, it is essential to first
evaluate the reliability of current climate science. The IPCC Assessment
Reports directly address this point by explaining in great details
how the Coupled Model Intercomparison Project (CMIP6) rigorously tests
global climate models (GCMs) against historical climate records. These
models undergo thorough a number of statistical validation in an attempt
to ensure that their confidence intervals align well with the observed
temperature trends. Additionally, climate models are cross-checked
with paleoclimate data from ice cores and tree rings, as well as insights
from oceanography, atmospheric physics, and geophysics. By integrating
evidence from multiple disciplines, the IPCC attempts to maintain
consistency between climate projections and real-world observations.

\begin{figure*}
\raggedright{}%
\noindent\fbox{\begin{minipage}[t]{1\textwidth - 2\fboxsep - 2\fboxrule}%
\begin{flushleft}
\textbf{Appendix A.}\\
\textbf{(Top) The figure illustrates that there are the approximately
1,000 coal-fired power plants across Asia currently under construction
or planned, particularly in China and India \citep{GCPT2024}.}\\
\textbf{(Bottom) The figure illustrates that as of 2023, the European
Union's 27 member states collectively contributed only 6.08\% of global
greenhouse gas emission, which continues to grow globally by approximately
2\% annually, despite reductions in Europe. Data from the ``JRC SCIENCE
FOR POLICY REPORT GHG EMISSIONS OF ALL WORLD COUNTRIES 2024'' \citep{EU2024}.}
\par\end{flushleft}
\begin{center}
\includegraphics[width=0.95\textwidth]{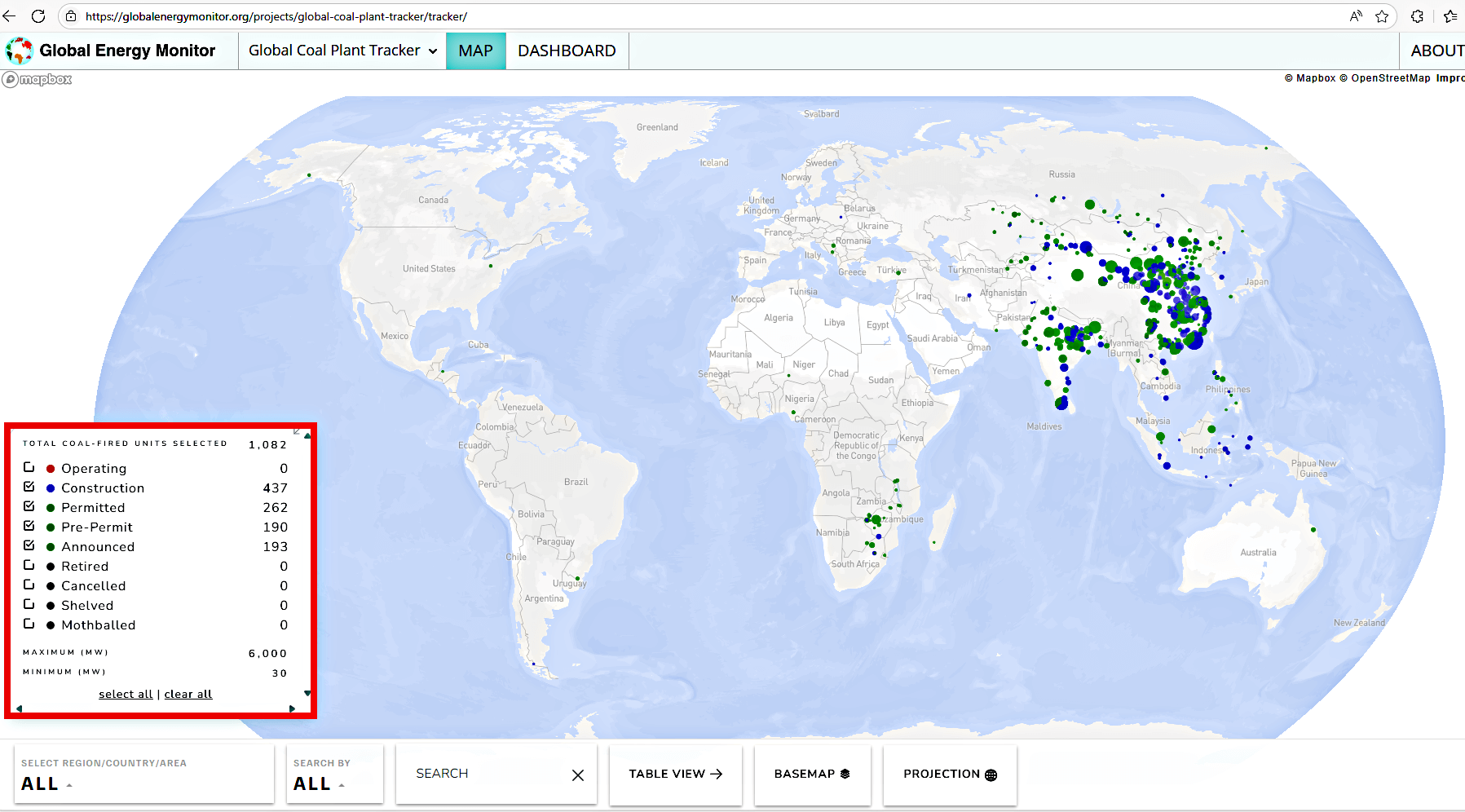}\\
\par\end{center}
\begin{center}
\includegraphics[width=0.95\textwidth]{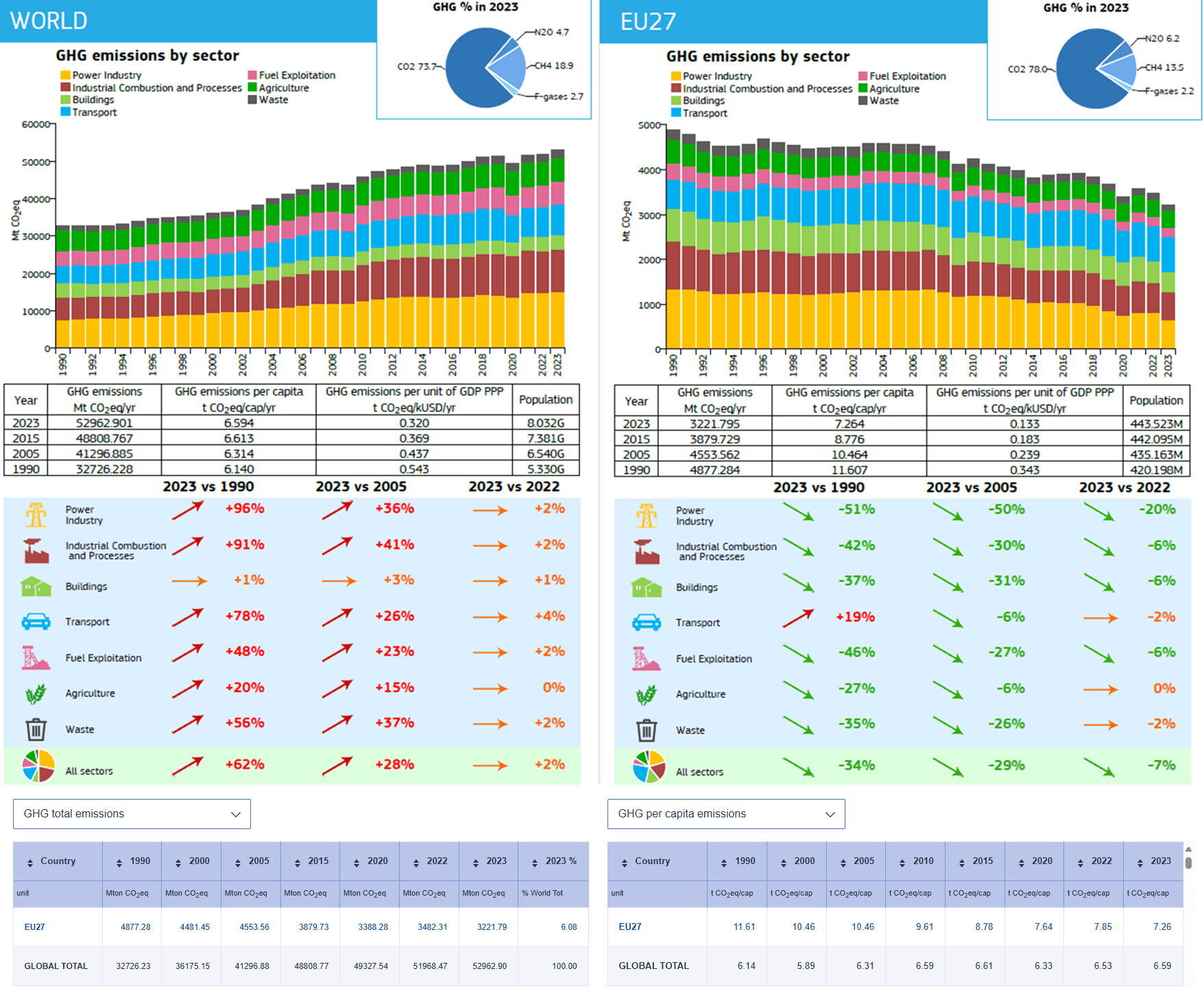}
\par\end{center}%
\end{minipage}}
\end{figure*}

However, unresolved major scientific questions regarding the science
of climate change persist. In fact, contrary to some popular perception,
climate science remains unsettled and highly debated. Recent research
highlights persistent uncertainties in the detection, attribution,
and prediction of climate change, and advocate for a more cautious
approach to this multifaceted complex issue.

Scientific challenges in climate detection, attribution, and modeling
stem from three primary issues:
\begin{enumerate}
\item the inherent uncertainty of what measurements really indicate complicates
the detection of climate change and its causative factors;
\item the anthropogenic contribution is superimposed to natural climate
variability, necessitating comprehensive understanding and accurate
modeling of the latter;
\item key physical processes, such as cloud formation and solar contributions
to climate dynamics, remain poorly characterized.
\end{enumerate}
Anthropogenic climate change becomes meaningful only if it is significant
relative to expected natural variability. Accurate attributions require
precise modeling of external forcing mechanisms and climate responses.
However, ongoing uncertainties in these areas continue to impede the
accurate attribution of observed climatic changes to either natural
or human influences \citep{Curry2023}.

Section 2 of this paper outlines the evolution of the ``\emph{Anthropogenic
Global Warming Theory}'' (AGWT) as presented in the IPCC Assessment
Reports published from 1990 to 2023 \citep{IPCC1990,IPCC1995,IPCC2001,IPCC2007,IPCC2013,IPCC2021}.
Sections 3 and 4 critically examine uncertainties surrounding the
reliability of global temperature records, the extent of solar influences
on climate change, and climate sensitivity to radiative forcing. Evidence
suggests that estimates of global surface warming since the pre-industrial
period may have been exaggerated \citep[e.g.:][]{ScafettaOuyang2019,Scafetta2021a,Katata2023,Soon2023}
and that solar and astronomical contributions to climate dynamics
may have been underestimated \citep[e.g.:][]{Connolly2023,Scafetta2023a}.
Additionally, contemporary GCMs remain insufficiently validated, and
climate sensitivity parameters may be lower than previously anticipated
\citep[e.g.:][]{Lewis2023,Scafetta2022a,Scafetta2023b}.

Empirical models incorporating natural variability are discussed in
Section 5 and suggest that global warming during the 21\textsuperscript{st}
century under realistic SSPs may be moderate \citep[e.g.:][]{Connolly2020,Scafetta2024}
and not as severe as projected by the \citet{IPCC2021,IPCC2023}.
Consequently, it is investigated whether the risks associated with
future climate change have been overestimated, underscoring the need
to address several physical uncertainties prior to develop and implement
effective climate policies.

\section{The Anthropogenic Global Warming Theory (AGWT)}

This section outlines the key arguments supporting the Intergovernmental
Panel on Climate Change's theory of anthropogenic global warming,
and traces the evolution of the AGWT from the 1990s to the most recent
assessment report \citep[AR6,][]{IPCC2021,IPCC2023}.

\subsection{The IPCC Assessment Reports from 1990 to 2023}

The IPCC's Sixth Assessment Report \citeyearpar[AR6,][]{IPCC2023}
claims definitive evidence linking human activities to the observed
global warming. According to the report: ``\emph{Global surface temperature
was 1.09 {[}0.95 to 1.20{]} }°C\emph{ higher in 2011--2020 than 1850--1900,
with larger increases over land (1.59 {[}1.34 to 1.83{]} }°C\emph{)
than over the ocean (0.88 {[}0.68 to 1.01{]} }°C\emph{)}''; further,
``\emph{It is unequivocal that human influence has warmed the atmosphere,
ocean and land}'' owing to ``\emph{observed increases in well-mixed
greenhouse gas (GHG) concentrations since around 1750 are unequivocally
caused by human activities}''. The report concludes: ``\emph{The
likely range of total human-caused global surface temperature increase
from 1850--1900 to 2010--2019 is 0.8}°C\emph{ to 1.3}°C\emph{, with
a best estimate of 1.07}°C''. These statements support the conclusion
that nearly 100\% of the global surface warming observed from the
pre-industrial period (1850--1900) to 2011--2020 is attributable
to anthropogenic drivers, which forms the basis of the Anthropogenic
Global Warming Theory (AGWT). However, the IPCC's attribution of natural
versus anthropogenic contributions to climate change has evolved significantly
over time.

The IPCC's First Assessment Report \citeyearpar[FAR,][]{IPCC1990}
acknowledged that in the 20\textsuperscript{th} century human activities
had begun influencing global climate, but precise quantification of
their contribution remained elusive due to uncertainties surrounding
natural variability. FAR highlighted the paleoclimatic temperature
record of the last millennium by \citet{Lamb1965}. This was a proxy
temperature reconstruction for England that could more broadly represent
the European and North Atlantic climates. This reconstruction identified
a pronounced Medieval Warm Period (MWP) (900--1300 AD), which was
estimated definitely warmer than the first half of the 20\textsuperscript{th}
century, followed by the Little Ice Age (LIA) (1300--1850 AD). On
the basis of such evidence, FAR noted that the warming observed since
1900 was ``\emph{broadly consistent with predictions of climate models,
but it is also of the same magnitude as natural climate variability}'',
and suggested that ``\emph{the observed increase could be largely
due to this natural variability; alternatively this variability and
other human factors could have offset a still larger human-induced
greenhouse warming}'' (FAR Executive Summary, p. xii).

The Second Assessment Report \citeyearpar[SAR,][]{IPCC1995} largely
corroborated FAR's findings, estimating that anthropogenic activities
could have accounted for approximately 50\% of the observed global
warming of the 20\textsuperscript{th} century with natural variability
contributing the remainder. SAR and FAR also noted a moderate global
cooling from the 1940s to the 1970s.

The IPCC's Third Assessment Report \citeyearpar[TAR,][]{IPCC2001}
posited that human activities likely accounted for around 70\% of
the observed warming since 1900, emphasizing the role of greenhouse
gas emissions (e.g., carbon dioxide, methane, nitrous oxide) from
industrial processes, agriculture, and deforestation. TAR introduced
a novel reconstruction of the Northern Hemisphere temperatures over
the past millennium \citep{Mann1999}, which depicted the controversial
“\emph{hockey-stick}” pattern showing a very modest \textasciitilde 0.2°C
cooling from MWP to LIA followed by \textasciitilde 1°C warming from
1900 to 2000. This graph suggested that the 20\textsuperscript{th}-century
warming was unprecedented and that the overall temperature pattern
of the last millennium correlated well with historical greenhouse
gas concentration records; such a correlation increased confidence
in climate models and in their climate attribution assessments \citep[cf.][]{Crowley2000}.
However, the hockey-stick graph contradicted earlier paleoclimate
research \citep[e.g.][]{Lamb1965} and sparked controversy \citep[e.g.][]{Soon2003,Soon2003b,vonStorch2004}
due also to methodological concerns \citep{McIntyre2003}, rapidly
becoming even a politically charged debate \citep[e.g.:][]{Deming2006,Gore2006}.

The Fourth Assessment Report \citeyearpar[AR4,][]{IPCC2007} proposed
again the hockey-stick graph \citep{Mann1999} and concluded that
``\emph{most of the observed increase in global average temperatures
since the mid-20\textsuperscript{\emph{th}} century is 'very likely'
}(i.e., >90\% probability)\emph{ due to the observed increase in anthropogenic
greenhouse gas concentrations}''.

Meanwhile, several novel paleoclimatic studies \citep[e.g.:][]{Moberg2005,Mann2008}
revealed a more pronounced MWP, sometimes comparable to the Current
Warm Period (CWP) of the last of the last decades of the 20\textsuperscript{th}
century \citep[e.g.:][]{Ljungqvist2010,Christiansen2012}. These records
were acknowledged in the IPCC Fifth Assessment Report \citeyearpar[AR5, ][its figure 5.7]{IPCC2013},
where the original hockey-stick temperature graph was abandoned. However,
the report maintained that \emph{``It is 'extremely likely' }(i.e.,
95--100\% probability)\emph{ that human influence was the dominant
cause of global warming between 1951 and 2010}''.

Finally, AR6 \citep{IPCC2021,IPCC2023} attributed 0.8--1.3°C of
the observed 0.95--1.20°C warming (from 1850--1900 to 2011--2020)
to anthropogenic emissions, and relied on the \citet{PAGES2k2019}
proxy model, which again attenuated the Medieval Warm Period although
more moderately than in the original hockey-stick temperature record
proposed by \citet{Mann1999}. However, representing the Common Era
temperature history through a single paleoclimatic reconstruction
has been recently criticized as reductive \citep{Esper2024}, indicating
that the debate remains ongoing. Accurate assessments of past climates
are necessary to validate the AGWT, as discussed below.

\subsection{The crucial role of the \textquotedblleft Global Climate Models\textquotedblright{}
(GCMs)}

Despite significant uncertainties in paleoclimate temperature reconstructions,
the IPCC's understanding of climate change attribution from the pre-industrial
period (1850--1900) to the present largely relies on computational
experiments conducted with sophisticated numerical climate models.
The principal characteristics and findings of these models are extensively
summarized in the IPCC's Fourth, Fifth, and Sixth Assessment Reports
\citep{IPCC2007,IPCC2013,IPCC2021}.

\begin{figure*}[!t]
\begin{centering}
\includegraphics[width=1\textwidth]{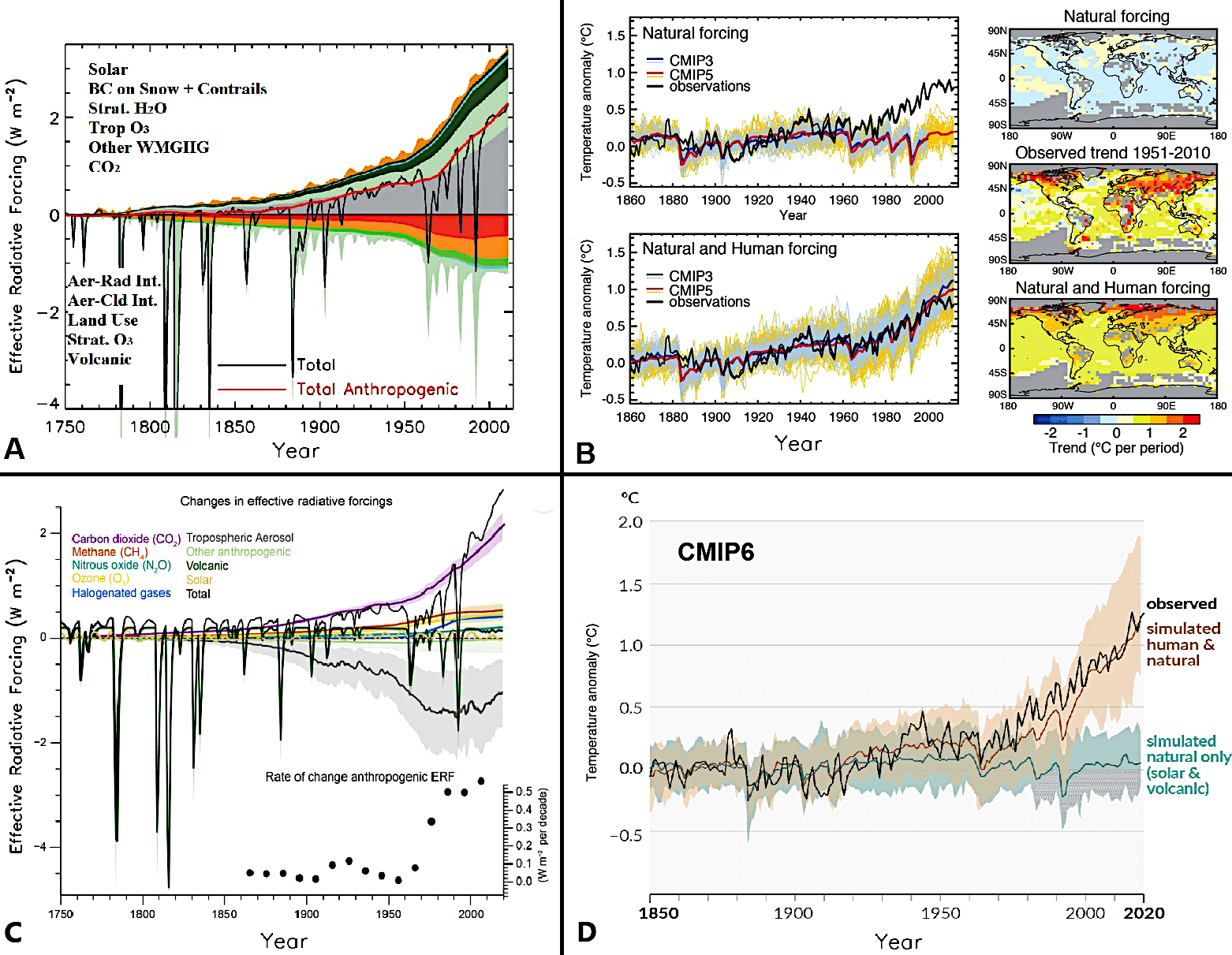}
\par\end{centering}
\caption{(A) Compilation of the radiative forcing functions utilized in the
CMIP5 GCMs \citep[adapted from ][Figure 8.18]{IPCC2013}. (B) Variations
in observed global surface temperature (black) alongside the CMIP3
and CMIP5 model simulations incorporating only natural forcing and
combined natural-anthropogenic forcing \citep[adapted from ][FAQ 10.1, Figure 1]{IPCC2013}.
(C) Compilation of the radiative forcing functions utilized in the
CMIP6 GCMs \citep[adapted from ][Figure 2.10]{IPCC2021}. (D) Observed
global surface temperature variations (black) alongside the CMIP6
model simulations incorporating only natural forcing and combined
natural-anthropogenic forcing \citep[adapted from ][Figure SPM.1]{IPCC2021}.
Notably, in both (B) and (D), the observational data necessary to
validate the GCM predictions that consider only natural forcings are
not reported because they do not exist.}
\label{Fig1}
\end{figure*}

Numerical climate models, commonly referred to as ``\emph{General
Circulation Models}'' or ``\emph{Global Climate Models}'' (GCMs),
are advanced computational tools designed to simulate the dynamics
of the climate system. These models capture interactions between the
ocean, atmosphere, and land, predicting variations in climate parameters
based on imposed radiative forcing functions. Radiative forcing functions
include alterations driven by greenhouse gas concentrations (e.g.,
carbon dioxide {[}CO\textsubscript{2}{]}, methane {[}CH\textsubscript{4}{]}),
atmospheric aerosols (e.g., industrial sulphate aerosols, black carbon),
land-use changes (e.g., deforestation, urbanization), solar radiation
fluctuations, and volcanic activity, which introduces aerosols and
gases into the atmosphere: see Figures \ref{Fig1}A and \ref{Fig1}C.
Each of these factors can perturb the energy balance of the climate
system, resulting in either warming or cooling effects. GCMs simulate
the system's integrated response to the combination of these forcing
functions.

The climate system's response to variations in radiative forcing is
mediated by mechanisms known as climate feedbacks. These feedbacks
may amplify or diminish the initial changes induced by the forcing;
they are categorized as positive (leading to enhanced warming) or
negative (leading to mitigated warming). Key feedback mechanisms include:
\begin{itemize}
\item \emph{Water Vapor Feedback} --- A positive feedback governed by the
Clausius-Clapeyron law, which links ocean evaporation rates to temperature
increases;
\item \emph{Albedo Feedback} --- A positive feedback arising from changes
in surface reflectivity due to ice and snow cover variations;
\item \emph{Cloud Feedback} --- Particularly challenging to quantify, as
cloud formation, type, and distribution are sensitive to warming;
certain clouds cool the surface by reflecting solar radiation, while
others trap emitted heat, making their net contribution highly uncertain;
\item \emph{Lapse Rate Feedback} --- A negative feedback involving modifications
to atmospheric temperature vertical gradients;
\item \emph{Carbon Cycle Feedback} --- Activated by warming-induced CO\textsubscript{2}
release from soils and oceans (per Henry's law), further increasing
atmospheric CO\textsubscript{2} concentrations;
\item \emph{Vegetation Feedback} --- Temperature and precipitation changes
alter vegetation cover, which influences carbon storage and surface
albedo.
\end{itemize}
While all available GCMs indicate that the positive feedbacks surpass
the negative ones thus amplifying the effects of radiative forcing,
large uncertainties associated with crucial feedback mechanisms ---
particularly those related to water vapor and cloud formation ---
remain substantial.

A doubling of atmospheric CO\textsubscript{2} from pre-industrial
levels (280 ppm) to 560 ppm is estimated to increase radiative forcing
by approximately $\Delta F_{2\times CO_{2}}=3.7$ W/m\textsuperscript{2}
at the Earth's surface. Under idealized assumptions --- neglecting
feedbacks and treating Earth as a black body governed by the Stefan-Boltzmann
law --- this forcing would theoretically result in $\sim$1°C warming
upon restoration of energy balance: $\Delta T_{2\times CO_{2}}=\Delta F_{2\times CO_{2}}/4\sigma T^{3}\approx1$°C
\citep{Rahmstorf2008}. However, the Coupled Model Intercomparison
Project Phase 6 (CMIP6) GCMs adopted in AR6 \citep{IPCC2021} estimate
equilibrium climate sensitivity (ECS) --- the projected warming at
equilibrium due to a CO\textsubscript{2} doubling --- ranging from
1.8°C to 5.6°C, reflecting substantial theoretical uncertainty in
the net climate feedback amplification \citep[cf. ][]{Hausfather2022,IPCC2021,Scafetta2024}.

The development of GCMs has been organized under the Coupled Model
Intercomparison Project \citep[accessed on April 10, 2025]{CMIP2024},
a global collaborative initiative aimed at advancing climate modeling
capabilities and understanding climate change processes. Initiated
in 1995 by the World Climate Research Programme's (WCRP) Working Group
on Coupled Modeling (WGCM), CMIP operates in phases:
\begin{itemize}
\item CMIP1 \& CMIP2 --- Developed during the 1990s, comprising 18 models;
\item CMIP3 --- Published in the mid-2000s, including 20 models used for
AR4 \citep{IPCC2007};
\item CMIP5 --- Conducted between 2010 and 2014, integrating 40 models
from over 20 international groups for AR5 \citep{IPCC2013};
\item CMIP6 --- Initiated in 2013 and ongoing, incorporating 55 models
that formed the basis for AR6 \citep{IPCC2021,Eyring2016}.
\end{itemize}
Through CMIP, the scientific community continues to refine the GCMs,
yet uncertainties in feedback processes and ECS highlight the need
for ongoing improvements in climate modeling physical reliability
and accuracy.

\begin{figure*}[!t]
\begin{centering}
\includegraphics[width=1\textwidth]{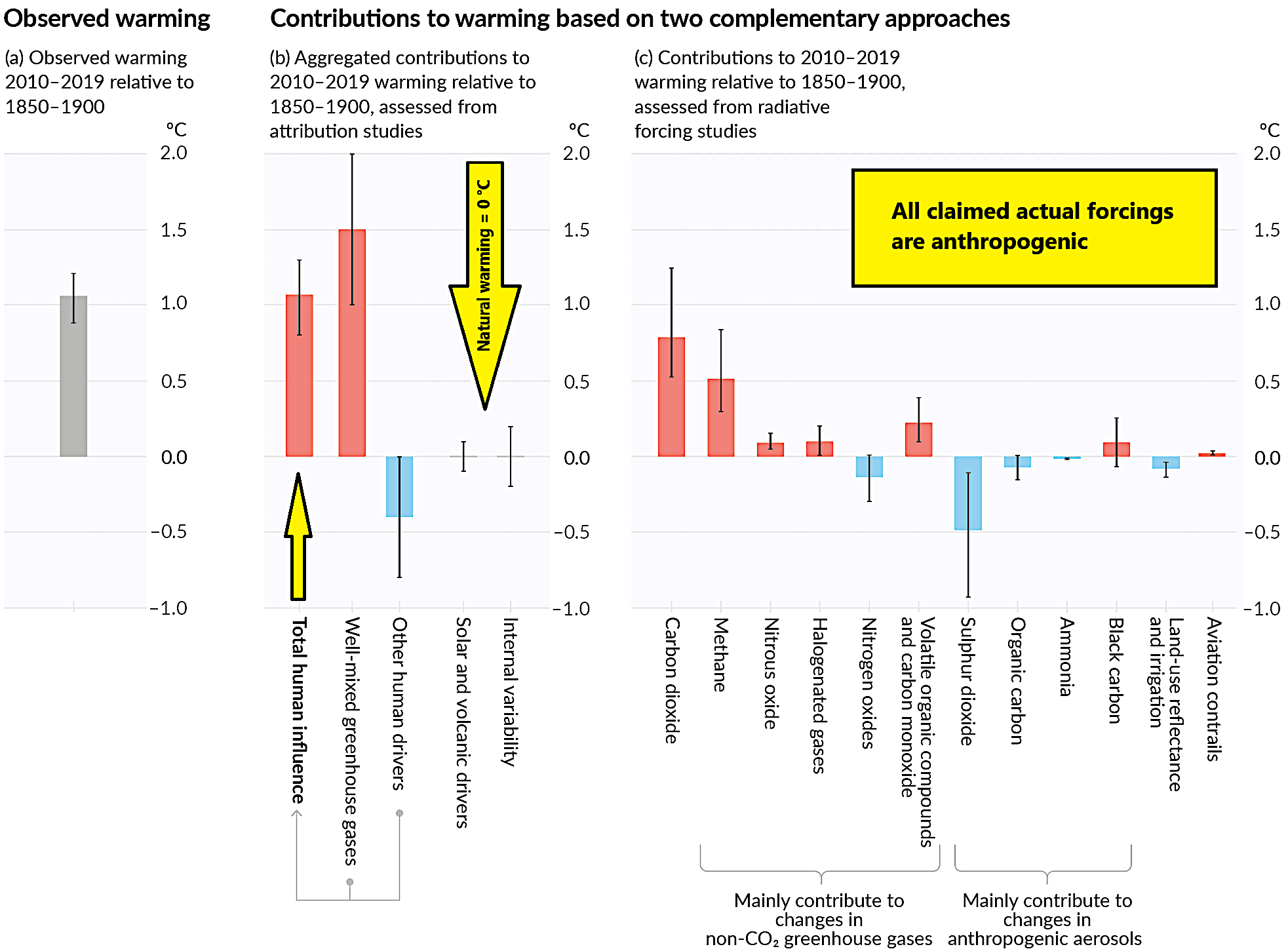}
\par\end{centering}
\caption{Attributions of warming associated with the radiative forcing functions
utilized in the CMIP6 GCMs from 1850--1900 to 2010--2019 \citep[adapted from][Figure SPM.2]{IPCC2021}.
Notably, natural components --- including solar and volcanic drivers,
as well as internal variability --- are estimated to have contributed
an average of 0 °C each. Thus, the observed warming is attributed
to anthropogenic drivers alone.}
\label{Fig2}
\end{figure*}

\subsection{Summary of the GCM results supporting the AGWT}

Figure \ref{Fig1} summarizes the findings of the GCMs developed during
the Coupled Model Intercomparison Project (CMIP) phases 3, 5 and 6.
Figures \ref{Fig1}A and \ref{Fig1}C illustrate the effective radiative
forcing functions utilized in the CMIP5 and CMIP6 GCMs, respectively
\citep{IPCC2013,IPCC2021}, while Figures \ref{Fig1}B and \ref{Fig1}D
present the observed global surface temperature record (black) alongside
the GCM simulations generated under two scenarios: one including only
natural forcings (solar + volcanic) and the other combining both natural
and anthropogenic forcings.

Figures \ref{Fig1}A and \ref{Fig1}C show that the combined contributions
of the solar and volcanic effective radiative forcings are substantially
smaller relative to the cumulative impact of anthropogenic forcings.
Solar forcing exhibits a modest 11-year cycle superimposed on an almost
negligible upward trend, while volcanic forcing is characterized by
sharp negative spikes resulting from major eruptions. These volcanic
spikes induce short-term cooling effects lasting only a few years
\citep{Zanchettin2022}.

The GCM simulations depicted in Figures \ref{Fig1}B and \ref{Fig1}D
strongly support the Anthropogenic Global Warming Theory (AGWT). When
GCMs are driven solely by solar and volcanic forcing functions ---
the only two natural components assumed to influence Earth's climate
--- the numerical simulations fail to produce any substantial warming
or cooling trends throughout the 20\textsuperscript{th} century.
This indicates that, absent anthropogenic contributions, the models
predict a relatively stable climate from the pre-industrial era (1850--1900)
to present. However, by incorporating anthropogenic radiative forcing
functions, the GCMs successfully replicate the observed global surface
warming trend. This comparison implies that, according to these GCMs,
nearly 100\% of the warming recorded since the 1850--1900 period
is attributable to anthropogenic factors, as explicitly concluded
in AR6 \citep{IPCC2021}.

Figure \ref{Fig2} explicitly illustrates the attribution results
derived from the CMIP6 models, showing the average warming contribution
of radiative forcing components from 1850--1900 to 2010--2019 \citep[its figure SPM.2]{IPCC2021}.
According to these GCMs, both natural (solar + volcanic) forcings
and internal climate variability contributed approximately 0°C to
the observed warming. Instead, nearly 100\% of the observed 1.07°C
global surface warming is attributed to anthropogenic forcings, with
well-mixed greenhouse gases contributing $\sim$1.5°C and anthropogenic
aerosols offsetting $\sim$0.4°C of the warming.

\begin{figure*}[!t]
\begin{centering}
\includegraphics[width=1\textwidth]{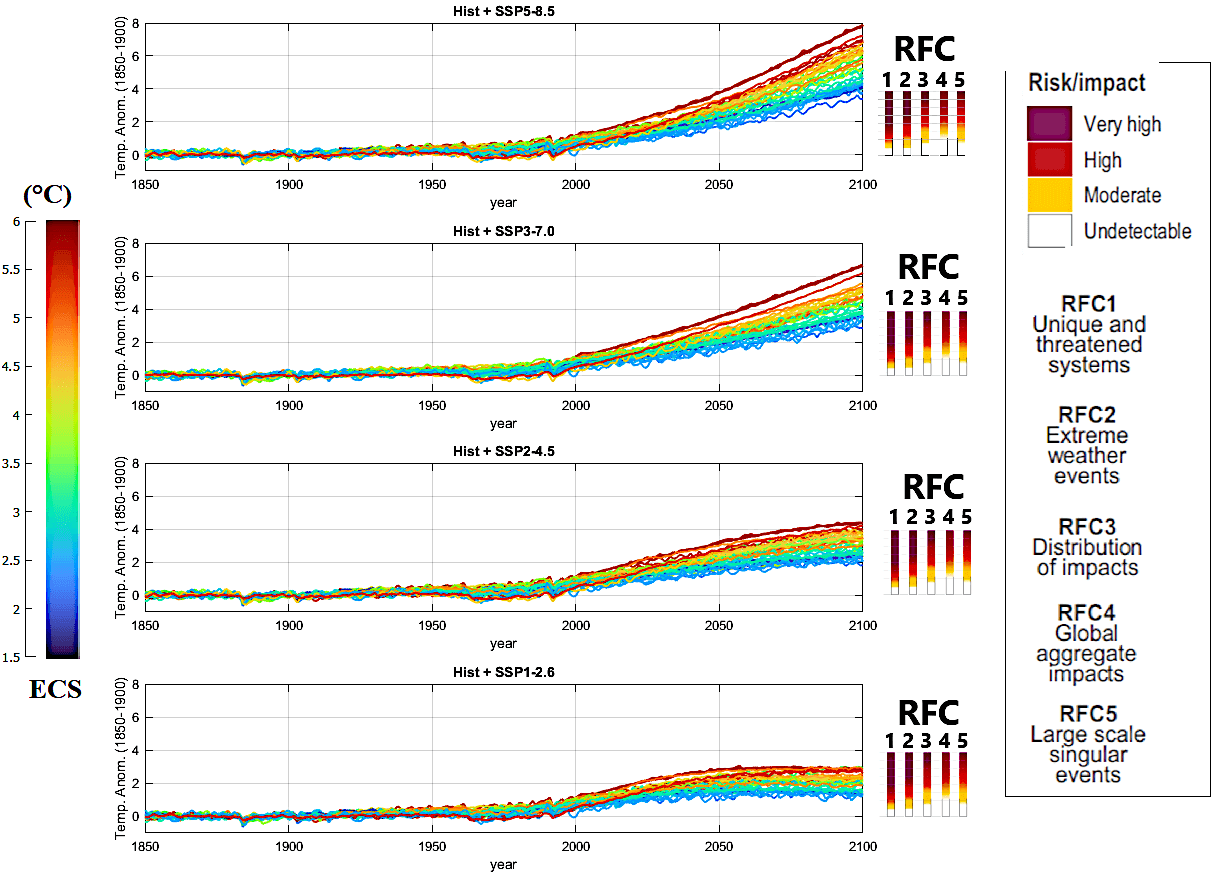}
\par\end{centering}
\caption{CMIP6 GCM ensemble mean simulations spanning from 1850 to 2100, employing
historical effective radiative forcing functions from 1850 to 2014
(see Figure \ref{Fig1}C) and the forcing functions based on the SSP
scenarios 1-2.6, 2-4.5, 3-7.0, and 5-8.5. Curve colors are scaled
according to the equilibrium climate sensitivity (ECS) of the models.
The right panels depict the risks and impacts of climate change in
relation to various global Reasons for Concern (RFCs) \citep{IPCC2023}.
\citep[Adapted from][]{Scafetta2024}.}
\label{Fig3}
\end{figure*}

The CMIP6 GCMs are also employed to simulate future climate scenarios
based on hypothetical radiative forcing functions derived from Shared
Socioeconomic Pathways (SSPs). The ones mainly adopted in the IPCC
AR6 are:
\begin{itemize}
\item SSP1-2.6 --- low greenhouse gas emissions, with robust adaptation
and mitigation measures leading to Net-Zero CO\textsubscript{2} emissions
between 2050--2075;
\item SSP2-4.5 --- intermediate emissions, where CO\textsubscript{2} levels
remain near current levels until 2050 and subsequently decline without
achieving Net-Zero by 2100;
\item SSP3-7.0 --- high emissions, with CO\textsubscript{2} concentrations
doubling by 2100 under minimal policy intervention;
\item SSP5-8.5 --- very high emissions, with CO\textsubscript{2} levels
tripling by 2075 under a worst-case scenario devoid of mitigation
measures.
\end{itemize}
Figure \ref{Fig3} shows the ensemble mean simulations of the CMIP6
GCMs spanning 1850--2100. These simulations adopt historical radiative
forcing functions from 1850--2014 (as illustrated in Figure \ref{Fig1}C)
and SSP-derived forcing functions for 2015--2100. The color-coded
curves correspond to the Equilibrium Climate Sensitivity (ECS) values
of the models. Results indicate that the global surface temperature
could rise significantly --- potentially up to 8°C relative to the
1850--1900 reference period --- if SSP5-8.5 is implemented and the
ECS value is high (5--6°C).

The right panels of Figure \ref{Fig3} illustrate the projected impacts
and risks of climate change associated with various global Reasons
for Concern (RFCs). These projections suggest that only SSP1-2.6,
which employs aggressive mitigation and Net-Zero policies, can limit
the 21\textsuperscript{st}-century warming to below 2°C relative
to pre-industrial levels, which is the threshold identified as critical
under the \citet{ParisAgreement}.

\subsection{Conclusion}

Over the span of approximately three decades, from the publication
of the First Assessment Report \citep[FAR,][]{IPCC1990} to the Sixth
Assessment Report \citep[AR6,][]{IPCC2021}, the Intergovernmental
Panel on Climate Change (IPCC) has significantly advanced its understanding
of the role of anthropogenic emissions in driving global warming.
In the 1990s the IPCC posited that both natural mechanisms and human
activities could have contributed roughly equally ($\sim$50\% each)
to the observed warming of the 20\textsuperscript{th} century. However,
since the years 2000s the prevailing scientific opinion has shifted,
and the \citet[AR6,][]{IPCC2021} now asserts that human activities
are almost exclusively responsible ($\sim$100\%) for the global warming
and climate change observed from 1850--1900 to 2011--2020.

The most recent assessment reports \citet{IPCC2021,IPCC2023} underscore
this conclusion with striking clarity. As shown in Figure \ref{Fig2},
the average contribution of natural factors --- solar and volcanic
forcing and internal natural variability --- to global warming during
the aforementioned period is estimated to be approximately 0°C. Consequently,
from the CMIP GCM perspective, concerns about future climate warming
due to additional anthropogenic greenhouse gas (GHG) emissions are
well-founded. However, this conclusion depends on the reliability
of global surface temperature records and the robustness of the physical
science underpinning global climate models (GCMs).

Under the worst-case scenario, the SSP5-8.5, which assumes very high
and accelerating GHG emissions without mitigation, the GCM projections
indicate alarming potential warming of 4°C to 8°C by 2080--2100.
Despite these concerning projections, the \citet[AR6,][its table 12.12, p. 1856]{IPCC2021}
expresses low confidence in attributing changes in the frequency,
severity, or extent of several climate impact drivers, such as: river
floods, frost, landslides, hydrological drought, drought, agricultural
and ecological drought, fire weather, mean wind speed, severe wind
storms, tropical cyclones, sand and dust storms, heavy snow and ice
storms, hail, snow avalanches, coastal erosion, coastal flooding,
marine heat waves, air pollution, and surface radiation. Conversely,
AR6 identifies moderate-to-high confidence in climate impact drivers
more directly linked to increasing atmospheric CO\textsubscript{2}
concentrations and global warming. These include: increases in mean
air temperature, extreme heat events, sea level, mean ocean temperature,
ocean acidity, and ocean salinity; decreases in cold spells, snow
cover, glaciers, ice sheets, permafrost, lake ice, river ice, sea
ice, and dissolved oxygen; regional variability in mean precipitation
(increasing in some regions and decreasing in others).

This evolution of understanding underscores the complexity of the
climate system and the importance of continuously refining projections
through improved climate modeling and robust observational datasets.

\section{Main challenges to the AGWT}

Scientific critiques of the Anthropogenic Global Warming Theory (AGWT)
emerge from diverse perspectives, reflecting significant concerns
about persistent physical uncertainties and the potential validity
of alternative explanations for climate change.

\subsection{Erroneous or unsatisfactory critiques}

First of all, it is crucial to differentiate scientifically substantiated
critiques from various misguided claims often circulating online,
which tend to oversimplify the physics of climate change. One such
incorrect argument asserts that additional CO\textsubscript{2} emissions
cannot affect the climate due to a supposed saturation of the infrared
(IR) absorption bands of such gas. This claim is demonstrably inaccurate,
as the CO\textsubscript{2} IR absorption bands have not reached their
maximum capacity yet. This fact is also acknowledged by \citet[their figure 7]{vanWijngaarden2023}
--- two noted skeptics of climate change alarmism --- who calculated
that doubling CO\textsubscript{2} concentrations from 400 to 800
ppm would reduce the upward IR heat flux by approximately 3 W/m\textsuperscript{2}
at the mesopause altitude (86 km) under standard atmospheric conditions
without clouds, assuming a surface temperature of 288.7 K. This result
is slightly lower and does not differ significantly from what already
reported in the scientific literature and in the IPCC reports, and
can be easily verified using the MODTRAN Infrared Light in the Atmosphere
model (available online at \href{https://climatemodels.uchicago.edu/modtran/}{https://climatemodels.uchicago.edu/modtran/},
accessed on April 10, 2025), which also estimates a downward IR heat
flux at the surface of about 3.6 W/m\textsuperscript{2} under identical
physical conditions.

In fact, the CO\textsubscript{2} infrared (IR) absorption bands are
quasi-saturated, leading to a logarithmic relationship between CO\textsubscript{2}
concentration and radiative forcing. For instance, a doubling of atmospheric
CO\textsubscript{2} concentration from 400 to 800 ppm results in
a typically-estimated increase in radiative forcing of approximately
3.7 W/m\textsuperscript{2}, yielding to a theoretical radiative forcing
function of the type: $\Delta F_{CO_{2}}\approx3.7\log_{2}(P_{CO_{2}}/400)$
W/m\textsuperscript{2} with a 10\% uncertainty \citep{Myhre1998,Etminan};
here $P_{CO_{2}}$ is the CO\textsubscript{2} atmospheric concentration.
This evidence directly contradicts the saturation hypothesis that
incorrectly infers that any increase in CO\textsubscript{2} levels
would result in a radiative forcing increase of 0.0 W/m\textsuperscript{2}.
As already explained in Section 2.2, the CMIP6 GCMs predict that a
radiative forcing of 3.7 W/m\textsuperscript{2} could induce, at
equilibrium, a warming from about 1.8 to 5.7°C, depending on the modeled
net feedback strength.

It is important to realize that the principal scientific challenge
lies not in debating the CO\textsubscript{2} saturation argument
but rather in accurately determining the climate sensitivity to radiative
forcings that depends on the sign and magnitude of the climate feedback,
the actual forcings of the climate system, and the quantitative detection
of climate change. Addressing these complex scientific issues requires
a detailed understanding of the climate system, of its astronomical
environment and of its feedback mechanisms, domains that remain highly
uncertain and form a critical focus of ongoing research, as elaborated
below.

\begin{figure*}[!t]
\raggedright{}{\bfseries{}%
\noindent\fbox{\begin{minipage}[t]{1\textwidth - 2\fboxsep - 2\fboxrule}%
\begin{flushleft}
\textbf{Appendix B.}
\par\end{flushleft}
\begin{flushleft}
\textbf{In relation to Section 3.1, the figure below illustrates the
infrared light spectra obtained using MODTRAN (available at the web
page: \href{https://climatemodels.uchicago.edu/modtran/}{https://climatemodels.uchicago.edu/modtran/})
in a standard atmosphere under the physical conditions shown in the
right-hand inset. The MODTRAN model simulates the emission and absorption
of infrared radiation in the atmosphere. }
\par\end{flushleft}
\begin{flushleft}
\textbf{The curves represent theoretical emission spectra at different
atmospheric CO\textsubscript{2} concentrations: (green) 0 ppm, (blue)
400 ppm, and (red) 800 ppm. }
\par\end{flushleft}
\begin{flushleft}
\textbf{The top-left panel displays the infrared light spectra observed
at the surface of the atmosphere (0 km) looking upward. The top-right
panel shows the spectra at the top of the atmosphere (100 km) looking
downward; this figure closely aligns with the results obtained by
\citet{vanWijngaarden2023}. }
\par\end{flushleft}
\begin{flushleft}
\textbf{The total energy flux from all infrared light is labeled as
Upward IR Heat Flux, $F$, measured in units of $W/m^{2}$. The quantities
$\Delta F$ in the upper inserts indicate the variation in radiative
forcing relative to the baseline value at 400 ppm. }
\par\end{flushleft}
\begin{flushleft}
\textbf{The bottom panel illustrates how radiative forcing changes
as a function of CO\textsubscript{2} concentration relative to the
pre-industrial level of 280 ppm. The blue curve represents $\Delta F$
values obtained with MODTRAN, while the red curve corresponds to the
traditionally used approximation function $\Delta F(\mathrm{CO_{2}})=3.7\log_{2}(\mathrm{CO_{2}}/280)$. }
\par\end{flushleft}
\begin{flushleft}
\textbf{The model effectively demonstrates the influence of wavelength-selective
greenhouse gases on Earth's outgoing infrared energy flux, indicating
that the spectral bands of the CO\textsubscript{2} present in the
atmosphere are not fully saturated yet.}\\
\par\end{flushleft}
\begin{center}
\includegraphics[width=1\textwidth]{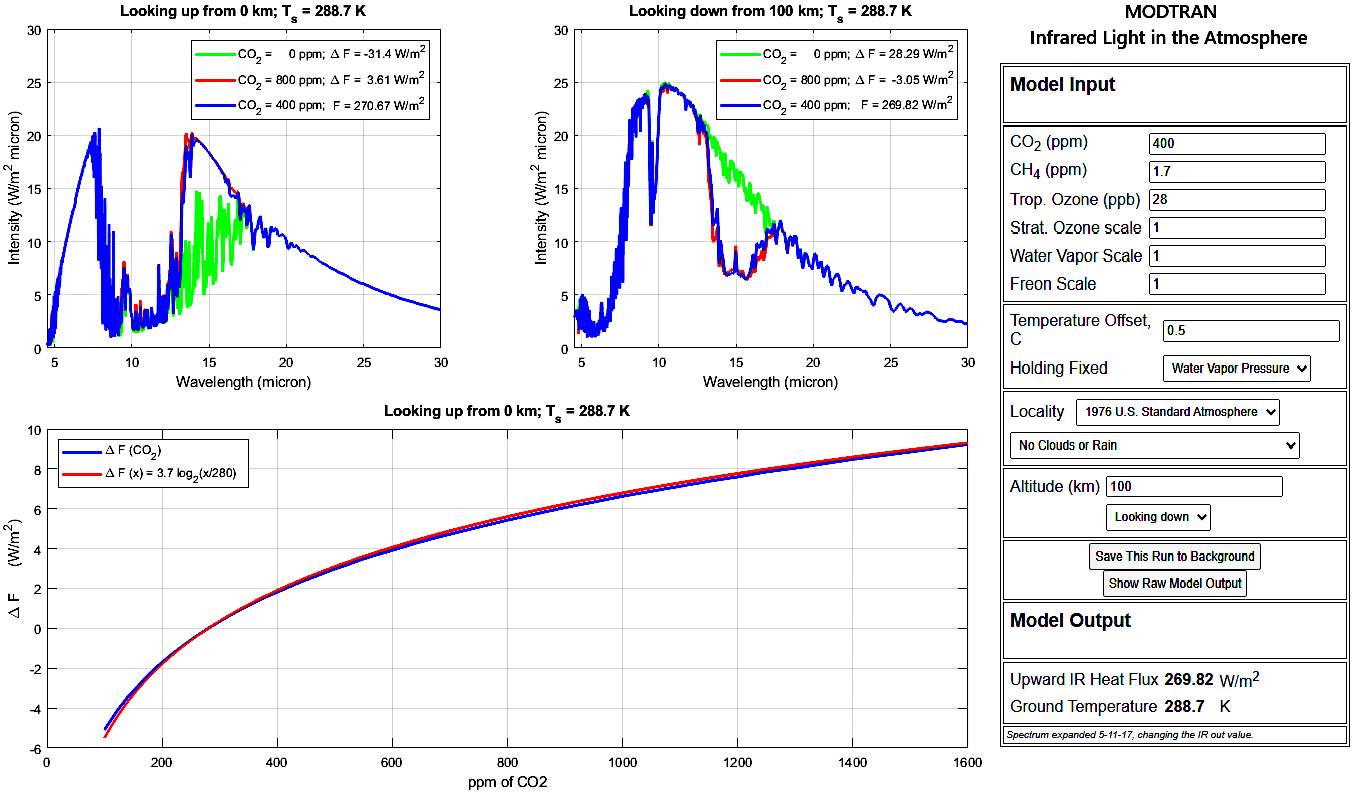}
\par\end{center}%
\end{minipage}}}
\end{figure*}

\citet{vanWijngaarden2023} expressed skepticism only toward climate
change alarmism by arguing that the net climate feedback should predominantly
be negative. This is a legitimate physical expectation that the authors
supported by citing the findings of \citet{Lindzen2001} and \citet{Lindzen2011}
obtained by comparing high-frequency deseasonalized fluctuations in
sea surface temperatures and the concurrent fluctuations in top-of-atmosphere
(TOA) outgoing radiation using ERBE and CERES satellite data from
1985 to 2008. However, \citet{vanWijngaarden2023} specifically invoked
also Le Chatelier's Principle --- which states that when a system
at equilibrium experiences a disturbance, it will adjust to counteract
the disturbance and establish a new equilibrium --- to suggest that
the climate system should inherently resists radiative perturbations.
Citing Le Chatelier's Principle to conclude that the net climate feedback
should be strictly negative appears to oversimplify the complexity
of the climate system, which incorporates both negative feedbacks
(such as increased cloud cover reflecting sunlight) and positive feedbacks
(like ice-albedo effects, where melting ice reduces reflectivity and
accelerates warming). It is the interplay between these positive and
negative feedbacks that ultimately determines the thermodynamic conditions
of the new equilibrium of the climate system. The new equilibrium
temperature will depend on the prevailing feedback mechanisms; whether
the net feedback is negative or positive just determines whether the
resulting temperature deviates downward or upward from the level expected
solely due to increased atmospheric CO\textsubscript{2}, which is
$\sim$1°C for CO\textsubscript{2} doubling \citep{Rahmstorf2008}.
Thus, in this regard, \citet{vanWijngaarden2023} appear to express
only an opinion against the possibility of a catastrophic runaway
greenhouse effect scenario where Le Chatelier’s Principle is not satisfied
because the new equilibrium is never reached. Such a catastrophic
scenario did not occur even during the Cambrian Period (\textasciitilde 540--485
million years ago) when the Earth's temperature was much higher than
today and atmospheric CO\textsubscript{2} levels were estimated to
be 4,000--5,000 ppm \citep{Royer2001}, that is ten times larger
than current levels. In this regard, \citet{vanWijngaarden2023} do
not appear to have actually developed a solid critique of the AGWT
advocated by the IPCC, which is perfectly compatible with Le Chatelier’s
Principle. Indeed, none of the CMIP GCMs used in IPCC reports predicts
a catastrophic runaway greenhouse effect scenario. Only critiques
grounded in empirical observations could offer a solid foundation
for challenging the AGWT.

In the following, Section 3 provides a summary of the primary scientific
critiques of the AGWT outlined in Section 2. It further sets the stage
for the exploration of alternative interpretations of climate change,
discussed in Sections 4 and 5. Together, these critiques underscore
the complexity of the debates surrounding the causes and consequences
of climate change and highlight the range of scientific opinions in
this evolving research field.

\subsection{The concept of \textquotedblleft\emph{consensus}\textquotedblright{}
in the climate change debate}

The concept of ``\emph{consensus}'' holds political significance
but is less meaningful within scientific discourse, where the core
principle is the verifiability of results through empirical evidence.
Yet, in the realm of climate change, an appeal to consensus is often
employed to bolster the validity of the Anthropogenic Global Warming
Theory (AGWT), its associated alarmist claims and Net-Zero policies.

For instance, assumptions that climate change science is settled have
led editors of Earth science and health journals to advocate for urgent
global action on climate change \citep{McNutt2015}, assessing that
``\emph{our planet is in crisis}'' \citep{Filippelli2021} and warning
that unmitigated climate change poses the 21\textsuperscript{st}-century
``\emph{greatest threat to global public health}'' \citep{Atwoli2021}.
These appeals align with the commitments under the Paris Agreement
adopted at the United Nations Conference of the Parties (COP21) in
2015 \citep{ParisAgreement}.

Despite claims of scientific consensus, skepticism persists among
scientists and the public due to unresolved uncertainties and concerns
surrounding these assertions. Studies suggesting overwhelming agreement
among climate scientists --- often citing figures such as 97\% or
even 99\% \citep{Cook2013,Lynas2021} --- are frequently interpreted
as evidence of consensus on AGWT. However, critics \citep[e.g.:][]{Montford2013,Legates2015}
challenge such conclusions, pointing to statistical flaws and misleading
interpretations, which rely on selective datasets and imprecise survey
methodologies.

\begin{figure*}[!t]
\begin{centering}
\includegraphics[width=1\textwidth]{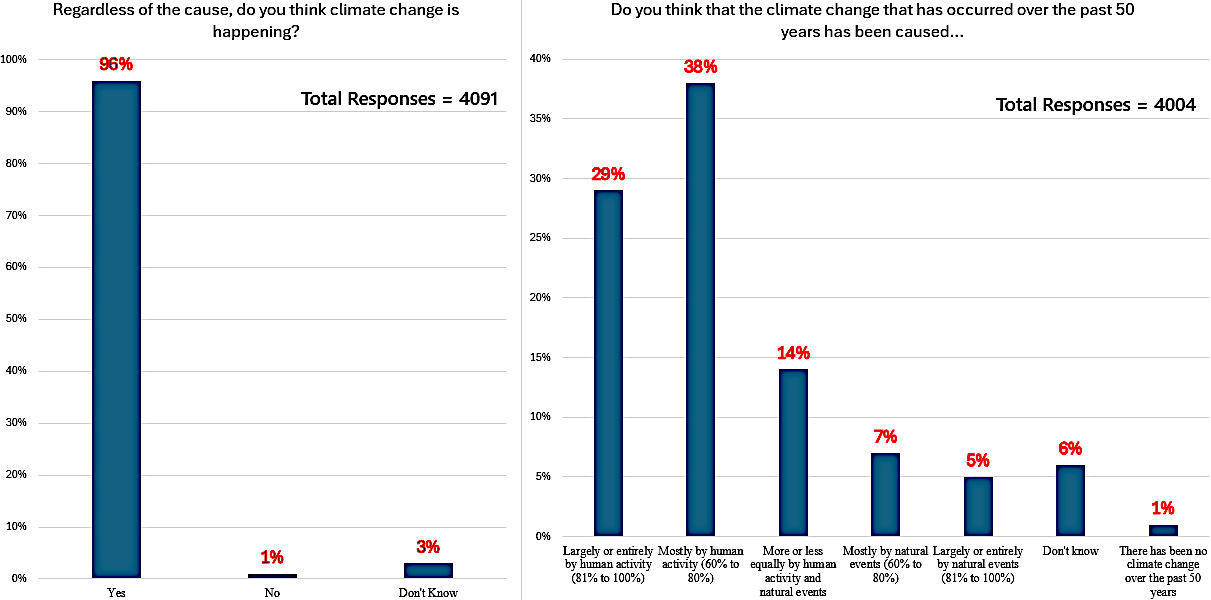}
\par\end{centering}
\caption{Survey conducted by the American Meteorological Society among its
members revealing a wide range of opinions on the causes of climate
change \citep[adapted from][pp. 5 and 8]{Maibach2016}. Only 29\%
of respondents agreed that global warming was predominantly or entirely
(81\% to 100\%) attributable to human activity --- a stance even
more moderate than the definitive 100\% attribution claimed by the
\citet[AR6,][]{IPCC2021,IPCC2023}, as shown in Figures \ref{Fig1}
and \ref{Fig2}.}
\label{Fig4}
\end{figure*}

A case in point is the ambiguous nature of the statement ``\emph{humans
are causing global warming}'' \citep{Cook2013}. This statement can
be interpreted to mean that nearly all observed warming ($\sim$100\%)
is due to anthropogenic activities, as asserted by the AGWT and explicitly
claimed by the \citet[AR6,][]{IPCC2021}. However, the same statement
could also mean that only a portion (\textasciitilde 50\%) of the
warming is attributable to human influence, reflecting earlier IPCC
reports such as FAR and SAR \citeyearpar{IPCC1990,IPCC1995}. Reliable
surveys must, therefore, distinguish between these divergent attribution
scenarios, as their implications for climate science and assessments
of future risks differ significantly.

When surveys adopt clearer attribution distinctions, the diversity
of scientific opinions on climate change becomes evident. Broadly,
there is general agreement among scientists on qualitative aspects
such as that global warming has occurred since 1850--1900, the greenhouse
effect is real, anthropogenic emissions have exacerbated it, and this
has contributed to warming over the past century. However, there is
no consensus on the precise quantitative attribution of natural versus
anthropogenic factors to the observed warming.

For example, Figure \ref{Fig4} summarizes a national survey by the
American Meteorological Society revealing a wide range of opinions
among respondents about the causes of climate change \citep[pp. 5 and 8]{Maibach2016}.
While 96\% agreed that climate change is occurring, only 29\% attributed
it mainly or entirely to human activities (81\% to 100\%), while 38\%
attributed it mostly to human activities (60\% to 80\%). Fourteen
percent attributed climate change equally to human and natural causes
(40\% to 60\%), and smaller percentages leaned toward predominantly
natural drivers or expressed uncertainty. This means that at least
71\% of the respondents disagree with the IPCC AR6's claim that humans
are responsible for 100\% of the observed global warming. Additionally,
two-thirds of respondents believed that aggressive mitigation policies
would only moderately or slightly reduce future warming \citep[p. 9]{Maibach2016}.
This diversity of opinions is not unexpected because it well aligns
with the range of views documented in the IPCC assessment reports
from FAR \citeyearpar{IPCC1990} to AR6 \citeyearpar{IPCC2021}.

\begin{figure*}[!t]
\begin{centering}
\includegraphics[width=0.9\textwidth]{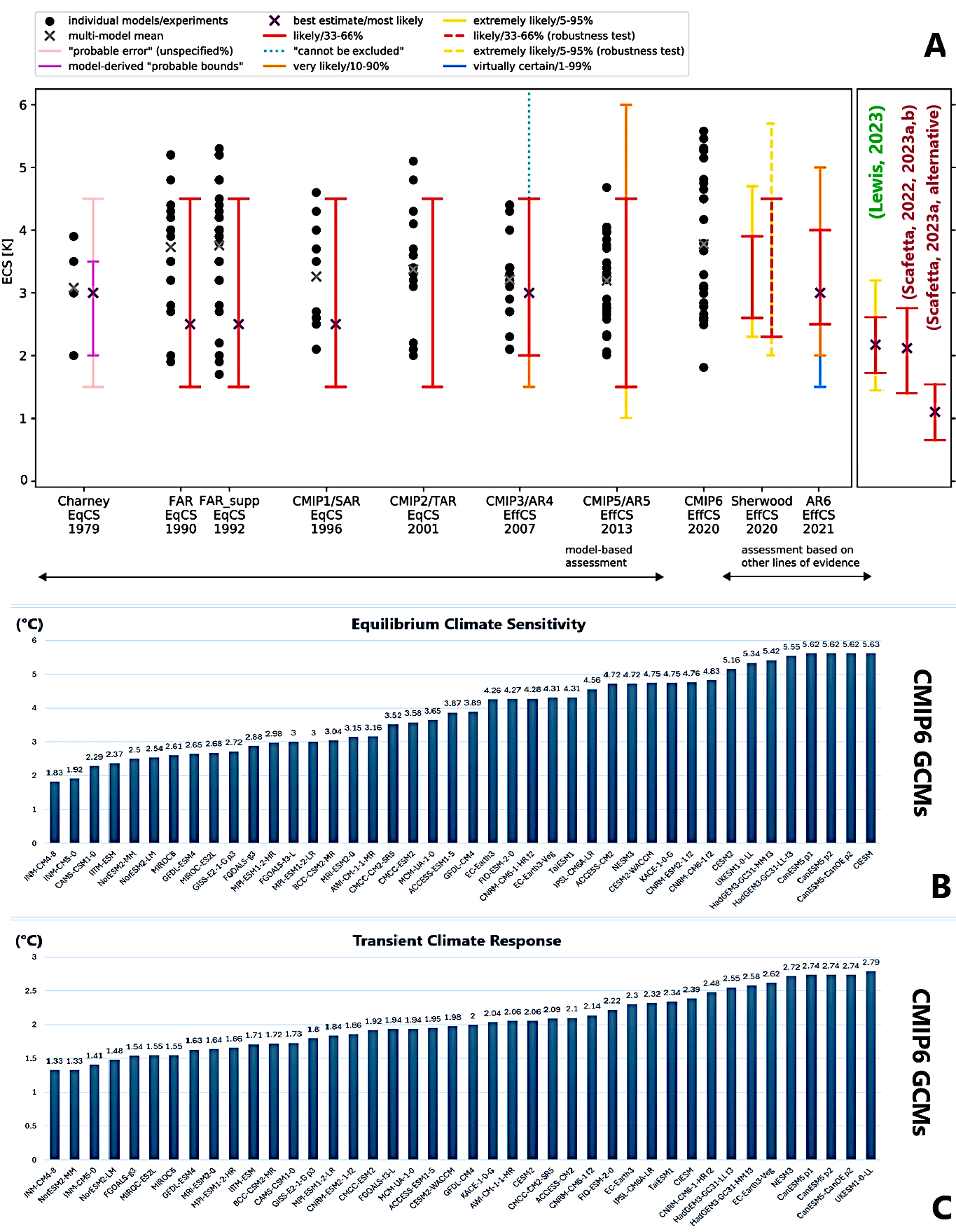}
\par\end{centering}
\caption{{[}A{]} Historical progression of the equilibrium climate sensitivity
(ECS) estimates from \citet{Charney1979} to AR6 \citep{IPCC2021}
\citep[adapted from][]{Undorf2022}, with the addition of the recent
estimates from \citet{Lewis2023} and \citet{Scafetta2022a,Scafetta2023a,Scafetta2023b}.
The lowest estimate --- ECS $=1.1\pm0.4$ °C --- assumes that the
climate system is hypersensitive to solar activity variations; see
Sections 4.2.3 and 4.3.2. {[}B{]} The equilibrium climate sensitivity
(ECS) and {[}C{]} the transient climate response (TCR) of the CMIP6
GCMs \citep{IPCC2021,Hausfather2022,Scafetta2024}.}
\label{Fig5}
\end{figure*}

Inded, although the IPCC findings have evolved to assert in AR6 that
anthropogenic factors are the sole cause of climate change since 1850--1900,
shifts in scientific conclusions do not necessarily reflect genuine
progress, which requires reducing underlying uncertainties. In science,
new findings that appear to challenge earlier results must rigorously
justify the invalidation of previous conclusions, not just propose
an alternative viewpoint. However, in geosciences, conflicting studies
often coexist, with earlier findings persisting despite newer apparently
contradictory research.

One prominent example of persistent uncertainty in climate science
is the assessment of the equilibrium climate sensitivity (ECS), a
crucial metric for gauging the response of the climate system to radiative
forcing. From \citet{Charney1979} to AR6 \citep{IPCC2021}, the ECS
uncertainty range has remained relatively unchanged --- if not broadened
\citep{Undorf2022} --- as illustrated in Figure \ref{Fig5}A. This
persistent uncertainty underscores unresolved methodological and scientific
challenges.

Consensus-based arguments also risk logical fallacies, such as appeals
to popularity or authority. In fact, many individuals, including researchers
in climate-related fields, may accept the IPCC conclusions without
critically engaging with the uncertainties highlighted within the
IPCC's own reports. Consequently, reliance on such forms of ``\emph{consensus}''
does little to resolve the crux of the climate debate regarding whether
human activity accounts for 100\% of observed warming between 1850--1900
and 2011--2020, as asserted by AR6 \citep{IPCC2021}, or whether
a more moderate contribution ($\sim$50\%) might better align with
actual scientific findings. If the latter opinion holds, the projected
21\textsuperscript{st}-century warming based on the CMIP5 and CMIP6
GCMs would be substantially lower, mitigating assessments of future
climate change risks and hazards \citep{Scafetta2013,Scafetta2024}.

The following subsections delve into key scientific open issues that
question the reliability of the GCMs and directly challenge the AGWT.
These open key issues continue to be debated in the scientific literature.

\subsection{The impossibility of testing the main prediction of the GCMs}

The scientific method necessitates that the hypotheses, including
physical models, are rigorously tested to ensure that their predictions
align with empirical evidence. When evidence supports a hypothesis,
the model may be retained; however, when results challenge it, the
model must be discarded or revised, requiring the formulation of a
new hypothesis or model.

As discussed in Section 2 and illustrated in Figures \ref{Fig1}B
and \ref{Fig1}D, the claim that human activity accounts for approximately
100\% of the observed warming from 1850--1900 to 2011--2020 is derived
solely from the climate simulations of the existing GCMs. These simulations
rely on specific radiative forcing functions, which are assumed to
be complete and accurate. However, the only tangible evidence is that
there is no data fully confirming the main GCM prediction. More specifically,
no data exists to demonstrate that, absent anthropogenic forcing,
the Earth's climate would have remained stable from 1850--1900 to
the present. Thus, such critical prediction of the GCMs cannot be
empirically validated, as there is no ``\emph{twin Earth}'' devoid
of human influence from which to obtain the necessary data. Consequently,
the absence of such data --- as also reflected in Figures \ref{Fig1}B
and \ref{Fig1}D --- highlights the key limitation in the validation
of these models.

Therefore, the AGWT remains a hypothesis contingent upon the reliability
of the current GCMs, which can only be ``assumed'' but not ``proven''
correct. The fact that the primary prediction of these models has
neither been experimentally validated nor is it likely to be so in
the future, leaves open the possibility that the present GCMs and
their climate attribution assessments (as depicted in Figures \ref{Fig1}
and \ref{Fig2}) may be fundamentally flawed.

While certain qualitative arguments, supported by paleoclimatic evidence
(e.g., the hockey-stick temperature reconstruction by \citealp{Mann1999})
suggest that the current warm period surpasses past warm periods such
as the Medieval Warm Period on a global scale, such evidence does
not confirm the validity of the AGWT's climate attribution assessments
proposed in the latest IPCC assessment report \citeyearpar[AR6,][]{IPCC2021}.
Hockey-stick reconstructions, as \citet{Crowley2000} posited, may
“\emph{enhance confidence}” that anthropogenic forcings significantly
contributed to current warming but cannot definitively prove they
are its sole ($\sim$100\%) cause. \citet{Crowley2000}, for example,
used hockey-stick temperature records and still estimated that ``\emph{about
25\% of the 20\textsuperscript{th}-century temperature increase can
be attributed to natural variability}'', which was primarily due
to increased solar activity during the 19\textsuperscript{th} and
20\textsuperscript{th} centuries.

Importantly, the GCMs themselves may be based on assumptions, mechanisms,
and specific forcings that could inadequately capture the complexity
of the Earth's climate system. Such limitations can result in overestimated
projections of anthropogenic warming and its impacts. For instance,
natural processes like solar forcing and internal climate variability
may not be accurately represented in the GCMs, a fact that could have
yielded to potentially underestimate their contributions to observed
climate variability. Indeed, natural processes may have had a larger
influence on climate variability than the 0°C warming predicted by
the CMIP6 models (Figure \ref{Fig2}).

Additionally, many GCM mechanisms depend on parameterization, calibration,
and tuning of free parameters. These parameters are adjusted to enable
the GCMs to approximate the observed climate system as closely as
possible \citep{McSweeney2018}. Tuning involves iterative tests to
identify parameter values that produce optimal agreement between model
outputs and observational data. However, such tuning does not guarantee
that the models are physically accurate. Parameter values can vary
significantly when model mechanisms and forcings are altered, as demonstrated
in various studies \citep[e.g.:][]{Golaz2019,Ma2022,Mauritsen2019}.
Thus, the apparent agreement between the GCM predictions and observed
warming from 1850--1900 to 2011--2020 under specific radiative forcing
functions, as shown in Figure \ref{Fig1}, may stem more from tuning
methodologies rather than from intrinsic physical accuracy.

\subsection{The GCMs' failure in capturing natural climate variability}

In scientific literature and in the same IPCC assessment reports there
is growing evidence of physical uncertainty that raises significant
questions about the reliability of the GCMs. A key challenge lies
in the inherent difficulty of rigorously validating these models as
also argued above. This section explores some of the most pressing
open questions currently under debate, including the pervasive issue
of the GCMs' inability to reconstruct major patterns in natural climate
variability.

\subsubsection{The \textquotedblleft\emph{hot model}\textquotedblright{} problem}

As shown in Figure \ref{Fig5}, the CMIP6 GCMs predict a wide and
inconsistent range of equilibrium climate sensitivity (ECS) and transient
climate response (TCR) values, spanning from 1.8°C to 5.6 °C and from
1.33°C to 2.79°C, respectively \citep[cf.][Figure 7.18]{Hausfather2022,IPCC2021}.
This large uncertainty range presents significant challenges to accurately
evaluating the risks and potential hazards of future climate change,
as suggested in Figure \ref{Fig3} \citep[cf.][]{Scafetta2024}.

The ECS uncertainty is attributed to variations in the mechanisms,
parameterizations, and free parameter settings employed by different
GCMs. These differences significantly influence the modeling of key
climate feedbacks, such as water vapor and cloud formation \citep{Lutsko2022}.
The discrepancies among GCM projections become especially pronounced
outside the historical period, where internal free parameter tuning
is not possible. For instance, under the same Shared Socioeconomic
Pathways (SSPs), the GCMs display divergent climate projections for
the 21\textsuperscript{st} century, as illustrated in Figure \ref{Fig3}
with curves color-coded by ECS values ranging from low {[}blue{]}
to high {[}red{]} values.

A major unresolved question in climate science is how to constrain
the ECS and TCR values to better approximate the true sensitivity
and response times of the climate system. While the GCMs estimate
ECS to fall within 1.8°C to 5.6°C, broader literature reports ranges
as wide as 0.5°C to 10°C \citep{Knutti2017}. \citet[AR6,][]{IPCC2021}
proposed a likely ECS range from 2.5°C to 4.0°C (17--83\% probability)
and a very likely range from 2.0°C to 5.0°C (5--95\% probability),
based on the assessment by \citet{Sherwood2020}, who used multiple
lines of evidence by emerging constraints and paleoclimate data.

However, such assessment remains contentious. For example, the reanalysis
of the same line of evidences proposed by \citet{Lewis2023} directly
challenged the assessment by \citet{Sherwood2020} by highlighting
a number of possible statistical errors and other shortcomings. \citet{Lewis2023}
estimated that the same evidence could suggest a significantly lower
ECS range, with a median of 2.16°C (17--83\% range: 1.75--2.7°C;
5--95\% range: 1.55--3.2°C). Other studies propose even lower values
centered between 0.5°C to 2°C \citep{Forster2006,Lindzen2011,Harde2014,Monckton2015}.
Similar low ECS estimates (ECS < 3.0°C) have been reported by \citet{Scafetta2013,Scafetta2022a,Scafetta2023a,Scafetta2023b,Scafetta2023c}.
See Figure \ref{Fig5}A. In general, \citet{Gervais2016}, \citet{Knutti2017}
and \citet{Rugenstein2023} highlighted that already in the early
2010s, several authors noted that climate sensitivity estimates from
climate models differed significantly from those based on observed
warming and radiative balance studies, with the latter suggesting
much lower values. Attempts to solve this conundrum --- for example
by assessing whether ECS may be variable under different time-scales
and/or physical conditions \citep{Rugenstein2023} --- are under
investigation.

ECS and TCR values are critical for understanding the climate's sensitivity
to radiative forcing and for estimating expected future warming under
different SSP scenarios (as shown in Figure \ref{Fig3}). Notably,
the GCM-derived ECS and TCR estimates cannot be treated as repeated
measurements of a single observable but rather as diverse theoretical
estimates. Consequently, the true ECS and TCR values may fall either
within or outside the ranges reported by the GCMs. For example, the
actual ECS and TCR values could potentially be lower than those predicted
by any GCM, as suggested in multiple studies \citep[cf.:][]{Knutti2017,Harde2014,Lewis2023,Lindzen2011,Monckton2015,Scafetta2023a,Scafetta2023b,Scafetta2024}.

\begin{figure*}[!t]
\begin{centering}
\includegraphics[width=1\textwidth]{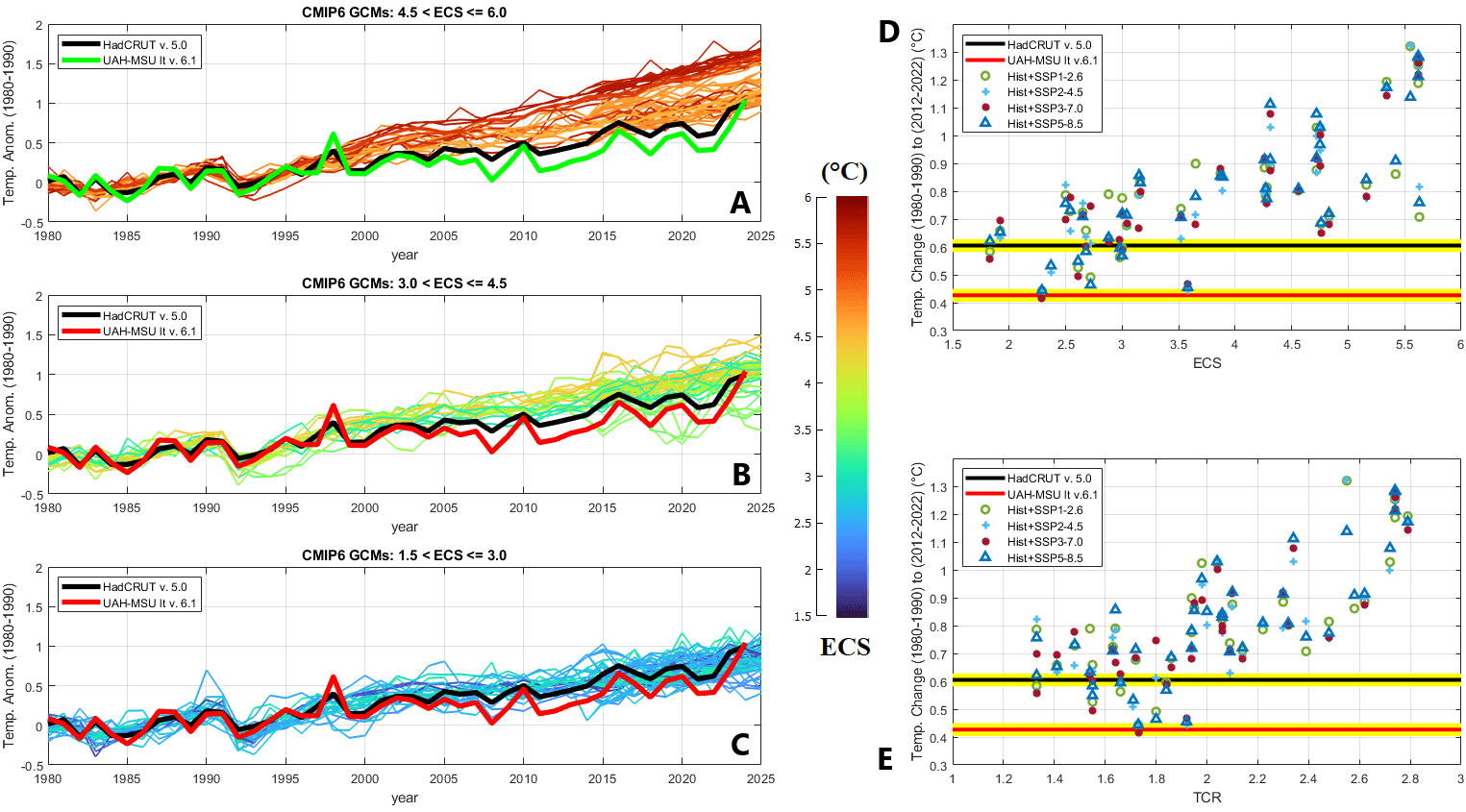}
\par\end{centering}
\caption{{[}A, B, C{]} The CMIP6 GCM simulations separated into three macro-GCM
ensembles based on their ECS values, as represented by a colored scale.
The synthetic temperature records are compared with the HadCRUT v5.0
(infilled data) global surface temperatures \citep{Morice2021} and
the UAH-MSU lt v6.1 satellite-based temperature dataset \citep{Spencer2017}.
{[}D, E{]} Temperature changes from 1980--1990 to 2012--2022, as
simulated by the 42 adopted GCMs, are evaluated against the warming
observed in the HadCRUT v5.0 (0.605 °C \textpm{} 0.02 °C) and UAH-MSU
lt v. 6.1 (0.427 °C \textpm{} 0.03 °C) records, against the ECS and
TCR values of the GCMs. \citep[Adapted from][]{Scafetta2024}.}
\label{Fig6}
\end{figure*}

Several lines of evidence indicate that the GCMs may inadequately
represent key aspects of climate change. For example, the CMIP6 GCMs
--- particularly those with ECS values exceeding 3°C --- tend to
overestimate the warming observed from 1980 to 2022 \citep{Scafetta2022a,Scafetta2023b,Scafetta2024}.
This persistent bias, which is undisputed for high-ECS GCMs, is widely
referred to as the “\emph{hot model}” problem \citep{Hausfather2022}.
Furthermore, the GCMs significantly overestimate warming trends between
1979 and 2014 across the vertical atmospheric profile, particularly
in the upper tropical troposphere \citep{Mitchell2020,McKitrick2018,McKitrick2020,Hudson2023}.
This problem is documented by the IPCC \citeyearpar[AR6,][its figure 3.10]{IPCC2021}
and, in the literature, it is referred to as the ``\emph{missing
hot-spot}'' conundrum. In fact, while models predict a strong warming
in the upper tropical troposphere, several observations suggest a
weaker or absent warming in that region. Potential reasons for the
missing hot-spot include data errors, natural variability, but, of
course, also climate model limitations.

Figures \ref{Fig6}A--C compare the CMIP6 GCM simulations (grouped
by low {[}1.5--3.0°C{]}, medium {[}3.1--4.5°C{]}, and high {[}4.6--6.0°C{]}
ECS values) against the HadCRUT5 global surface temperature record
\citep{Morice2021} and satellite-based UAH-MSU v. 6.1 record \citep{Spencer2017}.
The warm bias of the medium- and high-ECS models is statistically
significant. Figures \ref{Fig6}D--E further highlight these biases,
showing that the GCM-predicted warming from 1980--1990 to 2012--2022
exceeds observed warming by the HadCRUT5 ($0.605\pm0.02$ °C) and
UAH-MSU v6.1 ($0.427\pm0.03$ °C). Figure \ref{Fig6} updates \citet{Scafetta2023b,Scafetta2024},
where this issue is discussed in more details.

The analysis shown in Figure \ref{Fig6} does not consider the years
2023--2024, which were marked by a sudden temperature spike possibly
linked to the Hunga Tonga-Hunga Ha\textquoteleft apai eruption in
2022. This submarine volcanic eruption significantly increased stratospheric
water vapor by \textasciitilde 10\%, potentially exerting a positive
radiative forcing effect \citep{Jenkins2023,Schoeberl2024}. However,
even accounting for this anomaly, the warming estimates for 2014--2024
($\sim$0.70 °C for HadCRUT5 and $\sim$0.54°C for UAH-MSU) remain
below most of the GCM projections. In the future, the rapid cooling
that has been observing since the mid-2024 peak could further exacerbate
the “\emph{hot model}” problem rather than mitigate it.

\begin{figure*}[!t]
\begin{centering}
\includegraphics[width=1\textwidth]{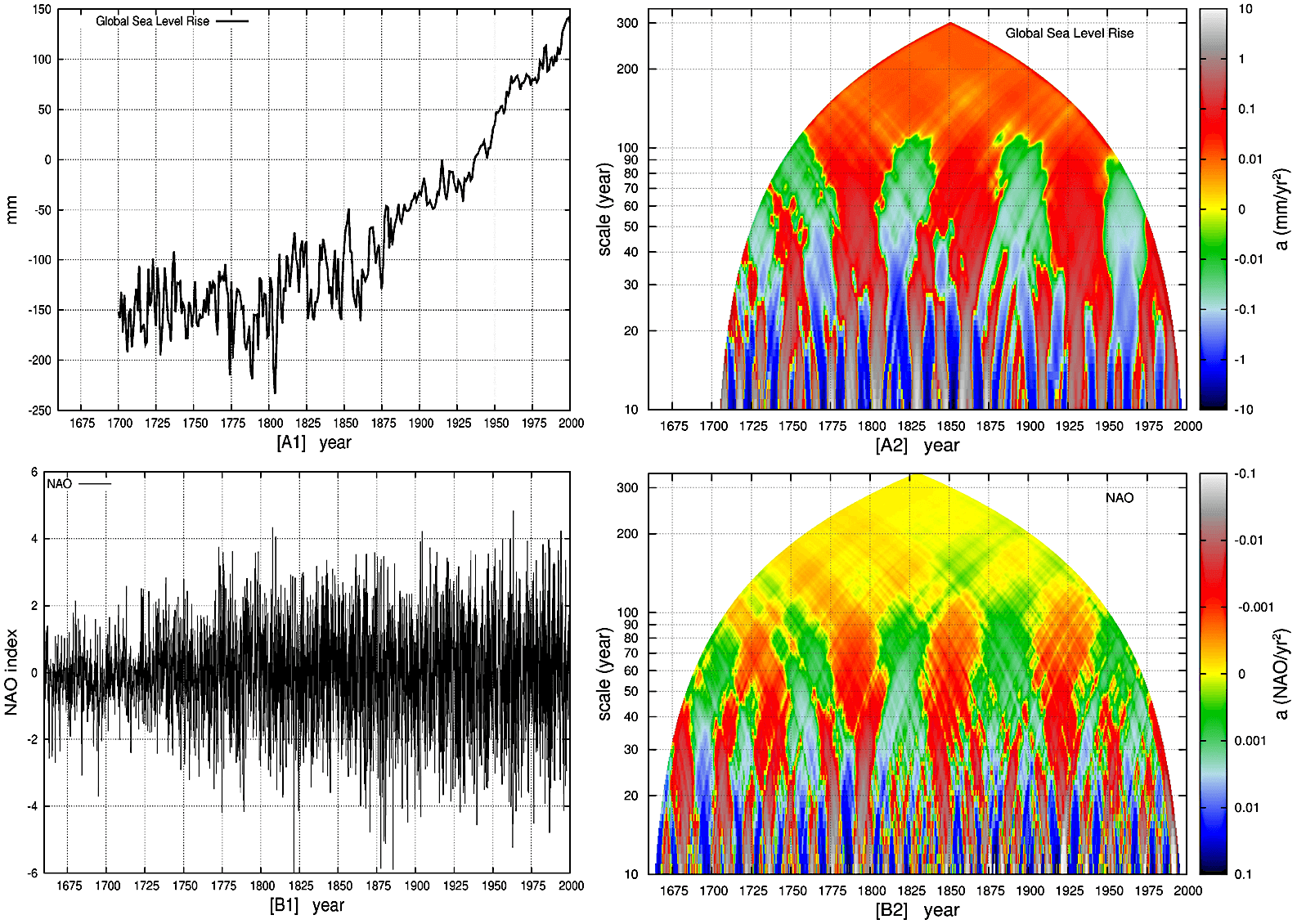}
\par\end{centering}
\caption{{[}A1{]} Global sea level record from \citet{Jevrejeva2008} (left),
alongside its multi-scale acceleration analysis (MSAA) colored diagram
(right). {[}B1{]} North Atlantic Oscillation (NAO) from \citet{Luterbacher2002}
(left), alongside its MSAA colored diagram (right). The diagrams highlight
a consistent 55--70-year oscillation since 1700, represented by alternating
green and red regions within the 30--110-year scale. \citep[Adapted from][]{Scafetta2014}.}
\label{Fig7}
\end{figure*}

\subsubsection{The 50--70-year cycle and its implications}

The CMIP3 and CMIP5 GCMs have been identified as failing to adequately
reconstruct observed climate oscillations on decadal to multidecadal
timescales \citep{Scafetta2012a,Scafetta2013}. Specifically, the
climate system exhibits a distinct quasi-60-year oscillation that
these models fail to replicate. This natural oscillation has resulted
in alternating warming (w) and cooling (c) phases superimposed on
a long-term secular trend: 1850--1880 (w), 1880--1910 (c), 1910--1940
(w), 1940--1970 (c), 1970--2000 (w), and possibly 2000--2030? (c)
\citep[e.g.:][]{Scafetta2014,WyattCurry2014}. Evidence of this oscillation
has also been recently observed in sunshine hours over central Europe
\citep{L=0000FCdecke2024}, suggesting a direct link between surface
temperatures and cloud cover.

A prominent manifestation of a quasi-60-year oscillation can be seen
in the Atlantic Multidecadal Oscillation (AMO) index, which tracks
the variability of the North Atlantic sea surface temperatures (SSTs)
\citep{Schlesinger1994}. The AMO index is calculated by removing
the linear trend from the SSTs observed within the region {[}0°N--70°N,
80°W--20°E{]}, thereby isolating the oscillation from the broader
secular warming trend that may still have both natural and anthropogenic
origins. This oscillation has been documented in climate data since
1700 \citep[e.g.][]{Scafetta2014}, it is recognized as a defining
feature of the North Atlantic ocean-atmosphere variability over the
past 8,000 years \citep{Knudsen2011}, and it is found also in the
paleoclimatic reconstruction of the Northern Hemisphere by \citet{Mann1999}
during the last millennium, as well as in other long-term climate
records \citep{Black1999,Davis2001,Neff2001,Agnihotri2003,Patterson2004,Klyashtorin2008,Camuffo2010,Chambers2012,Mazzarella2012,Scafetta2012c,WyattCurry2014}.

For instance, Figure \ref{Fig7} compares the global sea level record
\citep{Jevrejeva2008} and the North Atlantic Oscillation (NAO) record
\citep{Luterbacher2002} since 1700. The right panels feature their
multi-scale acceleration analysis (MSAA) plot introduced by \citet{Scafetta2014}.
The analyses highlight the 50--70-year oscillation evident in both
datasets. Such spectral coherence between two independent yet physically
coupled indices --- AMO and NAO --- underscores the assertion that
this quasi-60-year oscillation is a genuine dynamical feature of the
climate system, rather than an artifact.

\begin{figure*}[!t]
\begin{centering}
\includegraphics[width=0.87\textwidth]{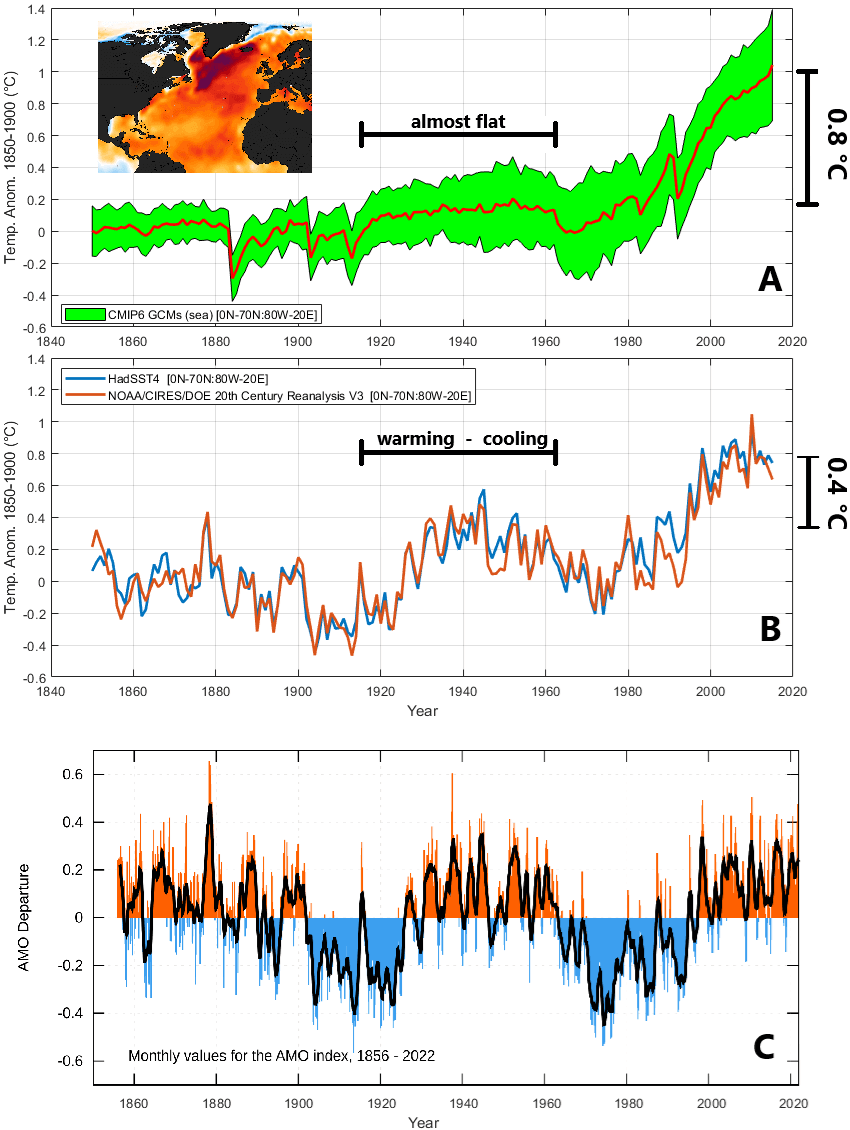}
\par\end{centering}
\caption{{[}A{]} CMIP6 GCM ensemble temperature simulation for the North Atlantic
Ocean surface ({[}0°N--70°N : 80°W--20°E{]}). {[}B{]} Sea surface
temperature records for the same North Atlantic region ({[}0°N--70°N
: 80°W--20°E{]}) derived from HadSST4 \citep{Kennedy2019} and the
NOAA/CIRES/DOE 20\protect\textsuperscript{th} Century Reanalysis
V3. The side panel shows the approximate warming from 1940 to 2020:
+0.8 °C based on GCM simulations; and +0.4°C based on observational
data. {[}C{]} Atlantic Multidecadal Oscillation (AMO) index, calculated
as the linearly detrended North Atlantic Ocean surface temperature
anomalies from 1856 to 2022. The insert in {[}A{]} depicts the approximate
North Atlantic region used to define the AMO index.}
\label{Fig8}
\end{figure*}

The warming observed from the 1910s to the 1940s, followed by the
cooling from the 1940s to the 1970s, clearly appeared in some figures
published in the IPCC FAR \citeyearpar[its figure 8.4]{IPCC1995}.
Notably, this oscillatory pattern shows no correlation with the anthropogenic
forcing functions, which have monotonically accelerated since 1750
(Figure \ref{Fig1}A and \ref{Fig1}C).

Figure \ref{Fig8} provides further evidence by comparing observed
oceanic surface temperatures in the region {[}0°N--70°N, 80°W--20°E{]},
as captured by: (a) CMIP6 GCM ensemble simulations; (b) HadSST4 records
\citep{Kennedy2019} and NOAA/CIRES/DOE 20\textsuperscript{th} Century
Reanalysis V3; and (c) the AMO index from NOAA. The data reveal that
the GCMs fail to replicate the amplitude of the natural oscillation
observed in the historical record. For instance, while GCMs predict
a monotonic warming trend ($\sim$0.2°C) from 1880 to 1960 --- with
occasional interruptions from volcanic events --- the data show a
distinct oscillatory pattern, alternating between warming and cooling
phases with amplitudes up to 0.4--0.5°C.

The existence of a natural 50--70-year cycle not modeled by the CMIP6
GCMs challenges the AGWT \citep[cf.][]{Gervais2016,Scafetta2013},
particularly because the observed warming from the 1970s to 2000s
is strikingly similar to the warming from the 1910s to 1940s. This
suggests that natural variability, such as the quasi-60-year oscillation,
may have contributed to the recent warming trends \citep[cf.:][]{Scafetta2012a,Scafetta2013},
contrary to the finding of the GCMs attributing all 1970--2000 warming
solely to anthropogenic forces. Figures \ref{Fig8}A--B estimate
the warming from 1940 to 2020 as +0.8°C for the GCMs and +0.4°C for
the observed data, which further demonstrates the discrepancy between
the GCM simulations and the empirical records.

\begin{figure*}[!t]
\begin{centering}
\includegraphics[width=1\textwidth]{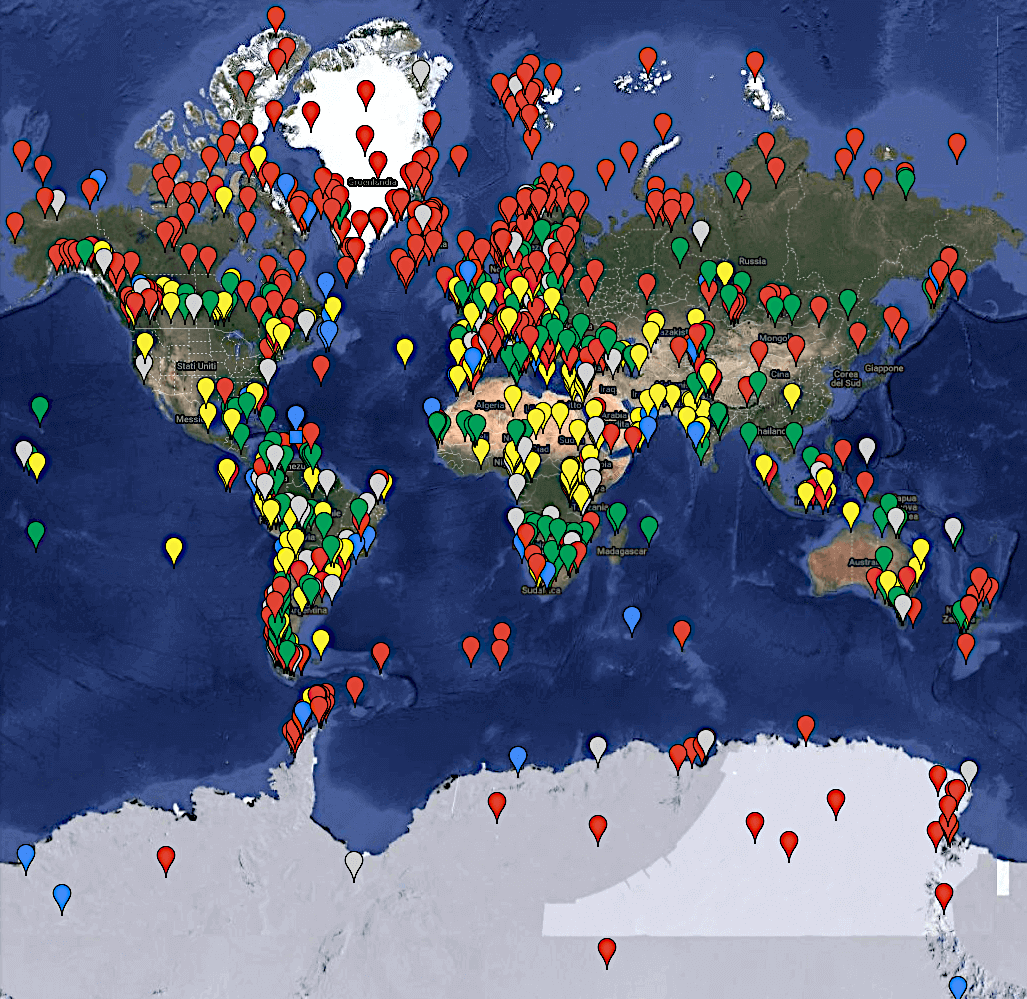}
\par\end{centering}
\caption{Climate reconstructions derived from 1,272 scientific studies on the
\textquotedblleft Medieval Warm Period\textquotedblright{} (MWP) spanning
from 1000--1200 AD \citep{Luning2022}. Each point represents the
location of a specific study where the color represents the observed
climate: warm (red), cool (blue), dry (yellow), wet (green), and no
trend or unclear (grey). The interactive map is accessible at \protect\href{http://t1p.de/mwp}{http://t1p.de/mwp}
(accessed on April 10, 2025).}
\label{Fig9}
\end{figure*}

Discrepancies between model predictions and observations spanning
decadals and longer timescales are significant. Indeed, climate models
are expected to reproduce changes occurring over periods exceeding
$\sim$15 years, whether driven by external forcings or internal variability.
For example, \citet{Knight2009} remarked that ``\emph{simulations
rule out (at the 95\% level) zero trends for intervals of 15 year
or more}'' implying that observing such prolonged data inconsistencies
 with the GCM expected warming rates signals physical flaws in the
models.

Similar multi-decadal discrepancies are evident in precipitation patterns
and trends. Both CMIP5 and CMIP6 GCMs struggle to model also such
climatic patterns accurately, as demonstrated by various studies \citep[e.g.:][]{Connolly2019,Li2022,Zhang2024,Hellmuth2025,Plavcova2025}.

\begin{figure*}[!t]
\begin{centering}
\includegraphics[width=1\textwidth]{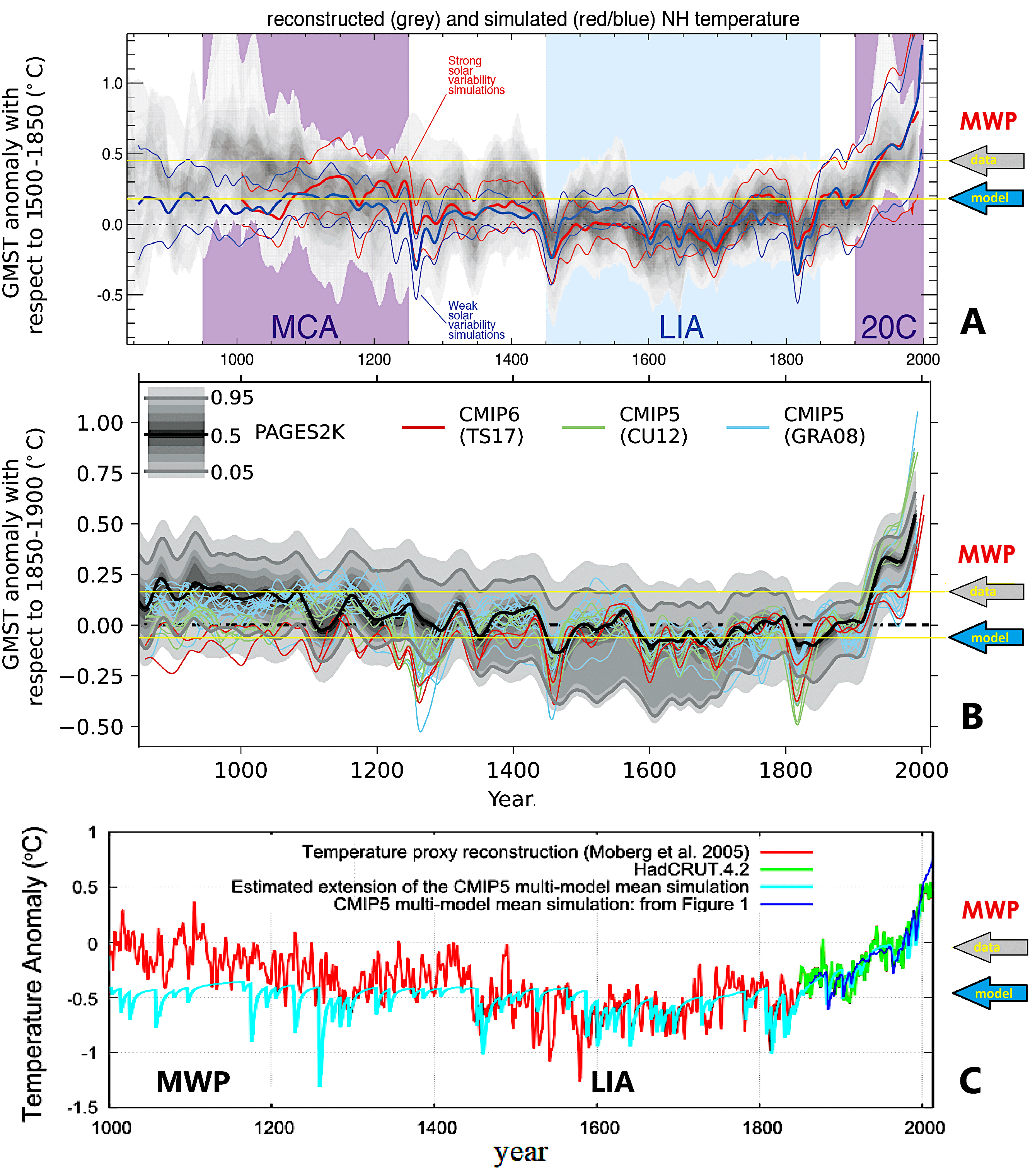}
\par\end{centering}
\caption{The Medieval Warm Period (MWP) temperature conundrum. {[}A{]} Comparison
of simulated versus reconstructed Northern Hemisphere temperature
anomalies over the last millennium, including the Medieval Climate
Anomaly (MCA), Little Ice Age (LIA), and 20\protect\textsuperscript{th}
century (20C) \citep[adapted from][its Figure 5.8]{IPCC2013}. {[}B{]}
Time series of global mean surface temperature anomalies from the
PAGES2K multi-proxy model relative to 1850--1900, incorporating the
simulations from the CMIP5 and CMIP6 GCMs \citep[adapted from][its Figure 3.2]{IPCC2021}.
{[}C{]} Comparison of the multi-model mean simulation (cyan) for the
CMIP GCMs against the temperature proxy reconstruction by \citet{Moberg2005}
(red), which is calibrated and extended with the HadCRUT temperature
record (green) beginning in 1850 \citep[cf.:][]{Scafetta2013,Scafetta2021b}.
Blue and gray arrows to the side indicate the approximate average
temperature during the MWP estimated by the proxy models and simulated
by the GCMs, spanning approximately from 900 to 1100 AD.}
\label{Fig10}
\end{figure*}

\subsubsection{The Medieval Warm Period}

Global climate records from the past 1500 years reveal two prominent
periods of anomalous temperatures predating the 20\textsuperscript{th}
century: the Medieval Warm Period (MWP), spanning approximately 900--1300
AD, and the Little Ice Age (LIA), occurring roughly from 1400 to 1850
AD \citep{Lamb1965}. The magnitude and spatial patterns of the MWP
have been a focal point of scientific debate, particularly following
the introduction of the hockey-stick temperature graph by \citet{Mann1999},
which minimized its significance. A wealth of evidence, however, supports
the existence of a distinct MWP and LIA \citep{Soon2003,Soon2003b,vonStorch2004},
with regions such as Europe, the North Atlantic \citep{Lasher2019}
and China \citep{Ge2017} having potentially experienced temperatures
comparable to or exceeding those of the modern warm period.

While discussions continue on whether the MWP was a global or predominantly
regional phenomenon \citep{Mann2009}, numerous studies --- including
the IPCC Assessment Reports --- acknowledge a Medieval Warm anomaly
relative to the subsequent LIA. For instance, the ``\emph{Medieval
Warm Period 1000--1200 AD}'' project \citep{Luning2022} summarizes
approximately 1300 studies, which predominantly support a warm MWP.
The resulting map is shown in Figure \ref{Fig9}, which highlights
that the number of studies and sites supporting a warm MWP (red points)
far outnumber those supporting a medieval cool period (blue points).

Climate reconstructions often rely on multi-proxy models, which are
inherently accompanied by considerable uncertainties \citep[AR5,][its figure 5.7]{IPCC2013}.
However, when the GCM predictions are compared to the available paleoclimatic
temperature records, significant discrepancies emerge during historically
warm periods including the MWP.

This issue is evident in Figure \ref{Fig10}A, which compares reconstructed
and simulated Northern Hemisphere (NH) temperature changes over the
past millennium \citep[AR5,][its figure 5.8]{IPCC2013}, and Figure
\ref{Fig10}B, which plots global mean surface temperature anomalies
from \citet{PAGES2k2019} against the CMIP5 and CMIP6 GCM simulations
\citep[AR6,][its figure 3.2]{IPCC2021}. These comparisons consistently
show that the GCMs fail to adequately reconstruct the MWP, as highlighted
by the colored and black arrows indicating the divergence between
the GCM ensemble simulations and the proxy-based temperature reconstructions.

This failure is particularly apparent in the CMIP6 models (Figure
\ref{Fig10}B), where simulations exhibit a climatic near-stationarity
between 850 and 1900 AD, apart from brief cooling periods triggered
by volcanic eruptions. Figure \ref{Fig10}C reinforces this point
by comparing an empirical simulation of the multi-model mean (cyan
curve) with the \citet{Moberg2005} reconstruction (red curve), calibrated
with the HadCRUT dataset (green curve) since 1850, as proposed by
\citet{Scafetta2013,Scafetta2021b}, who attributed this modeling
failure to inadequate representation of solar forcing in the CMIP5
and CMIP6 GCMs. The GCMs assume minimal secular solar variability
as Section 4.2 discusses in more detail.

The IPCC \citeyearpar[AR6,][pp. 433]{IPCC2021} explicitly acknowledges
``\emph{larger disagreements between models and temperature reconstructions}''
before the year 1300. Notably, AR6 emphasized the \citet{PAGES2k2019}
temperature reconstruction, which dampens the MWP anomaly compared
to alternative reconstructions \citep[e.g.][its figure 5.7]{IPCC2013}.
This decision to exclusively utilize the \citet{PAGES2k2019} paleoclimatic
proxy reconstruction has been recently criticized as reductive \citep{Esper2024}
because it oversimplifies the physical complexity of the Common Era
climate variability as highlighted in the scientific literature.

In conclusion, the CMIP6 GCMs predict an approximately stable natural
climate extending from 800 AD to the present when only natural (solar
+ volcanic) radiative forcings are applied (Figure \ref{Fig1}D).
However, this prediction conflicts with extensive paleoclimatic evidence
and historical records documenting pronounced cooling from the MWP
to the LIA. All recent multi-proxy reconstructions (e.g., \citealp[its figure 5.8]{IPCC2013};
\citealp{Esper2024}; \citealp{PAGES2k2019}) consistently capture
a significant temperature decline over this interval, although the
extent of this cooling depends on the adopted paleoclimatic record.
The failure of the GCMs to replicate such natural variability underscores
severe limitations in their ability to accurately reconstruct past
climate dynamics. Reconstructing the warm periods of the pre-industrial
era is an essential condition in ensuring the reliability of the GCM
assessments (Figure \ref{Fig2}) regarding the attribution of the
global surface warming reported from 1850--1900 to 2010--2019. This
is because the recent global surface warming may have been partially
driven by the same natural mechanisms (e.g. solar forcing) that caused
other warm periods over the past millennia.

\begin{figure*}[!t]
\begin{centering}
\includegraphics[width=1\textwidth]{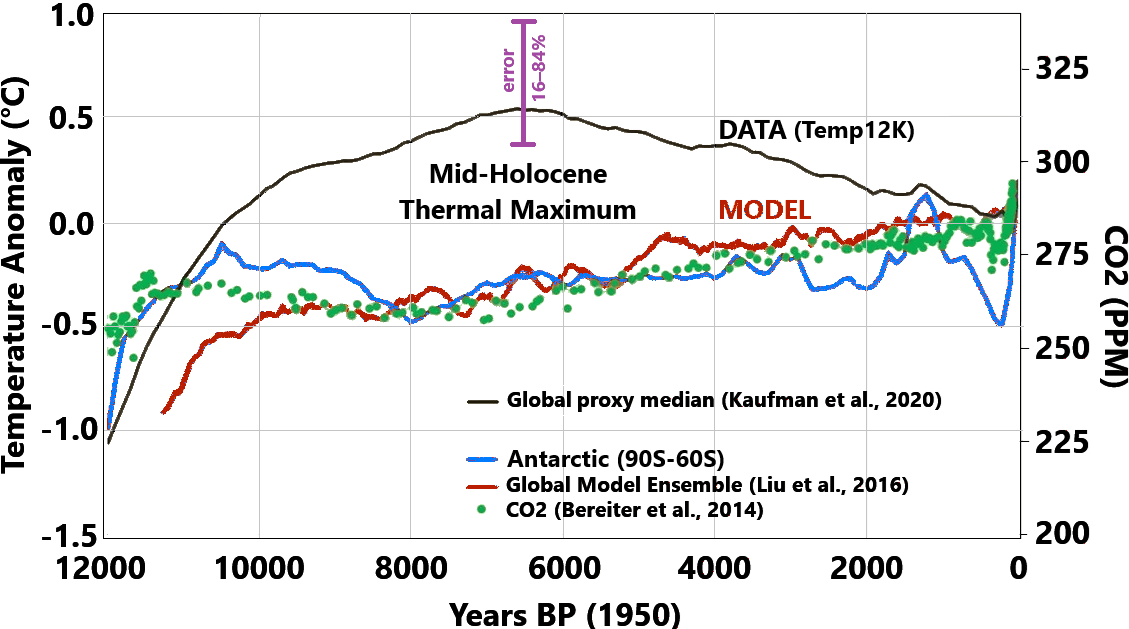}
\par\end{centering}
\caption{The Holocene temperature conundrum: (black line) global median temperature
proxy from \citet{Kaufman2020}; (red line) global climate model ensemble
from \citet{Liu2014}; (green dots) EPICA Dome-C CO\textsubscript{2}
record from \citet{Bereiter2014}; (blue line) Antarctic temperature
proxy from \citet{Kaufman2020}.}
\label{Fig11}
\end{figure*}

\subsubsection{The Holocene temperature conundrum}

The limitations of the GCMs in simulating natural climate change become
increasingly apparent when the entirety of the Holocene epoch is analyzed.
The mismatch between data and model predictions over this prolonged
period, termed the ``\emph{Holocene temperature conundrum}'' \citep{Liu2014},
raises further substantial concerns about the reliability of the existing
global climate models (Figure \ref{Fig11}).

The core of this conundrum lies in the gradual and near-monotonic
increase in atmospheric greenhouse gases --- carbon dioxide (CO\textsubscript{2})
and methane (CH\textsubscript{4}) --- likely driven by processes
such as oceanic degassing and permafrost melting (Figure \ref{Fig11},
green dots). Over this epoch, the climate models predict a quasi-monotonic
warming due to the rise in the greenhouse gas concentrations (Figure
\ref{Fig11}, red curve). However, paleoclimatic proxies derived from
natural archives --- including ice cores, sediment layers, and biological
indicators --- consistently suggest the occurrence of an early-to-mid
Holocene Thermal Maximum (HTM), which is also known as the ``\emph{Holocene
Optimum}''. The HTM contradicts the models' predictions.

Empirical evidence supports a significantly warm HTM, with summer
temperatures in East Greenland during the early Holocene estimated
to be at least 3°C higher than present \citep{Westhoff2022}. Similar
warming trends were documented in central Europe \citep{Zander2024,Melo2022},
the eastern Mediterranean \citep{Cruz-Silva2023}, in Japan \citep{Murata2025},
and in numerous other regions. Antarctica appears to be the notable
exception, showing a gradual warming rather than an HTM (Figure \ref{Fig11},
blue curve). Nevertheless, even this claim is debated, with alternative
studies suggesting the existence of the HTM in Antarctica as well
\citep{Jones2023}. On a global scale, several paleoclimatic reconstructions
suggest that the HTM was warmer than present-day temperatures, although
some estimates, such as those in Figure \ref{Fig11} \citep{Kaufman2020},
appear to mitigate its intensity.

Additional support for a warm HTM comes from other studies showing
that global mean sea levels were higher than today during the mid-Holocene
\citep{Creel2024}, and showing evidence of the African Humid Period
(also known as the ``\emph{Green Sahara}''), which occurred between
approximately 14,800 and 5,500 years ago \citep{deMenocalTierney2012,Kaufman2020}.
After the HTM, the surface temperatures declined throughout the late
Holocene, culminating in the minimum recorded during the Little Ice
Age. This dynamics correlates strongly with the changes in mean daily
insolation at 65°N on summer solstices due to variations in the orbital
parameters of the Earth (\citealp{Laskar2011}; see also the Milankovitch
orbital data viewer at \href{https://biocycle.atmos.colostate.edu/shiny/Milankovitch/}{https://biocycle.atmos.colostate.edu/shiny/Milankovitch/},
accessed April 10, 2025).

Efforts to address the Holocene temperature conundrum have included
investigating seasonal temperature variations since many paleoclimatic
proxies actually represent summer temperatures, as suggested by \citet{Bova2021}.
However, such interpretations have been challenged by other researchers
confirming the HTM also in annual temperature proxies \citep{Cartapanis2022,Dong2022,Kaufman2023}.

The Holocene temperature conundrum ultimately casts doubt on the accuracy
of either paleoclimatic reconstructions or the climate models, or
both. However, the large body of empirical evidence supporting a prominent
HTM implies that even the latest climate models significantly overestimate
the impact of greenhouse gases relative to other factors, such as
the climatic impact of variations in solar radiation input.

Additionally, the Holocene epoch appears marked by pronounced quasi-millennial
oscillations \citep{Alley2000,Bond2001,Kerr2001,Kutschera2020}. Current
climate models lack the capability to reconstruct these quasi-periodic
warm and cool periods, as evidenced by their failure to accurately
simulate the latest past warm and cool periods such as the cooling
from the Medieval Warm Period (MWP) to the Little Ice Age (LIA).

\begin{figure*}[!t]
\begin{centering}
\includegraphics[width=1\textwidth]{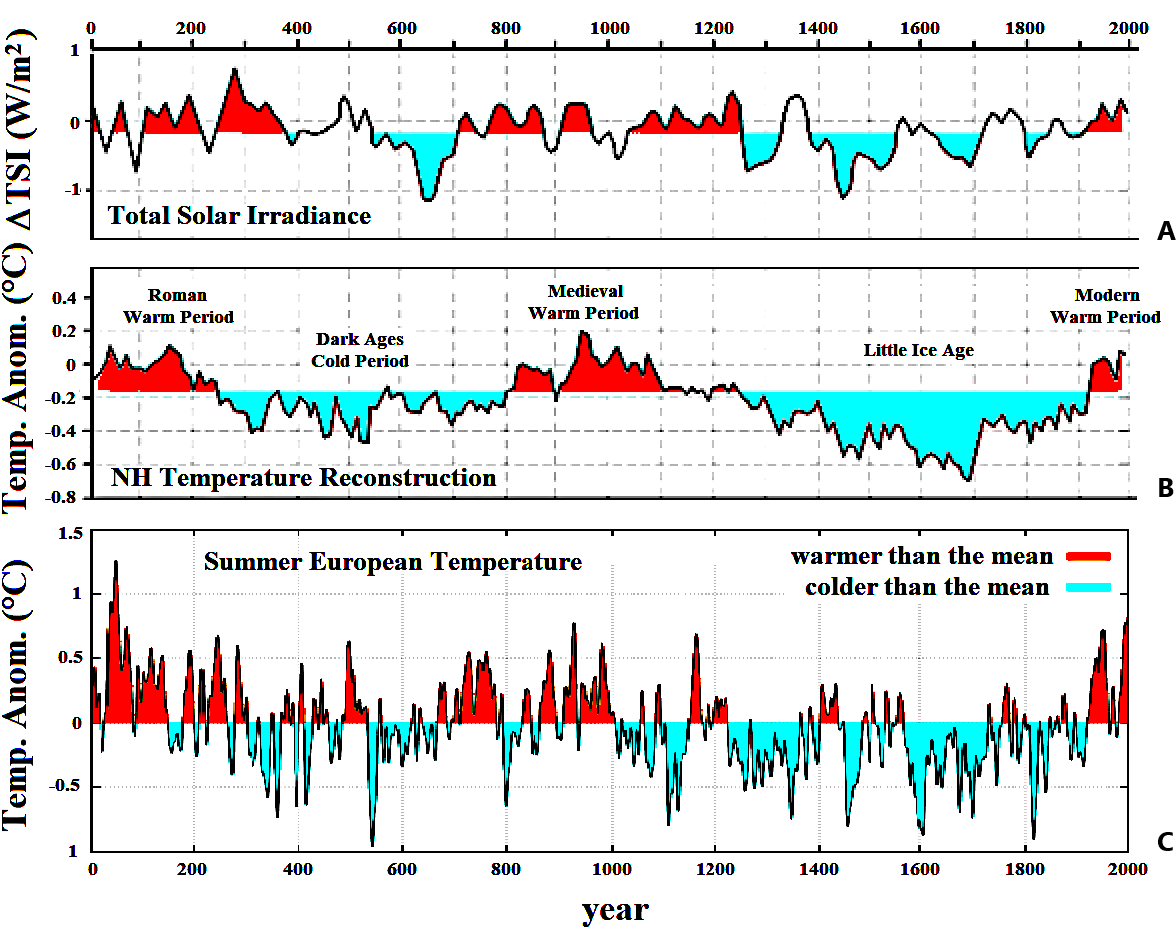}
\par\end{centering}
\caption{Synchronous quasi-millennial cycles observed in: {[}A{]} a reconstruction
of total solar irradiance \citep{Steinhilber2012}; {[}B{]} a reconstruction
of the Northern Hemisphere surface temperatures \citep{Ljungqvist2010};
{[}C{]} a reconstruction of the summer European temperatures \citep{Luterbacher2016}.}
\label{Fig12}
\end{figure*}

\subsection{Conclusion: assessing the CMIP GCMs versus natural climate variability}

The evidence presented above highlights critical limitations in the
Global Climate Models (GCMs), casting doubts on the robustness of
the Anthropogenic Global Warming Theory (AGWT). AGWT is heavily reliant
on the climate attribution analyses derived from the GCMs, yet substantial
empirical data challenges the validity of these models. Specifically,
the GCMs have consistently failed to reproduce natural climate variability
across multiple timescales throughout the Holocene, particularly during
key historical warm periods. Furthermore, their central prediction
--- that Earth's climate would have remained stable from 1850--1900
to 2011--2020 in the absence of anthropogenic forcing --- cannot
be validated because of lack of data. The notion that the discrepancies
between models and observational data may arise solely from data uncertainties
does not resolve the presented challenges, as scientific models must
ultimately be validated through empirical evidence. Consequently,
the AGWT promoted by the \citet{IPCC2021} cannot be considered supported
by empirical evidence.

The inability of the CMIP3, CMIP5 and CMIP6 GCMs to accurately reconstruct
natural climate variability --- including the Medieval Warm Period
(MWP), and other extended warm periods of the past such as the Holocene
Thermal Maximum (HTM) --- raises fundamental questions. Specifically,
concerns arise regarding whether these models employ appropriate natural
radiative forcings and incorporate essential physical mechanisms necessary
to accurately simulate past warm periods. This shortcoming suggests
that the observed warming from 1850--1900 to 2011--2020 may not
be exclusively attributable to anthropogenic activities because it
cannot be ruled out that it was partly induced by the same mechanisms
responsible for the other natural warm periods observed throughout
the Holocene; mechanisms that are absent or remain insufficiently
implemented in the current GCMs because of the models' failure in
reconstructing the warm periods of the past.

For example, a significant quasi-millennial oscillation has been documented
in the climate system over the past $\sim$9,000 years \citep[e.g.:][]{Bond2001,Kerr2001,Neff2001,Kutschera2020}.
This oscillation is evident in records such as the Greenland GISP2
ice core \citep{Alley2000} and many others. Notably, the Medieval
Warm Period (MWP) was preceded by the Roman Warm Period (RWP), which
occurred roughly 900--1000 years earlier \citep{Ljungqvist2010,Christiansen2012,Klimenko2014,Luterbacher2016,Kutschera2020,Li2023}.
The existence of this quasi-millennial cycle --- expected to peak
again in the second half of the 21\textsuperscript{st} century for
solar and astronomical reasons \citep{Scafetta2012b,ScafettaBianchini2023}
--- contrasts sharply with the AGWT's assertion that the warming
observed since 1850--1900 is solely (\textasciitilde 100\%) due
to human emissions.

Figure \ref{Fig12} illustrates this quasi-millennial cycle by comparing
two paleoclimatic temperature reconstructions of the Northern Hemisphere
and summer temperatures in Europe \citep{Ljungqvist2010,Luterbacher2016}
with a reconstruction of total solar irradiance based on the \textsuperscript{14}C
record \citep{Steinhilber2012}. The analysis highlights three warm
periods --- the Roman (0--300 AD), the Medieval (800--1300 AD),
and the contemporary (1900--2100 AD) warm periods --- which appear
equally warm, along with two cold periods: one during the Dark Ages
(DACP) (400--800 AD) and one, even colder, during the Little Ice
Age (LIA) (1300--1850 AD). This evidence suggests that the warming
observed in the 20\textsuperscript{th} century was partly driven
by the warm phase of a quasi-millennial natural cycle driven by solar
activity, which itself exhibits a similar cyclical variation. Further
discussion on this issue is provided in Section 4.2.

\section{Alternative approaches to the detection, attribution, and modeling
of climate change}

This section explores several critical and unresolved challenges concerning
the reliability of global surface temperature records used for detecting
climate change, as well as how the solar effect on the climate is
addressed in modeling frameworks.

\subsection{Challenges in assessing the reliability of global surface temperature
records}

The Intergovernmental Panel on Climate Change (IPCC) Sixth Assessment
Report \citeyearpar[AR6,][]{IPCC2023} states that the ``\emph{Global
surface temperature was 1.09 {[}0.95 to 1.20{]} }°C\emph{ higher in
2011--2020 than 1850--1900, with larger increases over land (1.59
{[}1.34 to 1.83{]} }°C\emph{) than over the ocean (0.88 {[}0.68 to
1.01{]} }°C\emph{)}''. Despite this, determining the precise magnitude
of the global warming between 1850--1900 and 2011--2020 remains
contentious due to the sensitivity of the outcomes to methodological
decisions, which has led to notable discrepancies among alternative
temperature datasets.

Global surface temperature anomalies are constructed from observational
records collected at meteorological stations distributed worldwide
and from sea surface temperature data obtained from buoys, ships,
and ocean reference stations. The processing methodology involves
calculating monthly mean temperatures at each station from daily maximum
and minimum values, from which annual averages are derived. These
averages are then normalized relative to a chosen baseline period
(e.g., 1961--1990) and aggregated to produce yearly global temperature
anomalies. However, significant methodological issues persist throughout
these stages.

A prominent concern is the uneven geographical distribution of the
observational networks, such as weather stations and shipping routes.
This issue is typically addressed through gridded averaging, wherein
the Earth's surface is divided into cells (e.g. 0.25°$\times$0.25°)
based on latitude and longitude, and temperature anomalies for each
cell are averaged to produce global estimates. Nonetheless, historical
underrepresentation of vast areas --- such as the Southern Hemisphere
and sparsely inhabited regions like deserts, polar regions, forests,
and expansive oceanic zones --- introduces substantial uncertainties.
For instance, prior to 1950, the majority of the temperature records
originated from North America, Europe, and East Asia, complicating
efforts to accurately reconstruct global temperature anomalies during
earlier historical periods \citep[cf.:][]{Lawrimore2011,Menne2018}.
Furthermore, gaps in station operation, changes in instrumentation,
and station relocations lead to missing data and biases, undermining
temperature trend analyses. Collectively, spatial and temporal inconsistencies
in local temperature records pose significant obstacles to accurately
assessing global temperature trends.

Another fundamental issue pertains to the suitability of local temperature
records for analyzing long-term climate trends. Weather stations are
primarily designed to monitor short-term meteorological variations
rather than long-term climatic changes. Over extended periods, environmental
transformations surrounding stations --- such as urbanization, deforestation,
or afforestation --- alongside station replacements and relocations,
can introduce systematic non-climatic biases into the data. These
biases, if predominantly aligned in one direction, could significantly
distort ensemble averages and, consequently, affect global temperature
anomaly computations \citep{Daleo2016,Watts2022}. As a result, the
integrity of the temperature records as indicators of long-term climatic
trends demands critical evaluation.

\begin{figure*}[!t]
\begin{centering}
\includegraphics[width=0.9\textwidth]{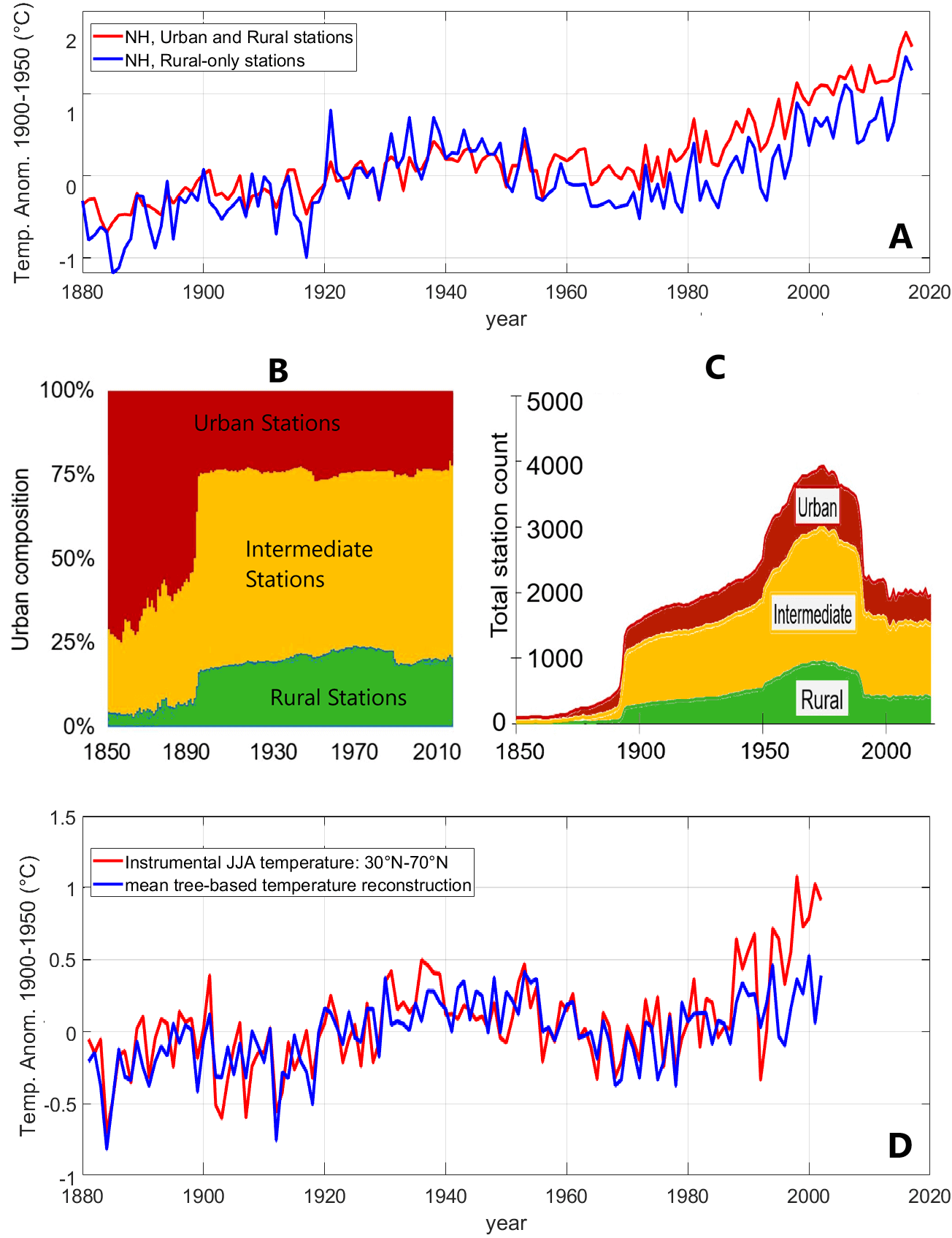}
\par\end{centering}
\caption{{[}A{]} Divergence observed in the Northern Hemisphere land surface
air temperature estimates between a reconstruction based on rural-and-urban
stations versus a reconstruction based exclusively on rural stations.
{[}B, C{]} Percentage and total count of stations categorized into
urban, intermediate, and rural subsets \citep[adapted from][]{Soon2023}.
{[}D{]} Divergence between summer (JJA) instrumental temperature averages
across land areas (30°N--70°N; red) and tree-ring-based mean temperature
reconstructions (blue), relative to the 1930--1960 baseline period
\citep[adapted from][]{Esper2018}.}
\label{Fig13}
\end{figure*}

\subsubsection{The urban heat island effect}

Urban areas are consistently warmer than their rural counterparts
due to several contributing factors. In cities, natural soil is frequently
replaced by heat-absorbing artificial materials, while high-rise buildings
disrupt airflow and create heat-trapping canyon effects. Additionally,
in urban landscapes rainwater drains, unlike in rural areas where
water moistens the soil and facilitates cooling via evaporation. Vegetation,
which is more prevalent in rural areas, utilizes sunlight for photosynthesis
rather than converting it into heat. Furthermore, anthropogenic activities,
including vehicular emissions, building heating and the use of air
conditioning systems, intensify urban heating. This cumulative phenomenon
is widely recognized as the ``\emph{urban heat island}'' (UHI) effect
\citep{McKitrick2007,Mohajerani2017,Stewart2012}.

As urbanization progresses, the temperature recorded by proximate
weather stations tends to increase, reflecting enhanced UHI effects.
This issue is particularly significant given that urban areas account
for less than 4\% of the Earth's land surface, yet a substantial proportion
of weather stations within the Global Historical Climatology Network
(GHCN) are situated in urban settings (see Figure \ref{Fig13}). Additionally,
the ongoing shift of weather stations from rural to urban environments
exacerbates the problem. The global marked increase in urbanization
since 1900 and, in particular, after 1950 underscores the growing
relevance of this issue \citep[cf.][]{Scafetta2021a}.

The IPCC Sixth Assessment Report \citeyearpar[AR6,][Chapter 2]{IPCC2021}
asserts that global surface temperature records are minimally affected
by UHI bias, estimating its contribution to global warming to be less
than 10\%, thereby suggesting it can be disregarded for first-order
analyses. However, a study by \citet{Helbling2023}, which analyzed
data from 118 countries spanning from 1960 to 2016, identified a strong
correlation between rising temperatures and urbanization rates. This
is just one of the recent evidence that has led some researchers to
argue that UHI contamination of climate records may have been underestimated,
advocating for a comprehensive reevaluation of global temperature
datasets and climate change mitigation strategies.

For example, \citet{Soon2023} examined the Northern Hemisphere land
surface temperatures from 1850 to 2018 using two distinct datasets:
one encompassing both rural and urban stations and another based exclusively
on verified rural stations. Their analysis revealed significant differences
in the long-term warming trends. Specifically, the combined rural-urban
dataset indicated a warming rate of 0.89°C per century since 1850,
whereas the rural-only dataset showed a markedly lower rate of 0.55°C
per century. This divergence, illustrated in Figure \ref{Fig13}A,
underscores the influence of urbanization on the global temperature
records used to assess climate change. Figures \ref{Fig13}B and \ref{Fig13}C
provide details on the composition and total number of stations employed
in the study.

\begin{figure*}[!t]
\begin{centering}
\includegraphics[width=1\textwidth]{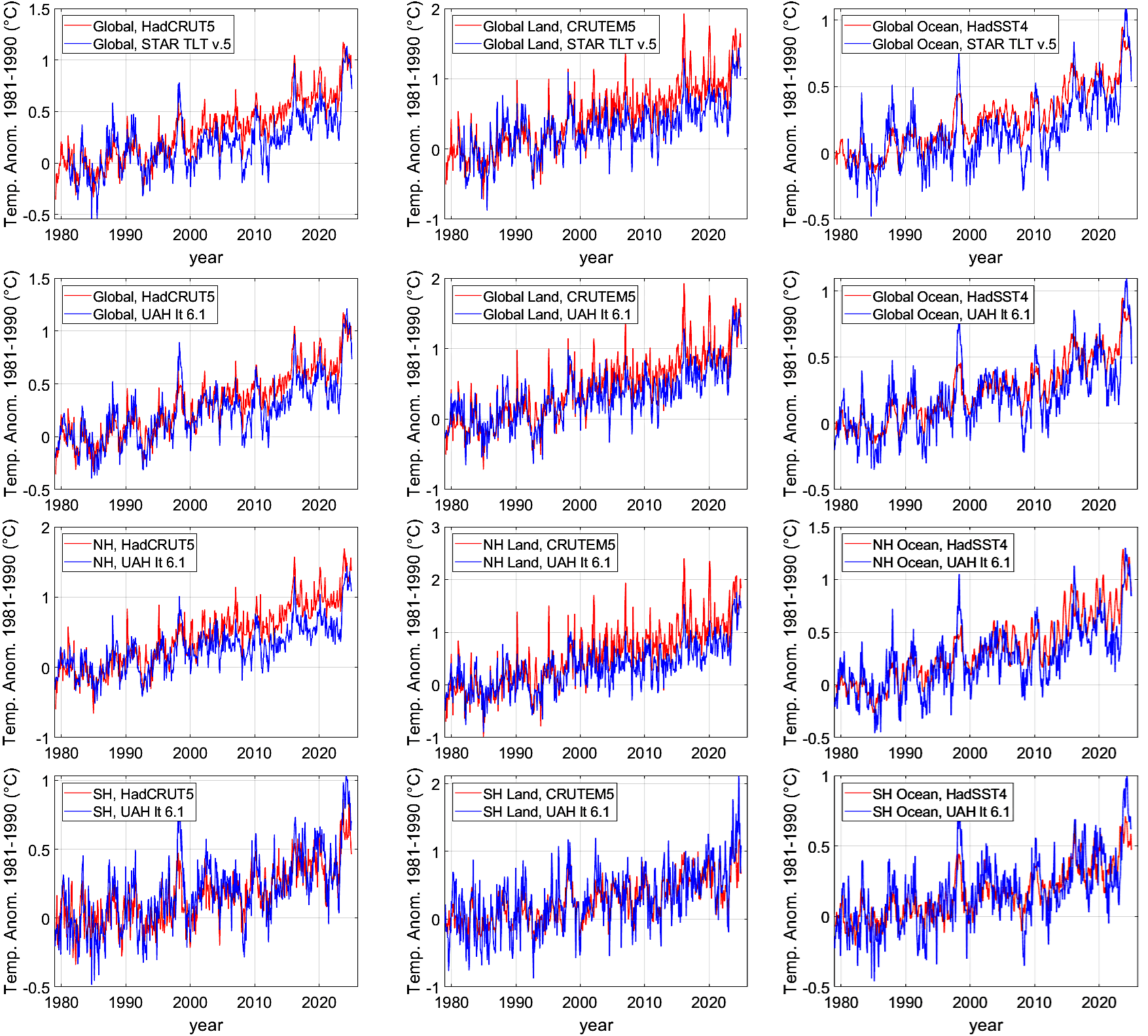}
\par\end{centering}
\caption{Comparison of surface temperature records --- HadCRUT5 (global),
CRUTEM5 (land), and HadSST (ocean) (red) \citep{Morice2021,Osborn2011,Kennedy2019}
--- with the satellite-based lower troposphere temperature records
NOAA-STAR v5.0 \citep{Zou2023} and UAH MSU v6.1 (blue) \citep{Spencer2017}.}
\label{Fig14}
\end{figure*}

\subsubsection{The divergence problems with the dendrochronological proxies and
the satellite records}

A similar pattern is observed in Figure \ref{Fig13}D, which compares
instrumental summer (June--August) temperatures averaged over land
areas between 30°N and 70°N latitude (red) with tree-ring-based mean
temperature reconstructions within the same latitudes (blue) \citep{Esper2018}.
Instrumental records exhibit a pronounced warming trend post-1980,
a phenomenon referred to as the “\emph{divergence problem}”, where
instrumental temperatures and dendrochronological proxies show conflicting
trends. This discrepancy raises questions about the reliability of
both global surface temperature records and dendrochronological proxies
as tools for reconstructing past climate variations \citep{Buntgen2021,Cai2023}.
For instance, \citet{Moberg2005} argued that tree-ring records are
more effective at capturing the high-frequency components of climate
signals while attenuating the secular and millennial low-frequency
modulation, potentially due to biological adaptations of the trees
to changing climatic conditions. Nevertheless, the substantial and
anomalous divergence observed after 1980 (as depicted in Figure \ref{Fig13}D)
together with the divergence observed with the record from the rural-only
stations (depicted in Figure \ref{Fig13}A), casts doubt on the accuracy
of the adopted climatic temperature records.

\begin{table*}[!t]
\begin{centering}
{\footnotesize{}%
\begin{tabular}{|l|c|ccc|ccc|}
\hline 
 & {\footnotesize Had-CRU} & {\footnotesize UAH 6.1} & {\footnotesize Had-CRU -- UAH} & {\footnotesize\%} & {\footnotesize STAR5} & {\footnotesize Had-CRU -- STAR} & {\footnotesize\%}\tabularnewline
{\footnotesize region} & {\footnotesize ( °C) ($\pm$ 95\%)} & {\footnotesize ( °C) ($\pm$ 95\%)} & {\footnotesize ( °C) ($\pm$ 95\%)} & {\footnotesize\% ($\pm$ 95\%)} & {\footnotesize ( °C) ($\pm$ 95\%)} & {\footnotesize ( °C) ($\pm$ 95\%)} & {\footnotesize\% ($\pm$ 95\%)}\tabularnewline
\hline 
{\footnotesize Global (Land+Ocean)} & {\footnotesize 0.73 $\pm$ 0.02} & {\footnotesize 0.58 $\pm$ 0.02} & {\footnotesize 0.15 $\pm$ 0.03} & {\footnotesize -21\% $\pm$ 04\%} & {\footnotesize 0.53 $\pm$ 0.02} & {\footnotesize +0.20 $\pm$ 0.03} & {\footnotesize -27\% $\pm$ 04\%}\tabularnewline
{\footnotesize Global (Land)} & {\footnotesize 1.03 $\pm$ 0.15} & {\footnotesize 0.79 $\pm$ 0.02} & {\footnotesize 0.24 $\pm$ 0.15} & {\footnotesize -23\% $\pm$ 14\%} & {\footnotesize 0.68 $\pm$ 0.02} & {\footnotesize +0.35 $\pm$ 0.15} & {\footnotesize -34\% $\pm$ 14\%}\tabularnewline
{\footnotesize Global (Ocean)} & {\footnotesize 0.58 $\pm$ 0.02} & {\footnotesize 0.49 $\pm$ 0.02} & {\footnotesize 0.09 $\pm$ 0.03} & {\footnotesize -16\% $\pm$ 05\%} & {\footnotesize 0.46 $\pm$ 0.02} & {\footnotesize +0.12 $\pm$ 0.03} & {\footnotesize -21\% $\pm$ 05\%}\tabularnewline
{\footnotesize NH (Land+Ocean)} & {\footnotesize 1.03 $\pm$ 0.03} & {\footnotesize 0.69 $\pm$ 0.02} & {\footnotesize 0.34 $\pm$ 0.04} & {\footnotesize -33\% $\pm$ 04\%} & {\footnotesize 0.72 $\pm$ 0.02} & {\footnotesize +0.31 $\pm$ 0.04} & {\footnotesize -30\% $\pm$ 04\%}\tabularnewline
{\footnotesize NH (Land)} & {\footnotesize 1.24 $\pm$ 0.15} & {\footnotesize 0.83 $\pm$ 0.02} & {\footnotesize 0.41 $\pm$ 0.15} & {\footnotesize -33\% $\pm$ 12\%} &  &  & \tabularnewline
{\footnotesize NH (Ocean)} & {\footnotesize 0.77 $\pm$ 0.02} & {\footnotesize 0.59 $\pm$ 0.02} & {\footnotesize 0.18 $\pm$ 0.03} & {\footnotesize -23\% $\pm$ 04\%} &  &  & \tabularnewline
{\footnotesize SH (Land+Ocean)} & {\footnotesize 0.42 $\pm$ 0.05} & {\footnotesize 0.46 $\pm$ 0.02} & {\footnotesize -0.04 $\pm$ 0.06} & {\footnotesize +10\% $\pm$ 14\%} & {\footnotesize 0.34 $\pm$ 0.02} & {\footnotesize +0.08 $\pm$ 0.06} & {\footnotesize -19\% $\pm$ 14\%}\tabularnewline
{\footnotesize SH (Land)} & {\footnotesize 0.62 $\pm$ 0.31} & {\footnotesize 0.70 $\pm$ 0.02} & {\footnotesize -0.08 $\pm$ 0.31} & {\footnotesize +12\% $\pm$ 49\%} &  &  & \tabularnewline
{\footnotesize SH (Ocean)} & {\footnotesize 0.40 $\pm$ 0.02} & {\footnotesize 0.42 $\pm$ 0.02} & {\footnotesize -0.02 $\pm$ 0.03} & {\footnotesize +5\% $\pm$ 07\%} &  &  & \tabularnewline
\hline 
\end{tabular}}{\footnotesize\par}
\par\end{centering}
\caption{Temperature increase from January 2014 to December 2024 relative to
the baseline period of January 1981 to December 1990, as determined
from surface temperature records (HadCRUT5-global, CRUTEM5-land, and
HadSST-ocean) compared to the satellite UAH-MSU v6.1 \citep{Spencer2017}
and NOAA-STAR v5.0 \citep{Zou2023} lower troposphere temperature
record. These records are illustrated in Figure \ref{Fig14}. Error
bars represent approximately the 95\% confidence interval. For UAH-MSU
and NOAA-STAR v5.0, the error bars are not explicitly reported but
assumed here to match the global surface temperature error ($\pm$
0.02 °C) across all regions.}
\label{Tab1}
\end{table*}

Figure \ref{Fig14} highlights critical concerns regarding the reliability
of the current global surface temperature records. It presents a comparative
analysis of the HadCRUT5 (global), CRUTEM5 (land), and HadSST (ocean)
surface temperature datasets \citep{Morice2021,Osborn2011,Kennedy2019}
with the satellite-based UAH MSU v.6.1 lower troposphere temperature
records \citep{Spencer2017}, which is available since December 1978.
Additionally, the figure delineates results by hemisphere, distinguishing
the Northern Hemisphere (39\% land, 61\% ocean, encompassing 90\%
of the global population) from the Southern Hemisphere (19\% land,
81\% ocean, hosting 10\% of the global population).

The warming of the lower troposphere should approximate the expected
surface warming across sufficiently large spatial scales. However,
the GCMs predict that the troposphere --- particularly at its top
--- should exhibit greater warming than at the surface \citep{Mitchell2020,McKitrick2020,Hudson2023},
as illustrated in the IPCC \citeyearpar[AR6,][its figure 3.10]{IPCC2021}.
Contrary to these predictions, Figure \ref{Fig14} reveals that satellite
observations report a markedly lower warming rate relative to global
surface temperature records. Specifically, the surface records exhibit
a excess warming of approximately 21\%, which equates to a global
excess warming of roughly +0.15°C from 1981--1990 to 2014--2024,
a significant discrepancy. Furthermore, the figure indicates pronounced
surface excess warming in the Northern Hemisphere: approximately 33\%
over land and 23\% over ocean. In contrast, the surface and satellite
temperature trends in the Southern Hemisphere --- across both land
and ocean --- align well within their uncertainties (see Table \ref{Tab1}).
Similarly, at the upper tropical troposphere the GCMs predict a ``\emph{hot-spot}''
that is not observed in the data \LyXZeroWidthSpace\citep{McKitrick2018,McKitrick2020,Mitchell2020}\LyXZeroWidthSpace .

The compatibility of surface and satellite temperature records in
the Southern Hemisphere, as evidenced in Figure \ref{Fig14}, supports
the assumption that the warming trend in the two records should be
comparable, implying that the substantial warming bias observed in
Northern Hemisphere land surface temperatures implies the influence
of non-climatic factors, notably urban heat island (UHI) effects.
Similarly, Northern Hemisphere sea surface temperatures display a
slight warming bias relative to the satellite measurements, potentially
attributable to inadequate monitoring of the extensive polar regions.
In such areas, missing data are reconstructed using models, which
may inadvertently introduce additional warming biases, as discussed
by \citet{Scafetta2021c}. The reduced land area and the lower population
density in the Southern Hemisphere likely mitigate the impact of non-climatic
warming biases unrelated to climate change. If the warming trend indicated
by the lower troposphere satellite temperature record offers a more
reliable estimate of the actual global surface temperature trend,
the above findings further question the accuracy of the CMIP6 GCMs,
as they would all appear to overestimate the actual warming, as noted
in Section 3.3.1 and in Figure \ref{Fig6}.

The UAH-MSU lower troposphere temperature record \citep{Spencer2017}
has been subject to debate, particularly following adjustments to
the Remote Sensing System (RSS) dataset in 2014. These revisions led
to the development of an alternative satellite-based temperature record
by \citet{Mears2016}, aligning more closely with the accepted surface-based
global temperature records. The NOAA-STAR v. 4.0 dataset also reflected
similar trends \citep{Santer2017}, seemingly in contradiction to
the UAH-MSU findings. However, the UAH-MSU-lt dataset has demonstrated
consistency with global warming trends derived from the Integrated
Global Radiosonde Archive (IGRA) and reanalysis datasets \citep{Christy2018}.
Additionally, recent updates to the NOAA-STAR records \citep{Zou2023}
have acknowledged prior errors, culminating in the NOAA-STAR v. 5.0
dataset. This latest dataset exhibits slightly lower warming from
1981--1990 to 2014--2024 relative to UAH-MSU (refer to Table \ref{Tab1}
and Figure \ref{Fig14}, top panels). Collectively, these observations
suggest that the global surface temperature records may be significantly
warm-biased due to improper data processing and non-climatic influences
at the surface, particularly over land.

\begin{figure*}[!t]
\begin{centering}
\includegraphics[width=1\textwidth]{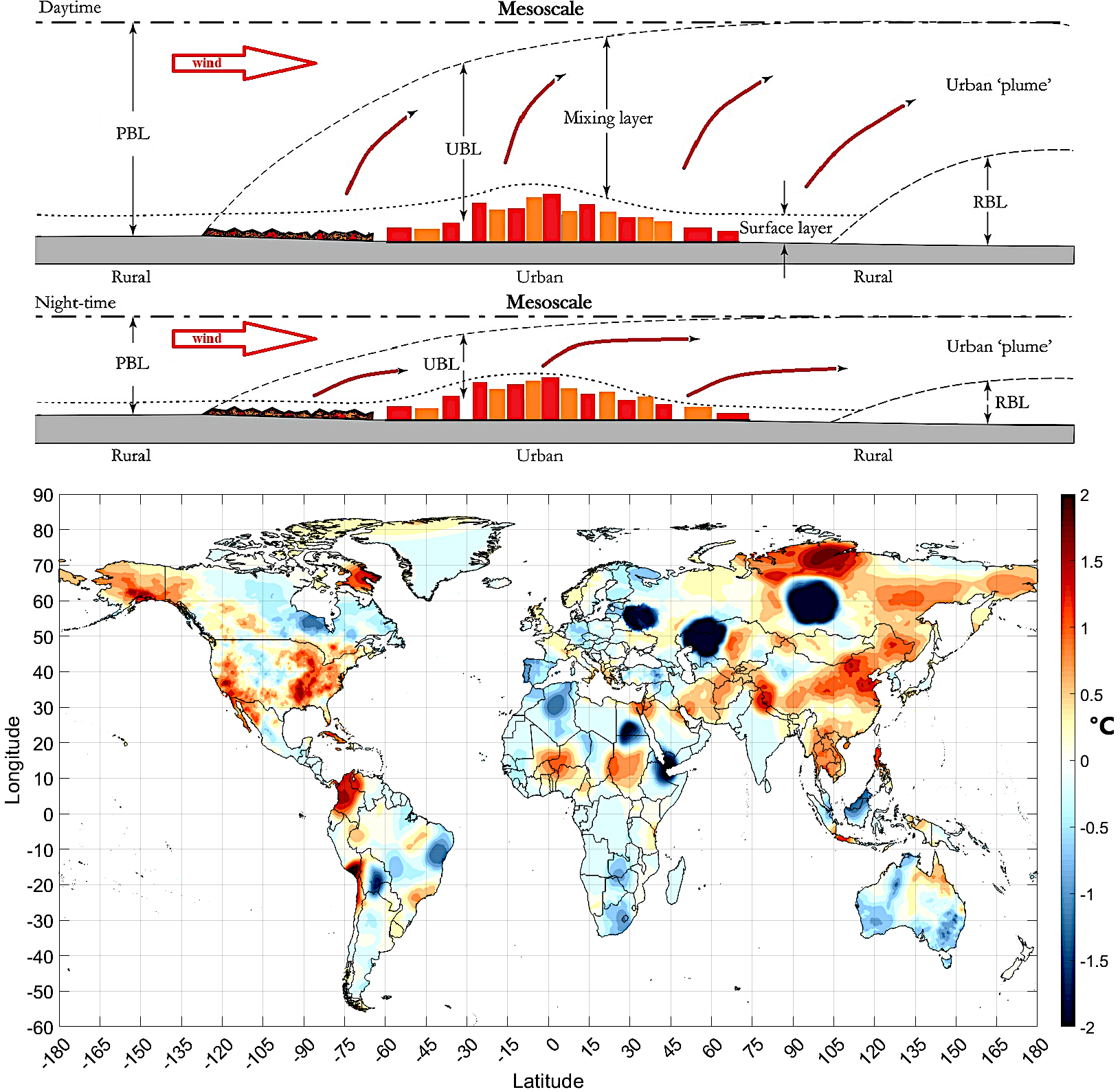}
\par\end{centering}
\caption{{[}Top{]} Diagram illustrating boundary layer structure over a city
and its surrounding areas: a high daytime boundary layer facilitates
heat dispersion, while a low nocturnal boundary layer retains hot
air near the surface. {[}Bottom{]} Global map of the divergence between
$T_{\text{min}}$ and $T_{\text{max}}$ comparing the periods 1945--1955
and 2013--2023, based on the CRU-TS4.08 land temperature data \citep{Harris2020}.
\citep[cf.:][]{ScafettaOuyang2019,Scafetta2021a}.}
\label{Fig15}
\end{figure*}

\subsubsection{The excessive Tmin warming relative to the Tmax warming}

An alternative method for identifying non-climatic warming biases
in surface temperature records has been explored by \citet{ScafettaOuyang2019}
and \citet{Scafetta2021a}. Their analysis revealed substantial regions
within global land surface temperature datasets that exhibit anomalously
high nighttime warming (Tmin) relative to daytime temperatures (Tmax).
This finding diverges from the predictions of the GCMs, which typically
suggest only moderate additional nighttime warming in comparison to
daytime warming.

Daytime surface warming is effectively mitigated by convective vertical
air movements, facilitated by a relatively elevated boundary inversion
layer. In contrast, nighttime warming forces a horizontal dispersion
of warm air near the surface due to the typically lower position of
the boundary inversion layer (cf. Figure \ref{Fig15}, top). In urban
environments, this meteorological behavior expands urban heat island
(UHI) effects during night, resulting in a stronger influence on nearby
thermometers at night compared to during day. Consequently, as urban
areas continue to grow, urbanization amplifies the warming trend in
Tmin compared to Tmax \citep[cf.][]{Kim2022}.

Figure \ref{Fig15} illustrates the global spatial divergence between
Tmin and Tmax across the periods 1945--1955 and 2013--2023, utilizing
the CRU-TS4.08 land temperature data \citep{Harris2020}. Prominent
orange-red regions in the Northern Hemisphere, often forming localized
warm clusters, indicate intense nighttime warming, which are often
near urban centers. Conversely, the Southern Hemisphere exhibits a
lower variability. Interestingly, even small urban centers can produce
significant UHI effects \citep{Cardoso2017,Pinho2000}. The darkest
wide areas may reflect anomalies in the dataset, which could be potentially
linked to regions exhibiting minimal warming in Tmin or influenced
by phenomena such as urban cool island (UCI) effects as in arid areas,
in deforestation zones, or in regions where permafrost may be melting
\citep[cf.:][]{ScafettaOuyang2019,Scafetta2021a}.

\subsubsection{Urban blending and homogenization uncertainties}

The above findings appear to underscore critical failures of the adopted
“\emph{homogenization}” processes. These are designed to correct biases
in meteorological records caused by environmental, instrumental, or
methodological changes \citep{Menne2018}. Homogenization algorithms
aim to enhance the accuracy of temperature datasets by identifying
and rectifying such inconsistencies, thus enabling accurate assessment
of true climate trends. For instance, the Global Historical Climatology
Network (GHCN) provides temperature datasets in both non-homogenized
and homogenized formats, with the latter used for constructing global
climate records. Despite their intended function, these algorithms
exhibit limitations. Studies have highlighted inconsistencies in detecting
breakpoints within temperature records \citep{Squintu2020}. For example,
\citet{ONeill2022} found that the homogenized temperature records
 can change each time the raw temperature records are reprocessed
by such homogenization algorithms.

Moreover, \citet{Katata2023} identified evidence of “\emph{urban
blending}” or aliasing of trend biases within homogenized records.
The blending artifact occurs when homogenization algorithms inadvertently
transfer portions of the warm bias from heavily affected urban stations
to less affected nearby stations, thereby distorting temperature records.
As urbanization expands globally, these blending effects may have
contributed to artificial warming biases in global surface temperature
datasets. This phenomenon could explain the above observed discrepancies
when comparing surface temperature records against rural-only datasets,
satellite-based records, tree-ring chronologies, or the observed excessive
divergence between daytime and nighttime temperature trends (Figures
\ref{Fig13}, \ref{Fig14}, and \ref{Fig15}).

\begin{figure}[!t]
\begin{centering}
\includegraphics[width=1\columnwidth]{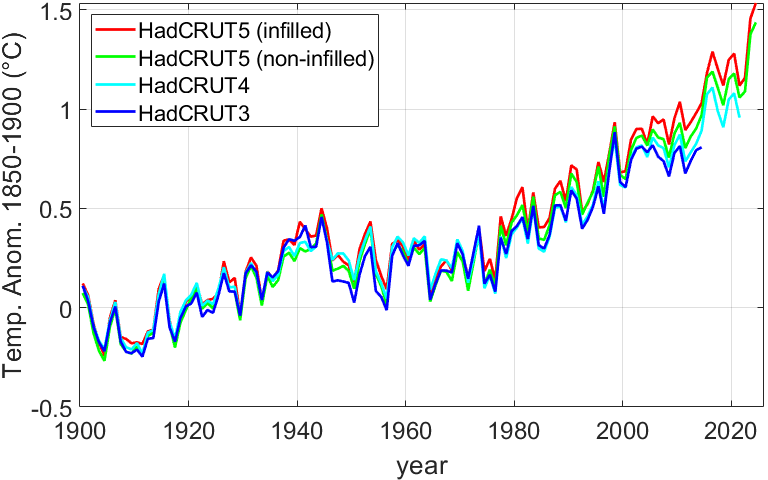}
\par\end{centering}
\caption{Comparison of global temperature records: HadCRUT3 (discontinued in
2014; \citealp{Brohan2006}), HadCRUT4 (discontinued in 2021; \citealp{Morice2012}),
and HadCRUT5, including both the infilled and non-infilled versions
\citep{Morice2021}.}
\label{Fig16}
\end{figure}

\subsubsection{The \textquotedblleft disappearance\textquotedblright{} of the 2000--2014
temperature \textquotedblleft\emph{pause}\textquotedblright{} or “\emph{hiatus}”}

Global temperature datasets used for climate change analyses are subject
to continuous updates, often reflecting substantial changes between
successive versions. For example, the transition from HadCRUT3 (discontinued
in 2014) \citep{Brohan2006} to HadCRUT4 (discontinued in 2021) \citep{Morice2012}
and subsequently to HadCRUT5 \citep{Morice2021} resulted in notable
temperature alterations during the overlapping periods. In particular,
Figure \ref{Fig16} highlights increasingly larger warming trends
in these updated records after 2000, with increases up to 20\% by
2014 \citep[cf.][]{Scafetta2023c}. These modifications are likely
to be attributable to the incorporation of additional input data,
adjustments in homogenization algorithms, and the inclusion of model-based
synthetic data to fill uncovered regions. Such changes may have been
also influenced by efforts to mitigate the discrepancies observed
between the GCM predictions and earlier datasets like HadCRUT3, where
the data-model divergences had become increasingly apparent \citep{Scafetta2012a,Fyfe2012}.
In fact, the latest revisions of the global surface temperature records
effectively removed the well documented temperature “\emph{pause}”
or “\emph{hiatus}” observed from 2000 to 2014 in several records,
whose existence was acknowledged also by the IPCC \citeyearpar[AR5,][]{IPCC2013}.
However, this 2000--2014 hiatus is still well visible in the latest
ocean and satellite-based temperature records (Figure \ref{Fig14}),
raising further questions about the reliability of the latest global
surface temperature records, in particular, over land.

In summary, while substantial evidence supports a global surface warming
between 1850--1900 and 2011--2020, several factors --- including
comparisons of sea and land records with model projections \citep{Scafetta2021a}
--- indicate that the actual global surface temperature warming may
be 15--25\% lower than what reported by the \citet{IPCC2021}. The
concerns surrounding the reliability of current temperature records
merit further investigation and scrutiny.

\begin{figure*}[!t]
\begin{centering}
\includegraphics[width=1\textwidth]{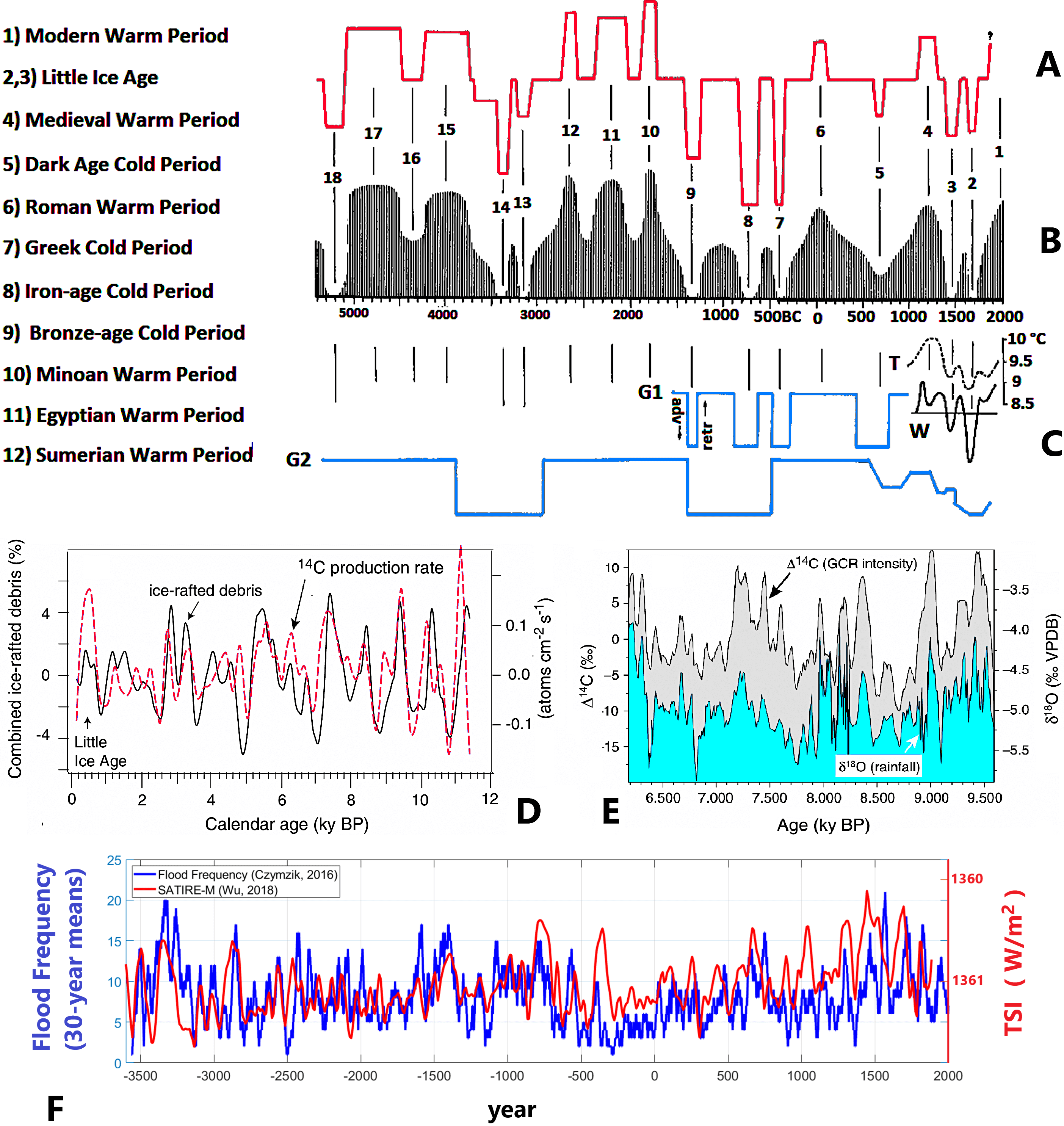}
\par\end{centering}
\caption{{[}A{]} Schematic deviations in the \protect\textsuperscript{14}C
cosmogenic record \citep[adapted from][]{Lin1975}. {[}B{]} Interpretation
of the curve {[}A{]} as representing the long-term envelope of solar
activity. {[}C{]} Four estimates of historical climate: (G1) periods
of Alpine glacier advance and retreat \citep{LeRoyLadurie}; (G2)
global glacier fluctuations \citep{Denton1973}; (T) mean annual temperature
estimates in Central England \citep{Lamb1965}; (W) winter-severity
index for the Paris-London region \citep{Lamb1972}. The left top
side shows a list of documented warm and cold periods in human history.
(Adapted from \citealp{Eddy1977}). {[}D, E{]} Comparison of the \protect\textsuperscript{14}C
cosmogenic record with two climate records (adapted from \citealp{Bond2001,Neff2001,Kirkby2007}).
{[}F{]} Correlation between reconstructed total solar irradiance \citep{Wu2018}
and the flood layer frequency in central Europe (30-year moving window)
(\citealp{Czymzik}: $r=-0.4$, $p<0.0001$).}
\label{Fig17}
\end{figure*}

\subsection{Climate and the changing Sun conundrum}

The correlation between solar activity and climate change throughout
the Holocene has been extensively documented in scientific literature,
particularly since the second half of the 20\textsuperscript{th}
century. This period marked the availability of several long-term
solar and climate proxy records \citep[e.g.:][]{Bray1968,Eddy1976,Eddy1977}.
Figures \ref{Fig17}A, \ref{Fig17}B and \ref{Fig17}C illustrate
this foundational understanding by reproducing, with adaptations,
a figure published by \citet{Eddy1977}, which demonstrates that the
Holocene \textsuperscript{14}C cosmogenic record --- a widely used
proxy for variations in solar activity --- is significantly correlated
with multiple climatic indices. These include: (1) the timing of Alpine
and global glacier advances and retreats \citep{LeRoyLadurie,Denton1973};
(2) an estimated mean annual temperature in England; and (3) reconstructed
winter severity indices for the Paris-London region, dating back to
the Medieval Warm Period \citep{Lamb1965,Lamb1972}. Notably, historical
records extensively document similar alternating warm and cold periods,
which are closely linked to the rise and decline of civilizations
\citep{Diamond2005,Fagan2008}. For instance, the Viking settlements
in Greenland thrived during the Medieval Warm Period --- a time of
prolonged high solar activity --- but were abandoned with the onset
of the Little Ice Age, which was characterized by long periods of
low solar activity \citep{Eddy1976,Lasher2019}.

Recent studies have further substantiated the presence of a significant
correlation between solar and climate variability throughout the Holocene,
including in the last century \citep[e.g.:][and numerous others]{Hoyt1997,Kerr2001,Bond2001,Neff2001,Scafetta2004,Kirkby2007,Steinhilber2012,Scafetta2012b,Soon2013,Czymzik,Connolly2023,ScafettaBianchini2023,Scafetta2023a,Xiao2024}.
Figures \ref{Fig17}D--F provide examples of this connection. However,
as discussed in Section 2, the \citet[AR6,][]{IPCC2021} aligns with
the CMIP6 GCMs' prediction that the solar influence on climate change
is negligible, particularly from 1850--1900 to the present (Figures
\ref{Fig2} and \ref{Fig3}). This raises a fundamental question:
how can the IPCC's claim be reconciled with the paleoclimate evidence
that indicates a robust correlation between solar activity and climatic
records throughout the Holocene?

This apparent paradox has been addressed in numerous recent studies
\citep{Scafetta2019,Scafetta2023a,Connolly2024,Connolly2023}. One
key finding suggests that the CMIP6 GCMs underestimate the solar contribution
to climate change by a large factor because these models rely only
on a specific TSI forcing function characterized by a very low secular
variability \citep{Matthes2017}, which, however, may be fundamentally
flawed. Additionally, these models may likely exclude significant
solar-climate physical mechanisms simply because they are still poorly
understood.

The CMIP6 GCMs primarily attribute the Sun's influence on climate
to variations in its luminosity, as reported in the total solar irradiance
(TSI) and solar spectral irradiance (SSI) forcings. While orbital
changes also affect solar irradiance reaching Earth, these factors
are categorized as orbital forcings and are predominantly relevant
on multi-millennial or longer timescales. The issues related to these
topics warrant further examination, as outlined below.

\begin{figure*}[!t]
\begin{centering}
\includegraphics[width=1\textwidth]{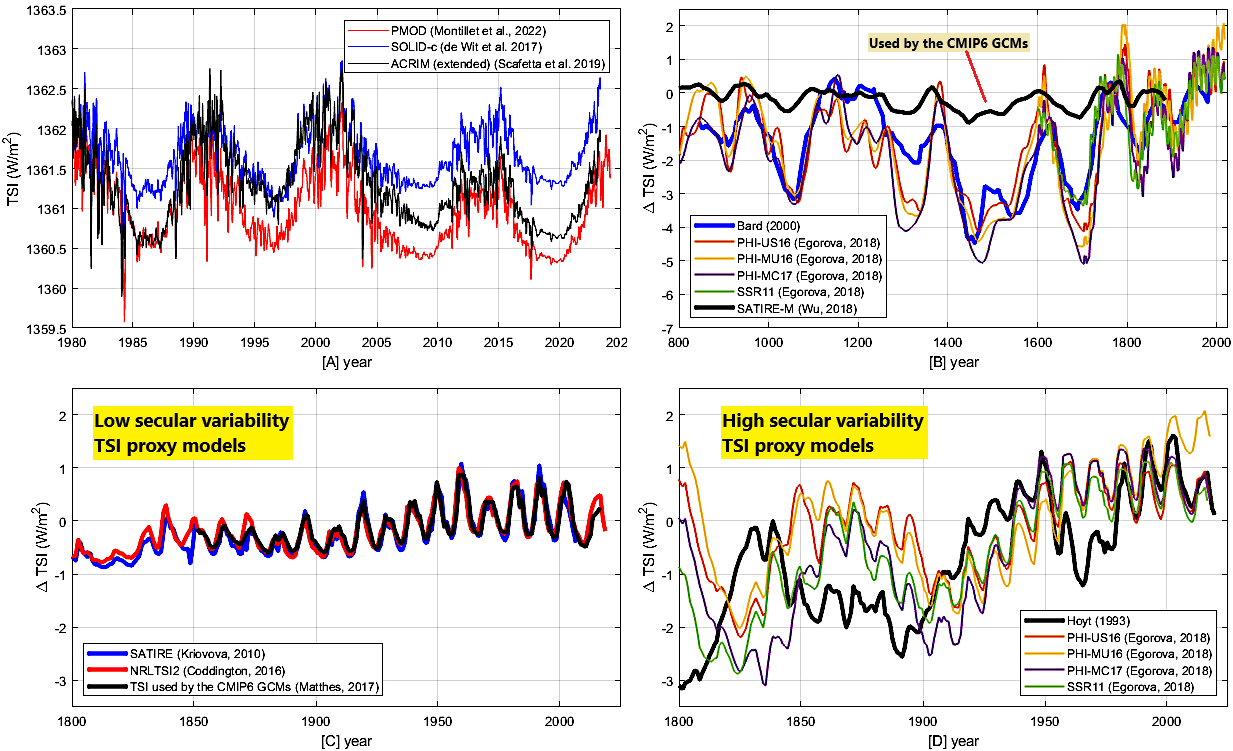}
\par\end{centering}
\caption{{[}A{]} Compilation sample of available total solar irradiance (TSI)
satellite composites. {[}B, C, D{]} Compilation of alternative TSI
proxy records. Data sources are found in \citet{Soon2023}, \citet{Connolly2024}
and \citet{Wu2018}.}
\label{Fig18}
\end{figure*}

\subsubsection{The TSI satellite composites (1978 to date)}

Decadal and long-term variations in solar activity remain largely
uncertain due to the limitations in the available data, as solar irradiance
can only be accurately measured by satellites since 1978 \citep{Hoyt1997,Willson2003}.

Satellite-based total solar irradiance (TSI) records have been the
subject of considerable debate, as their composites are not unique;
for a detailed discussion, see \citet{Scafetta2019} and \citet{Connolly2024}.
Scientific studies highlight that, in addition to the well-documented
11-year solar cycle, various TSI composites are possible, and they
exhibit distinct multidecadal trends. Such differences depend on the
specific TSI satellite records utilized, the adjustments applied to
the officially published data, and the selection of the overlapping
periods. For example, while some TSI composites indicate a slight
decrease or negligible trend from 1980 to 2023, others suggest a modest
positive trend, and others even suggest a pronounced increase from
1980 to 2000 followed by a stabilization or slight decline from 2000
to 2023. Figure \ref{Fig18}A illustrates examples of these TSI satellite
composites; additional details are provided in \citet{Connolly2024}.

Notably, the TSI composites constructed using the 1979--2013 ACRIM
TSI satellite composite \citep{Willson2003} depict an increase from
1980 to 2000, followed by either a slight decrease or a near-stationary
trend depending on the dataset used. The NASA-JPL ACRIM-based TSI
composites show a strong correlation with the global surface temperature
changes from 1980 to 2014, including a rapid warming trend from 1980
to 2000, and a subsequent plateau from 2000 to 2014. These observations
suggest a significant solar contribution to climate change \citep{Scafetta2008}.

By contrast, the CMIP6 GCMs adopted the solar forcing model proposed
by \citet{Matthes2017}, which averages two distinct TSI models: NRLTSI2
(an empirical model proposed by \citealp{Kopp2016}, and \citealp{Coddington2016})
and SATIRE (a semi-empirical model proposed by \citealp{Yeo2014}).
The TSI forcing proposed by \citet{Matthes2017} posits a slight decline
in TSI from 1980 to 2020, which explains why the CMIP6 GCMs attribute
no significant warming to solar influence during this period.

A critical scientific question arises regarding the accuracy of the
TSI forcing proposed by \citet{Matthes2017}, as its declining trend
from 1980 to the present conflicts with the ACRIM-based TSI satellite
composites, which is a composite constructed directly from the published
TSI satellite data. The ACRIM-based composites contradict the NRLTSI2
and SATIRE TSI models. Among the few TSI composites compatible with
NRLTSI2 are those that make use of the satellite TSI records modified
by PMOD \citep{Frohlich2012}. However, even the PMOD-adjusted composites
fail to align with the substantial downward trends in successive solar
minima predicted by the SATIRE model, as highlighted also by \citet{deWit2017}.

The modifications applied by PMOD to the published TSI satellite records
rely on subjective decisions by Dr. Fröhlich, who adopted TSI proxy
models and hypothetical sensor degradation corrections, which were
revised multiple times. This methodology has been criticized, as it
departs from the standard scientific approach, which typically involves
adjusting models to fit experimental published data rather than altering
published data based on model assumptions. In general, published experimental
data can be further adjusted only based on strict experimental considerations
alone. For a comprehensive discussion, see \citet{Scafetta2019} and
\citet{Connolly2024}. The PMOD adjustments of the published TSI satellite
records remain contentious and were never endorsed by the TSI satellite
experimental teams led by other researchers such as Drs. Richard C.
Willson for ACRIM and Douglass V. Hoyt for Nimbus7 \citep{Scafetta2019}.
Consequently, the unresolved ACRIM-PMOD controversy continues to carry
significant implications for both solar and climate science \citep{Connolly2024}.
Despite this, the IPCC \citeyearpar[AR6,][]{IPCC2021} has recently
chosen to disregard this debate, even though it was acknowledged in
the earlier IPCC assessment reports.

\subsubsection{The long-scale TSI proxy models}

To investigate long-term variations in solar activity, researchers
employ proxies such as sunspot numbers and cosmogenic isotope records
of \textsuperscript{14}C and \textsuperscript{10}Be. In the absence
of direct Total Solar Irradiance (TSI) measurements prior to 1978
and due to uncertainties in the TSI satellite composites after 1980,
modeling TSI changes over centuries and millennia remains highly challenging.
While sunspot numbers and cosmogenic records can identify alternating
periods of high and low solar activity, the precise magnitude of solar
luminosity variations remains unknown, rendering rigorous validation
of solar models unfeasible. Consequently, various TSI proxy models
have been proposed in recent decades. Selected examples of these models
are illustrated in Figures \ref{Fig18}B, \ref{Fig18}C, and \ref{Fig18}D.

Notably, the TSI model developed by \citet{Matthes2017}, which underpins
the TSI forcing for the CMIP6 GCMs, shows minimal secular variability,
as it remains nearly constant from 1850 to 2020 (Figure \ref{Fig18}C).
For millennial-scale simulations, the CMIP6 GCMs use the TSI forcing
derived from the SATIRE-M model \citep{Wu2018}, which similarly shows
a modest secular modulation (Figure \ref{Fig18}B). Consequently,
it is unsurprising that the CMIP6 GCMs attribute negligible climate
effects to solar forcing (Figures \ref{Fig1}, \ref{Fig2}, and \ref{Fig10}).

In contrast, other TSI proxy models documented in the literature display
significantly larger secular variability (Figures \ref{Fig18}B and
\ref{Fig18}D) \citep[e.g.:][]{Hoyt1993,Bard2000,Egorova2018,Penza2022}.
These models suggest a marked increase in solar activity since the
17\textsuperscript{th} century and continuing during the 20\textsuperscript{th}
century, which correlates well with the observed global warming since
the Little Ice Age. An increase of solar activity during the 19\textsuperscript{th}
and 20\textsuperscript{th} century is also supported by records of
the solar coronal magnetic field \citep{Lockwood1999}. This solar-climate
correlation is evident over centuries and millennia and is particularly
robust when specific TSI models are chosen \citep{Scafetta2007,Scafetta2023a,Soon2023}.

Given the absence of direct measurements of TSI secular variability,
the proposed TSI proxy models derived also from assumptions that changed
over time. For instance, \citet{Lean1995} estimated that at the Maunder
Minimum (1645--1715) TSI could have been 0.25\% (approximately 3.5
W/m\textsuperscript{2}) lower than the contemporary TSI satellite
baseline, currently estimated at approximately 1361 W/m\textsuperscript{2}.
However, this same estimate was later reduced by \citet{Lean2005}
to 0.1\% (approximately 1.3 W/m\textsuperscript{2}). More recently,
\citet{Yeo2020} asserted that the Sun's darkest possible state could
not exceed a reduction of $2.0\pm0.7$ W/m\textsuperscript{2} relative
to the modern baseline. These revised claims, supported partly by
the low multi-decadal variability of the TSI composites constructed
using the modified PMOD TSI satellite records underlie, the modest
secular variability presented by TSI models such as NRLTSI2 and SATIRE
(Figure \ref{Fig18}C). However, indirect evidence from luminosity
changes in Sun-like stars \citep{Judge2020} and the direct correlation
between climate records and the TSI models with large secular variability
suggest a much larger TSI variability over multi-centennial timescales
\citep{Schmutz2021}.

In fact, \citet{Schmutz2021} presented a compelling argument: the
strong correlation between climate and solar activity records over
secular and millennial scales (Figure \ref{Fig18}B) convincingly
challenges the assumption that solar activity has been characterized
by a minimal multidecadal and secular variability, as those manifested
by the NRLTSI2 and SATIRE TSI models. As illustrated in Figure \ref{Fig10},
when the GCMs are driven by TSI proxy models with low multidecadal
and secular variability, they just fail to capture the cooling transition
from the Medieval Warm Period to the Little Ice Age, casting doubt
on both the validity of these GCMs and their underlying solar forcing
assumptions.

\subsubsection{A solar-induced corpuscular forcing of the climate}

\citet{Schmutz2021} estimated that the emergence of the Maunder Minimum-type
cold climate excursions characteristic of the Little Ice Age necessitates
a total solar irradiance (TSI) reduction of approximately 10 W/m\textsuperscript{2}
from present. This magnitude of TSI change exceeds by more than a
factor of 10 the variability of the TSI functions employed in the
CMIP6 GCMs, but it also is two to three times greater than TSI variability
of the high-variability TSI proxy models (Figure \ref{Fig18}B). Consequently,
the climate system appears to exhibit an amplified sensitivity to
TSI variations \citep{Soon2000,Ziskin2012}. This over-sensitivity
is observed also in analyses of the climate response to the 11-year
solar cycle, whose amplitude is more reliably constrained than the
TSI long-term secular modulation \citep[e.g.:][]{Shaviv2008,Scafetta2023a}.
This raises a critical question: what accounts for the observed amplification
of the solar signal within the climate system? The resolution of this
puzzle may require acknowledging the existence of additional mechanisms
through which solar activity influences climate beyond direct TSI
forcing. In particular, a still poorly understood form of solar corpuscular
forcing has been hypothesized as a potential major contributor.

More specifically, cosmogenic isotope records (\textsuperscript{14}C
and \textsuperscript{10}Be), which serve as proxies for reconstructing
past solar activity, are generated by cosmic rays interacting with
the Earth's atmosphere. Solar magnetic activity modulates the cosmic
ray flux, which increases during periods of low solar activity and
decreases when solar activity is high. Additionally, cosmic ray flux
may experience episodic enhancements due to supernova events. Several
studies have proposed that atmospheric ionization induced by cosmic
rays plays a significant role in the formation of cloud condensation
nuclei, thereby exerting a direct influence on the cloud formation
processes \citep{Shaviv2003,Svensmark1997,Kirkby2007}. A comprehensive
review of these findings is available in \citet{Svensmark2019}.

A decline in solar activity is associated with increased cosmic ray
flux, which in turn leads to a greater atmosphere ionization and,
therefore, a more extended cloud cover, which increases the Earth's
albedo by reflecting more incoming shortwave radiation back into space.
This mechanism could results in a notable surface cooling since also
a very small increase in cloud cover may reduce the solar forcing
at the surface by several watts per square meter \citep{vanWijngaarden2025}.
Conversely, higher solar activity reduces cosmic ray flux, diminishing
cloud cover and thereby lowering the albedo of the Earth, leading
to a surface warming.

The cosmic ray hypothesis posits that variations in solar magnetic
activity may be even more significant than changes in solar luminosity
in driving climate changes via direct corpuscular forcing of the cloud
system \citep{Easterbrook2019}. Empirical evidence supporting this
hypothesis includes findings by \citet{Czymzik}, who identified a
robust correlation between cosmogenic records and flood frequency
across Europe during the Holocene, suggesting a direct link to cloud
formation \ref{Fig17}F.

Despite its potential explanatory power, the cosmic ray hypothesis
remains controversial. The CLOUD experiment at CERN, designed to investigate
this mechanism, concluded that cosmic rays are insufficient to induce
significant nucleation of cloud condensation nuclei \citep{Kirkby2011}.
However, observations of strong correlations between cosmic ray flux
and cloud cover during Forbush decreases --- which are sudden drops
in cosmic ray flux typically caused by solar coronal mass ejections
--- suggest that the conclusions of the CLOUD experiment may require
re-evaluation \citep{Svensmark2016,Svensmark2019,Matsumoto2022}.
Additional recent evidence for a pronounced 11-year solar cycle signature
in cloud records has been documented by \citet{Miyahara2023}. Moreover,
\citet{Svensmark2022} showed that the correlation between the cosmic
ray flux record and climatic data extend over the last 3.5 billion
years, where ocean nutrient proxy data were found to covary with supernova
frequency. These latest results further reinforce the hypothesis that
cosmic rays can significantly drive climate changes.

Another potential corpuscular forcing mechanism involves interplanetary
dust falling on Earth, which remains an area requiring further investigation.
Ionized interplanetary dust may exert direct influence on cloud nucleation
and contribute to climate variability. For instance, \citet{Scafetta2020}
identified a 60-year cycle in meteorite fall frequency that exhibits
strong coherence with the 60-year climatic oscillation (Figures \ref{Fig7}
and \ref{Fig8}). This cycle, which has persisted for centuries, appears
to be linked to the 60-year modulation of Jupiter's orbital eccentricity,
primarily governed by its gravitational interaction with Saturn.

\begin{figure}[!t]
\begin{centering}
\includegraphics[width=1\columnwidth]{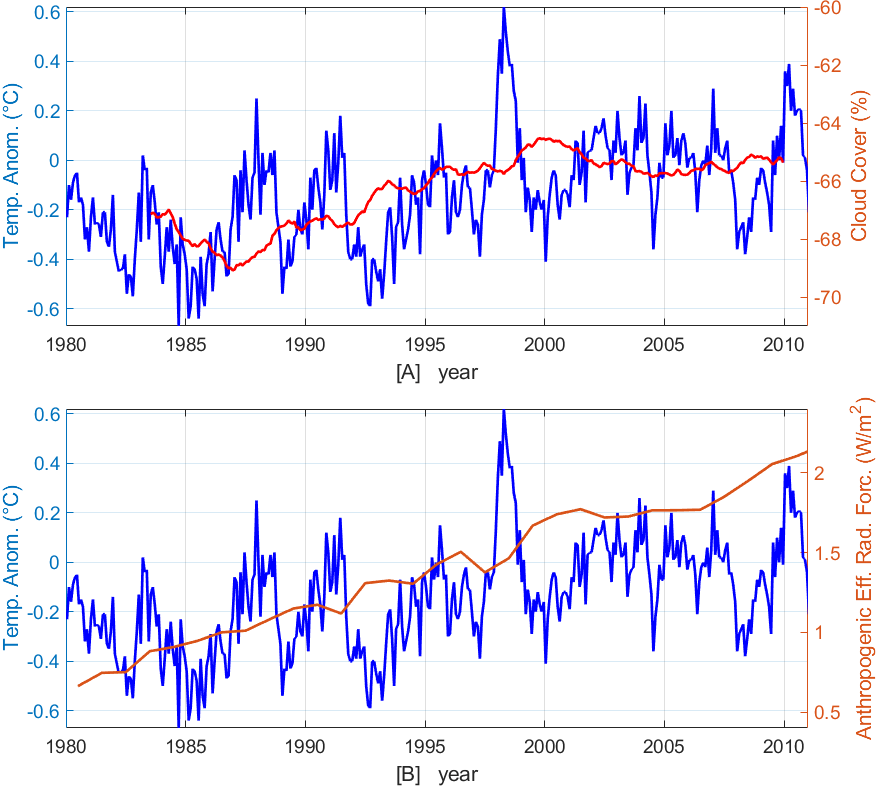}
\par\end{centering}
\caption{Comparison of the UAH-MSU lt v6.1 global satellite temperature record
(blue) with: {[}A{]} the global cloud cover percentage data from the
International Satellite Cloud Climatology Project (ISCCP); {[}B{]}
the total anthropogenic effective radiative forcing employed in the
CMIP6 GCMs.}
\label{Fig19}
\end{figure}

The prospect that solar activity could indirectly influence the climate
through corpuscular forcings and space weather dynamics --- processes
which are not accounted for in the current GCMs --- merits further
scrutiny. Variability in cloud cover has been proposed as a primary
driver of climate change, as illustrated by \citet[its figure 19]{Scafetta2013},
who demonstrated a strong negative correlation between the ISCCP global
monthly cloud cover and the global temperature records from 1983 to
2010.

Figure \ref{Fig19}A highlights this negative correlation by plotting
the UAH MSU lt v 6.1 global satellite temperature record (blue) against
the ISCCP global cloud cover percentage record from July 1983 to December
2009 (inverted scale). Both curves exhibit a well-correlated increase
from 1980 to 2000 followed by a more constant level from 2000 to 2010.
In contrast, Figure \ref{Fig19}B juxtaposes the same temperature
record with the total anthropogenic effective radiative forcing function
adopted by the CMIP6 GCMs (Figure \ref{Fig1}C), which exhibits a
continuous linear increase from 1980 to 2010. This comparison underscores
that global temperature records align more closely with variations
in cloud cover than with the anthropogenic forcing record.

\section{Modeling climate change using empirical and semi-empirical models}

As discussed above, several fundamental challenges complicate the
detection, attribution, and modeling of climate change. Global surface
temperature records may exhibit significant warming biases, while
numerical climate models may contain unresolved physical uncertainties.
Additionally, the actual solar forcing --- recognized as the dominant
natural driver of climate change on millennial timescales --- remains
highly uncertain because the long-term solar activity variability
is not well known and the mechanisms through which solar phenomena
influence climate may include both radiative forcing and a poorly
understood corpuscular forcing. In response to these concerns, various
proposals advocate for the use of empirical and semi-empirical models
to enhance climate change modeling.

\subsection{From reductionist to holistic approaches in climate modeling}

The Intergovernmental Panel on Climate Change Assessment Reports \citeyearpar{IPCC2007,IPCC2013,IPCC2021}
have relied predominantly on computer-based global climate models
(GCMs). These models exemplify a reductionist scientific approach,
which seeks to simplify complex systems by focusing on individual
components. While reductionism can offer valuable insights, it also
presents several limitations such as:
\begin{enumerate}
\item \emph{loss of context} --- by isolating individual components, reductionist
methodologies may overlook critical interactions, leading to an incomplete
understanding of the climate system.
\item \emph{oversimplification} --- certain essential processes, such as
cloud formation, remain poorly represented in the GCMs, affecting
their accuracy in climate simulation and projections;
\item \emph{emergent properties} --- some systemic properties only manifest
when the climate system is analyzed holistically and cannot be discerned
by examining isolated components;
\item \emph{interconnected systems} --- complex climate feedback mechanisms
and relationships may be inadequately captured by reductionist models,
leading to misleading conclusions;
\item \emph{application limits} --- insights derived from reductionist
models may not be readily transferable to real-world applications
where a broader, integrative understanding is required.
\end{enumerate}
Given these constraints, a more comprehensive approach incorporating
holistic methodologies is necessary to achieve an improved understanding
of complex climate dynamics. Holistic models, particularly empirical
and semi-empirical frameworks, offer several advantages such as:
\begin{enumerate}
\item \emph{reliance on observational data} --- these models emphasize
empirical data, enabling a direct modeling of real-world climate patterns,
thereby addressing limitations present in numerical climate models
such as the CMIP6 GCMs;
\item \emph{simplicity and transparency} --- compared to complex GCMs,
empirical models tend to be more accessible and interpretable;
\item \emph{robust statistical frameworks} --- these models apply rigorous
statistical techniques to quantify the influence of both anthropogenic
and natural climate drivers;
\item \emph{direct attribution of climate signals} --- by comparing observed
trends with historical records, empirical models help to link specific
climate variations to their underlying physical drivers;
\item \emph{policy and adaptation applications} --- empirical models provide
clear evidence that can inform climate policy and adaptation strategies
more effectively than complex numerical models.
\end{enumerate}
However, empirical models also have inherent limitations:
\begin{enumerate}
\item \emph{data dependence} --- the reliability of empirical models is
contingent on the quality and completeness of the available data,
which can sometimes lead to inaccuracies;
\item \emph{simplification assumptions} --- while useful, empirical models
may overlook complex climate interactions, leading to potential misrepresentations;
\item \emph{nonlinearity and complexity} --- the inherently nonlinear behavior
of the climate system poses challenges for empirical modeling, sometimes
resulting in oversimplifications;
\item \emph{regional specificity} --- findings based on one geographic
region may not be universally applicable, complicating broader climate
assessments.
\item \emph{attribution uncertainty} --- while empirical models provide
valuable insights into climate change attribution, they may be less
robust than physically based models for quantifying the contributions
of individual factors.
\end{enumerate}
Thus, the study of complex natural systems necessitates the integration
of both holistic and reductionist methodologies. Since each approach
has inherent strengths and limitations, comparing numerical and empiric
models should allow for a more accurate representation of climate
processes. Let us now discuss some of the empirical modeling of the
climate system proposed in scientific literature.

\subsection{Empirical and semi-empirical approaches in climate science}

Several types of empirical and semi-empirical models are commonly
employed in climate science such as:
\begin{enumerate}
\item \emph{Empirical Climate Models} (ECMs) --- ECMs use statistical techniques
such as multiple linear regression and spectral analysis to identify
relationships between climate change and external drivers. Examples
include models assessing El Niño-Southern Oscillation dynamics, volcanic
aerosol contributions, solar radiation, and anthropogenic influences.
These models provide preliminary attribution of observed temperature
changes.
\item \emph{Energy Balance Models} (EBMs) --- EBMs represent climate dynamics
by balancing incoming solar radiation with outgoing thermal emissions.
They range from simple zero-dimensional (0-D) models, which treat
the Earth as a single thermal unit, to more complex one-dimensional
(1-D) and multi-dimensional formulations that account for latitudinal
and spatial variations.
\item \emph{Statistical Downscaling Models} (SDMs) --- SDMs refine broad-scale
climate projections from the GCMs to more localized resolutions, enabling
assessments of regional climate impacts.
\item \emph{Machine Learning Models} (MLMs) --- Machine learning methodologies
have recently been introduced in climate research, offering promising
capabilities for identifying nonlinear relationships and predicting
climate patterns from large datasets. While MLMs have notable potential,
challenges related to interpretability and data requirements remain.
\end{enumerate}
Each of these modeling approaches possesses unique strengths and limitations,
and they are frequently utilized in combination to provide a more
comprehensive understanding of climate change.

The following subsections provide an overview of empirical models
presented in the literature that attribute climate change to both
anthropogenic and natural drivers by directly analyzing the climatic
records. This discussion will primarily focus on studies published
after the release of the IPCC's Sixth Assessment Report \citeyearpar{IPCC2021}.
These models employ diverse mathematical methodologies and underscore
the significant role of solar influences in climate change, though
the extent of this contribution varies depending on underlying assumptions
and the specific methodologies used by model developers. For further
technical details and statistical validation, readers are encouraged
to consult the cited references.

\subsection{A simple linear regression model}

Regression models are widely regarded as the simplest and most used
analytical tools in climate change attribution studies. They rely
on statistical techniques such as multi-linear regression analysis
to explore relationships between climate variables and a set of predictors.
For instance, \citet{Chylek2014} utilized this approach to model
the global surface temperature record by incorporating known radiative
forcing functions of greenhouse gases (GHG), aerosols (AER), solar
irradiance (SOL), volcanic activity (VOLC), as well as indices for
the El Niño Southern Oscillation (ENSO) and the Atlantic Multidecadal
Oscillation (AMO) as explanatory variables. They proposed the following
multi-linear regression model:
\begin{align}
T(t) & =a_{0}+a_{1}\cdot F_{GHG}(t)+a_{2}\cdot F_{AER}(t)+a_{3}\cdot F_{SOL}(t)+a_{4}\cdot F_{VOLC}(t)+a_{5}\cdot I_{ENSO}(t)+a_{6}\cdot I_{AMO}(t)\label{eq:1}
\end{align}
where the free parameters $a_{0...6}$ represent six regression coefficients.
In their analysis, \citet{Chylek2014} adopted the climate forcin

g functions derived from the GISS ModelE \citep{Hansen2007}, which
utilizes an earlier version of the NRLTSI TSI model based on the PMOD-modified
TSI satellite composite \citep{Lean2000}. They concluded that anthropogenic
forcing and the positive phase of the AMO collectively accounted for
approximately two-thirds and one-third, respectively, of the global
warming observed since 1975.

Despite its apparent utility, the proposed model has several potential
limitations. First, its reliance on a potentially inaccurate solar
forcing function likely skewed the results because as the post-1975
warming, which the model attributed to AMO, could have also be partially
attributed to solar activity variations if TSI records compatible
with the ACRIM TSI satellite composite were used \citep{Scafetta2008,Scafetta2009,Connolly2023}.
Second, the climate system processes the radiative forcing functions
through inherently non-linear mechanisms, and thus the assumed linear
relationships between $T(t)$ and its radiative forcing components
may oversimplify reality. Furthermore, ENSO and AMO are subsystems
of the climate system and, therefore, they cannot necessarily be considered
physically independent of the forcing functions.

Multi-linear regression analysis can also produce ambiguous or erroneous
results when significant collinearity exists among predictors or when
non-linear effects are neglected. For example, \citet{Benestad2009}
applied a multilinear regression model to global surface temperatures
incorporating the ten radiative forcing functions used by the GISS
model. Their result suggested that solar forcing may have contributed
approximately 0.1°C to the 0.65°C warming observed from 1900 to 2000,
but they acknowledged the lack of robustness for such finding.

In fact, \citet{Scafetta2013b} demonstrated that multilinear regression
models using GISS forcing functions could effectively reproduce global
temperature records even if the well-mixed GHG forcing function were
ignored because of its collinearity with the other predictors. Additional
analyses by \citet{Scafetta2006a,Scafetta2006b,Scafetta2007} and
later \citet{Scafetta2009} indicated that the solar influences may
have contributed to 50\% or more of the warming between 1900 and 2000.
These findings were derived using analytical methods such as wavelet
frequency decomposition and simplified zero-dimensional energy balance
models, which account for some non-linear effects. Models such as
the one proposed in Eq. \ref{eq:1} may underestimate solar contributions
due to the climate system's tendency to attenuate high-frequency signals
while amplifying low-frequency components, a consequence of its thermal
inertia. In this context, \citet{Benestad2009} identified a significant
solar signature, though it was attenuated relative to the finding
of \citet{Scafetta2006a,Scafetta2006b,Scafetta2007} largely due to
methodological errors in wavelet analysis, as later highlighted by
\citet{Scafetta2013b}.

\subsection{A simple ECM based on a zero-dimensional EBM}

Empirical regression models are most effective when they balance simplicity
and complexity. They should incorporate the minimum number of physically
relevant predictors, chosen carefully to avoid collinearity. Additionally,
these models should simulate some non-linearity. For example, radiative
forcing functions are not inherently linear predictors of climate
behavior, as the climate system processes them by attenuating the
high-frequency components relative to the low-frequency ones due to
its thermal capacity.

Taking into account the above considerations, \citet{Scafetta2023a}
proposed modeling the global surface temperature record, $T(t)$,
using the following equation:
\begin{equation}
T(t)=T_{A}(t)+T_{V}(t)+T_{S}(t)+\xi(t)=T_{AVS}(t)+\xi(t),\label{eq:2}
\end{equation}
where $T(t)$ is derived from four components: anthropogenic, $T_{A}(t)$;
volcano, $T_{V}(t)$; solar, $T_{S}(t)$; plus a fast-fluctuating
component, $\xi(t)$, simulating fast fluctuations such as the ENSO
signal. The three main contributors can be estimated using a 0-D energy
balance equation with a given time response, which can be generally
modeled as:

\begin{equation}
C\,\frac{dT(t)}{dt}=\frac{1-a}{4}\,S-\sigma\varepsilon\,T(t)^{4}+G.\label{eq:3}
\end{equation}
On the left side, $T(t)$ is the mean global temperature in Kelvin,
$C$ represents the effective heat capacity of the Earth's surface
and atmosphere. On the right side, in the first term $a$ denotes
the albedo, $S$ is the incoming solar radiation adjusted for the
planetary sphericity (the denominator ``4''); the second term represents
the outgoing longwave radiation, which depends on the temperature
itself; finally, $G$ incorporates the effect of the greenhouse gases.

By redefining the constants and simplifying the formulation, the above
equation can be approximated for each forcing component using differential
equations based on the temperature anomalies, $\Delta T(t)=T(t)-T_{0}$,
which are relative to a given mean temperature $T_{0}$:
\begin{eqnarray}
\frac{\Delta T_{A}(t)-\Delta T_{A}(t-\Delta t)}{\Delta t} & = & \frac{k_{A}F_{A}(t)-\Delta T_{A}(t-\Delta t)}{\tau}=\alpha_{A}\cdot F_{A}(t)-\beta\cdot\Delta T_{A}(t-\Delta t)\label{eq:4}\\
\frac{\Delta T_{V}(t)-\Delta T_{V}(t-\Delta t)}{\Delta t} & = & \frac{k_{V}F_{V}(t)-\Delta T_{V}(t-\Delta t)}{\tau}=\alpha_{V}\cdot F_{V}(t)-\beta\cdot\Delta T_{V}(t-\Delta t)\label{eq:5}\\
\frac{\Delta T_{S}(t)-\Delta T_{S}(t-\Delta t)}{\Delta t} & = & \frac{k_{S}F_{S}(t)-\Delta T_{S}(t-\Delta t)}{\tau}=\alpha_{S}\cdot F_{S}(t)-\beta\cdot\Delta T_{S}(t-\Delta t)\label{eq:6}
\end{eqnarray}
Here, $F_{A}(t)$, $F_{V}(t)$, and $F_{S}(t)$ represent the anthropogenic,
volcanic, and solar effective forcings, respectively. The coefficients
$\alpha_{A}=k_{A}/\tau$, $\alpha_{V}=k_{V}/\tau$, $\alpha_{S}=k_{S}/\tau$,
and $\beta=1/\tau$ depend on the sensitivity parameters $k_{A}$,
$k_{V}$ and $k_{S}$ and the characteristic time response $\tau$,
which is directly related to the Earth's effective heat capacity.
For simplicity, $\tau$ is assumed constant for all forcings.

Under equilibrium conditions, the temperature change $\Delta T_{2\times CO_{2}}$,
with corresponds to doubling atmospheric CO\textsubscript{2} concentration
(for example, from 280 ppm to 560 ppm), is the equilibrium climate
sensitivity

\begin{equation}
ECS=\Delta T_{2\times CO_{2}}=k_{A}\,\Delta F_{2\times CO_{2}}=3.7\,k_{A}.\label{eq:7}
\end{equation}
For the transient climate response (TCR), which is defined as the
temperature average over 20 years around the time of CO\textsubscript{2}
doubling under a linear increase scenario (1\% annually over 70 years),
the calculation gives:
\begin{equation}
TCR=3.7\,k_{A}-\frac{3.7}{70}\,k_{A}\tau\left(1-e^{-70/\tau}\right)=ECS\left[1-\frac{\tau}{70}\left(1-e^{-70/\tau}\right)\right].\label{eq:8}
\end{equation}
By iterating, these equations can be reformulated as:
\begin{equation}
\frac{\Delta T(t)-\Delta T(t-\Delta t)}{\Delta t}=\alpha_{A}\cdot F_{A}(t)+\alpha_{V}\cdot F_{V}(t)+\alpha_{S}\cdot F_{S}(t)-\beta\cdot\Delta T(t-\Delta t),\label{eq:9}
\end{equation}

where $\Delta t$ is set to 1 year. Using multilinear regression analysis,
the regression coefficients ($\alpha_{A}$, $\alpha_{V}$, $\alpha_{S}$
and $\beta$) can be evaluated, and can be used to reconstruct the
climatic signatures of each forcing and compute $\Delta T_{AVS}(t)$,
the deterministic component of $T(t)$. Unlikely Eq. \ref{eq:1},
the model expressed by Eq. \ref{eq:9} accounts for the climate system's
non-linear processing of the forcing functions due to its heat capacity,
which leads to a timescale-dependent amplitude and time-lag in the
response of the climate system to the input forcing function.

\citet{Scafetta2023a} demonstrated that if the climate sensitivity
is set to be equal for all forcings ($\alpha_{A}=\alpha_{V}=\alpha_{S}$)
--- indicating that the climate is equally sensitive to all types
of forcings --- the solar contribution to the global warming from
1850--1900 to 2011--2022 is minimal, particularly when it is adopted
the low-secular-variability TSI model used by the CMIP6 GCMs. Under
this scenario, the ECS was estimated to be $2.1\pm0.7$ °C (17--83\%
range), which is consistent with the low ECS range proposed by the
\citet{IPCC2021}. In this circumstance, the result suggests that
most of the observed warming could be attributable to the anthropogenic
forcing, although the ECS was estimated to be lower than 3.0°C. This
empirically derived ECS range optimally fits the findings of \citet{Lewis2023}
and \citep{Scafetta2021c,Scafetta2022a,Scafetta2023b}.

\begin{figure*}[!t]
\begin{centering}
\includegraphics[width=1\textwidth]{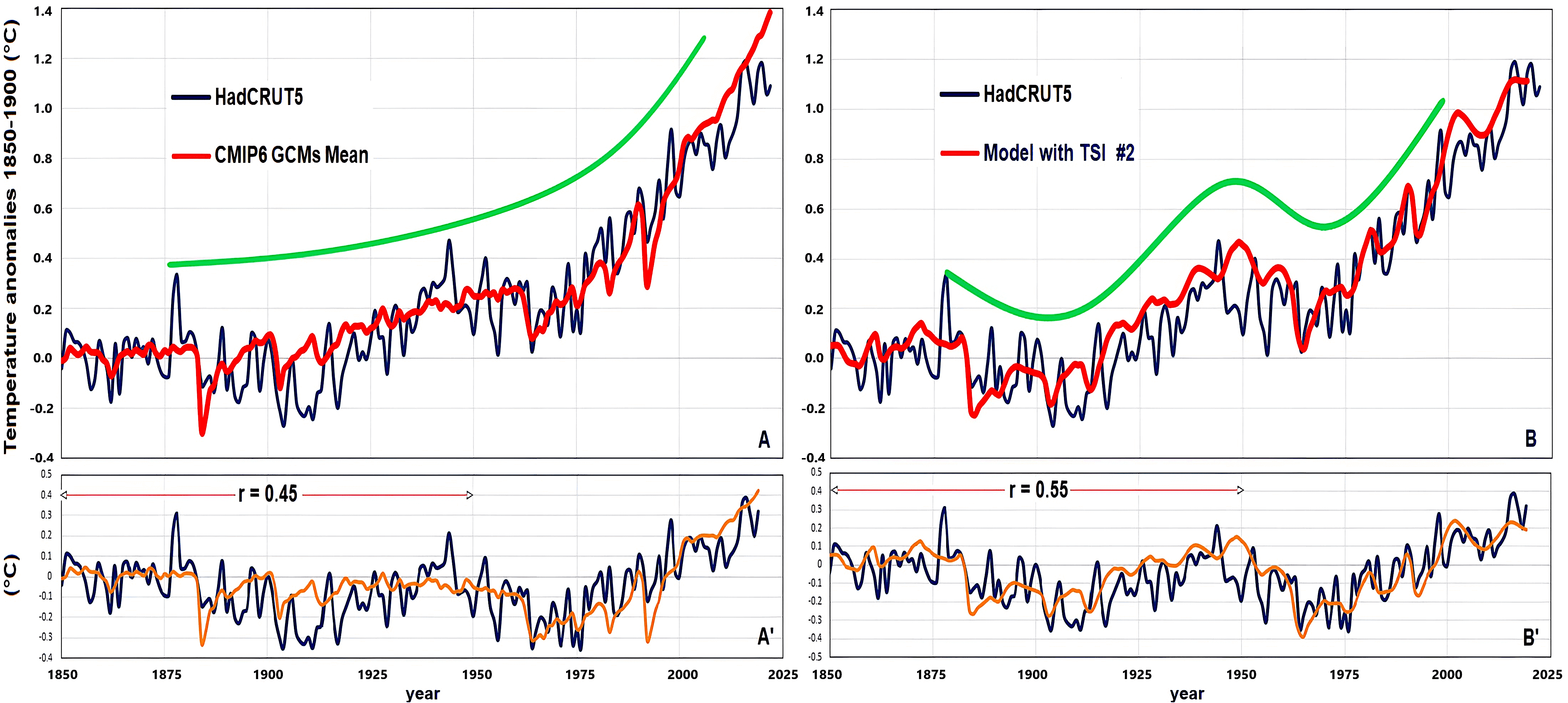}
\par\end{centering}
\caption{{[}A{]} Comparison of the HadCRUT5 global surface temperature record
with the CMIP6 GCM ensemble average. {[}A'{]} Both records are detrended
using the function $f(t)=a(x-1850)^{2}$, correlation coefficient
of $r=0.45$ for the period 1850--1950. {[}B{]} Comparison of the
HadCRUT5 global surface temperature record with the ECM derived from
Eq. \ref{eq:2} utilizing a medium-high variability Total Solar Irradiance
(TSI) proxy model and assuming the condition $\alpha_{A}=\alpha_{V}\protect\neq\alpha_{S}$.
{[}B'{]} The same detrending analysis as in {[}A'{]} is applied, yielding
an improved correlation coefficient of $r=0.55$. Adapted from \citet{Scafetta2024}.}
\label{Fig20}
\end{figure*}

Alternatively, assuming $\alpha_{A}=\alpha_{V}\neq\alpha_{S}$, which
means that the climate sensitivity to solar variations can be different
from that to the other radiative forcings, \citet{Scafetta2023a}
found that the climate could be also 4 to 6 times more responsive
to TSI forcing than to anthropogenic or volcanic forcing. In this
circumstance, the result implies an ECS that could be as low as $1.1\pm0.4$
°C (17--83\% range), suggesting that solar activity may directly
influence the Earth's albedo by physical mechanisms currently absent
in the GCMs. Under this scenario, if high-secular-variability TSI
models are adopted, solar activity changes could explain $\sim$50\%
or more of the observed warming from 1850--1900 to 2011--2022.

Figures \ref{Fig20}A and \ref{Fig20}B, respectively, compare the
HadCRUT5 global surface temperature record (black) against the CMIP6
GCM ensemble mean simulation and the ECM output (red) curves. The
latter was obtained using high-secular-variability TSI models under
the condition $\alpha_{A}=\alpha_{V}\neq\alpha_{S}$. The green curves
in both panels A and B highlight the distinct multi-decadal modulation
pattern evident in the red curves of the two types of models. In panel
A, the red curve demonstrates a steadily increasing trend, reflecting
the anthropogenic forcing function. In contrast, the red curve in
panel B exhibits a quasi 60-year oscillatory behavior that closely
resembles the observed climate temperature record. This distinction
becomes more evident in panels A' and B', which display the detrended
curves. In Figure \ref{Fig20}A', the correlation coefficient for
the period 1850 to 1950 is $r=0.45$, whereas in Figure \ref{Fig20}B',
the correlation coefficient is notably higher at $r=0.55$. More specifically,
the CMIP6 GCM ensemble average simulation inadequately captures the
warming observed from the 1900s to the 1940s and the subsequent cooling
from the 1940s to the 1970s. In contrast, these patterns are more
accurately reproduced by the ECM derived from Eq. \ref{eq:2} under
the outlined assumptions.

\subsection{Astronomically-based empirical harmonic models of natural climate
variability}

There is another widely utilized empirical approach to model climate
change, which again considers three primary components: anthropogenic,
volcanic, and other natural factors. However, the latter component
is conceptualized as made of a complex harmonic signal. The modeling
is based on the hypothesis that these other natural climatic components
derive from astronomical, solar and lunar forcings that could exhibit
approximate harmonic behavior. This empirical methodology involves
analyzing the periodogram of the climate records, identifying relevant
frequencies, and generating constituent harmonic functions using multiple
linear regression analysis. The harmonics are subsequently combined
to empirically reconstruct the hypothesized harmonic natural patterns
of the climate system.

One notable advantage of this approach lies in its ability to extend
harmonics for forecasting purposes. For example, it is well known
that seasonal variations can be approximately modeled using annual
and semi-annual cycles. However, issues arise when the selected frequencies
lack physical justification. Misinterpretation of artifacts in the
periodogram as genuine physical frequencies may lead to harmonic models
that perform well within the calibration interval, and yet they could
diverge substantially from reality outside the regression period.
Furthermore, periodograms of climatic records may only approximate
true physical frequencies, particularly in the low-frequency domain
where error margins are larger.

To mitigate the possibility of using non-physical frequencies, \citet{Scafetta2010,Scafetta2013,Scafetta2021b}
proposed selecting only those frequencies that can be theoretically
derived from astronomical considerations, which is a methodology analogous
to that employed in ocean tidal modeling since the seminal work of
Sir William Thomson (Lord Kelvin) in 1879 (for a recent application
of this methodology see \citealp{Ardalan2007}). This approach appears
justified, because temperature records present periodograms spectrally
coherent with planetary, solar, and lunar harmonics \citep{Scafetta2010,Scafetta2014b,Scafetta2021b}.
Solar records themselves exhibit planetary harmonics across annual
to multi-millennial scales \citep{Scafetta2012b,Scafetta2014b,Scafetta2016,ScafettaN2020,ScafettaBianchini2023}.
These harmonics include the 11-year solar cycle, the Eddy quasi-millennial
cycle, and the Bray-Hallstatt oscillation (\textasciitilde 2318 years)
identified in radiocarbon and climate records spanning the Holocene.
In general, it appears that the natural harmonics of the solar system
synchronize both solar and climate oscillations, although the mechanisms
underlying such planetary synchronization phenomenon remain under
investigation \citep[e.g.:][]{Scafetta2012d,ScafettaB2022,Stefani2024}.

\begin{figure*}[!t]
\begin{centering}
\includegraphics[width=1\textwidth]{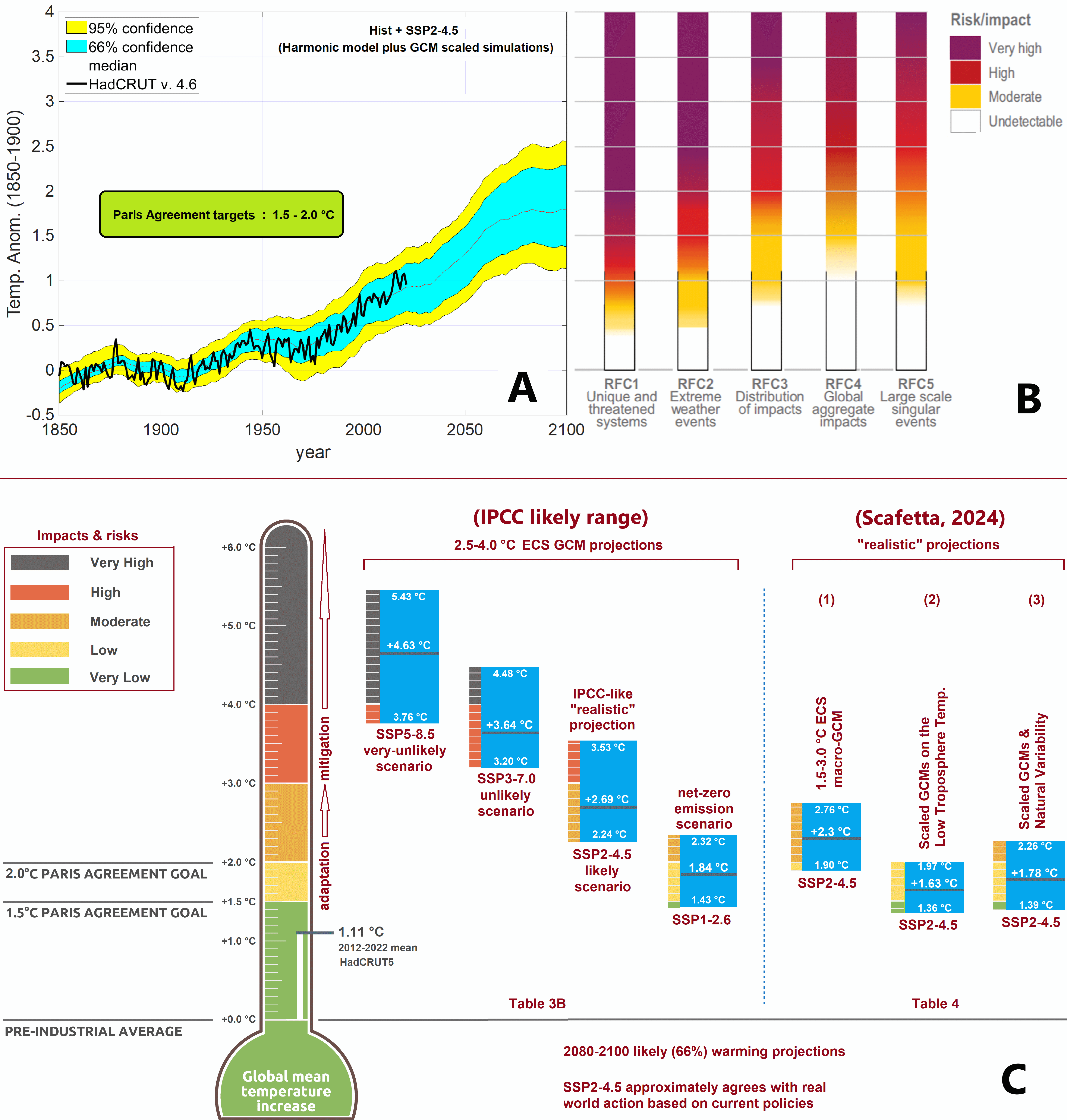}
\par\end{centering}
\caption{{[}Top{]} Comparison of the harmonic empirical global climate model
under the SSP2-4.5 scenario with the HadCRUT4.6 record (1850--2021)
\citep{Morice2012} alongside the burning ember diagrams representing
the five primary global Reasons for Concern (RFCs) under low-to-no
adaptation scenarios, as reported by the \citet{IPCC2023} AR6. {[}Bottom{]}
Summary and analysis of the projected impacts and risks of global
warming for the 2080--2100 period compared to the climate \textquotedblleft thermometer\textquotedblright{}
projections from \citet{CAT2024}. Adapted from \citet{Scafetta2024}.}
\label{Fig21}
\end{figure*}

\citet{Scafetta2021b} extended the model first proposed in \citet{Scafetta2010}
by hypothesizing that the climate system incorporates 13 major harmonics
with periods spanning approximately 3 to 1000 years. Anthropogenic
and volcanic contributions were derived from the ensemble mean of
the CMIP6 GCM simulations, utilizing half the climate sensitivity
estimated by these models, which means that the ECS was estimated
to be about 1.5-2.0°C, to incorporate a natural variability described
by the adopted harmonic constituent model. The proposed equation is
as follows

\begin{equation}
T(t)=T_{0}+\sum_{i=1}^{13}A_{i}\sin\left[2\pi\left(f_{i}\cdot(t-2000)+\alpha_{i}\right)\right]+0.5\cdot GCM(t),\label{eq:10}
\end{equation}
where $A_{i}$, $f_{i}$ and $\alpha_{i}$ per $i=1,...,13$ are coefficients
discussed in detail by \citet{Scafetta2021b,Scafetta2024}.

Figure \ref{Fig21}A shows the output of the above model extended
to 2100 using the SSP2-4.5 scenario, which is considered to be the
most realistic one \citep{Hausfather2020,Pielke2022}. Figure \ref{Fig21}B
presents burning ember plots for the top five global reasons for concern
(RFCs), based on the IPCC \citeyearpar[AR6,][]{IPCC2023} findings
under minimal adaptation assumptions. The comparison of panels A and
B indicates that the projected temperatures throughout the 21\textsuperscript{st}
century are likely to remain within the safe adaptation threshold
indicated by the \citet{ParisAgreement} limiting global warming to
less than 2°C relative to the 1850--1900 pre-industrial period. By
contrast, the CMIP6 GCM projections forecast significantly higher,
and potentially hazardous, warming levels (Figure \ref{Fig3}). Similar
conclusions were reached by \citet{Connolly2020} using an alternative
and more simple empirical model with low climate sensitivity to radiative
forcing under business-as-usual scenarios.

Figure \ref{Fig21}C compares the IPCC predictions with the outputs
of alternative empirical climate modeling under three different assumptions
based on the considerations discussed in the above sections:
\begin{enumerate}
\item using only the CMIP6 GCMs with ECS < 3°C;
\item rescaling the GCMs to fit the satellite low troposphere temperature
record;
\item using the empirical model expressed by Eq. \ref{Fig11}.
\end{enumerate}
Detailed explanations are found in \citet{Scafetta2024}.

The empirical approaches exhibit more robust alignment with the historical
data and project a moderate climate warming throughout the 21\textsuperscript{st}
century under realistic development scenarios, such as the SSP2-4.5.
These projections suggest that adaptation policies may be sufficient
to address future climate challenges without the necessity of adopting
economically disruptive Net-Zero mitigation strategies to meet the
climate targets of the \citet{ParisAgreement}.

\begin{figure*}[!t]
\begin{centering}
\includegraphics[width=1\textwidth]{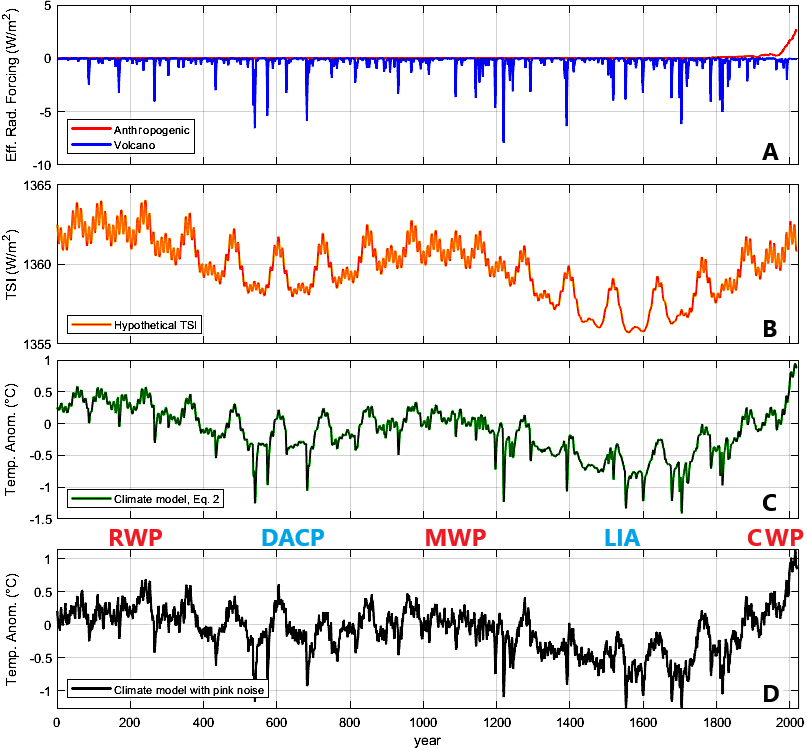}
\par\end{centering}
\caption{{[}A{]} Extended effective anthropogenic and volcanic forcings utilized
by CMIP6 GCMs, backdated from 1 to 1750 AD. {[}B{]} Synthetic model
of total solar irradiance constructed using specific identified climatic
harmonics \citep{Scafetta2012b,ScafettaN2020}. {[}C{]} Climate output
generated using the ECM based on Eq. \ref{eq:2}, incorporating the
parameters obtained by processing the TSI-2 and HadSST4 records from
\citet{Scafetta2023a}. {[}D{]} Same as {[}C{]}, but with pink noise
added to simulate the internal variability of the climate system.}
\label{Fig22}
\end{figure*}

\subsection{A synthetic solar-based empirical climate model from 1 to 2017 AD}

Let us finally discuss a synthetic empirical climate model spanning
from 1 to 2017 AD that may provide relevant insights into climate
patterns over the past two millennia. Figure \ref{Fig22} illustrates
an ideal application of the empirical climate model (ECM) formulated
in Eq. \ref{eq:2}, which integrates the following components:
\begin{itemize}
\item \emph{anthropogenic effective radiative forcing} (Figure \ref{Fig22}A)
--- zero anthropogenic forcing is applied from 1 to 1749 AD, while
the forcing values adopted by the CMIP6 GCMs are utilized for the
period from 1750 to 2017;
\item \emph{volcanic effective radiative forcing} (Figure \ref{Fig22}A)
--- the volcanic forcing values used in the CMIP6 GCMs for the period
1750--2017 are extended back from 1 to 1749 AD using the volcanic
aerosol optical depth records provided by \citet{Toohey2017};
\item \emph{solar effective radiative forcing} (Figure \ref{Fig22}B) ---
this forcing is derived from a synthetic total solar irradiance (TSI)
model that simulates the major features of the proxy solar records
(Figure \ref{Fig18}B). The function is calculated by subtracting
1361 W/m\texttwosuperior{} and dividing by 8 to align with the TSI
effective forcing scale used in the CMIP6 GCMs.
\end{itemize}
The synthetic TSI model serves as an idealized representation of solar
activity variations by reproducing five long harmonics whose frequencies
and phases are strictly based on astronomical considerations \citep{Scafetta2012b,Scafetta2014b,ScafettaN2020,ScafettaBianchini2023,ScafettaMilani2025}.
Additionally, the model incorporates a simulated 11-year solar cycle,
with amplitude variations linked to solar activity, that is it diminishes
during grand solar minima such as the Maunder Minimum of the 17\textsuperscript{th}
century. The periods of the five harmonics are: $P_{1}=60.95$ years,
$P_{2}=114.78$ years, $P_{3}=129.95$, years, $P_{4}=983.40$ years,
and $P_{5}=2318$ years. The chosen secular harmonics --- $P_{1}$,
$P_{2}$, and $P_{3}$ --- account for alternating solar maxima and
minima, such as the Maunder and Dalton minima, while the longer millennial
oscillations --- $P_{4}$ and $P_{5}$ --- correspond to the Eddy
and the Bray-Hallstatt cycles observed in both solar and climate records
\citep{McCracken2014,Scafetta2016}. The amplitude of these harmonics
approximates the variability found in high-secular-variability TSI
models, as those shown in Figure \ref{Fig18} \citep{Bard2000,Egorova2018}.

The ECM model (Eq. \ref{eq:2}) used the following parameters, estimated
by \citet{Scafetta2023a} using a solar variability model and the
HadSST4 record: $\alpha_{A}=\alpha_{V}=0.083$ °Cm\textsuperscript{2}/yW,
$\alpha_{S}=0.427$ °Cm\textsuperscript{2}/yW, and $\beta=0.303$
1/y\textsuperscript{}. These values imply an ECS equal to $1\pm0.3$
°C, and an enhanced sensitivity to TSI forcing by a factor of $5.1\pm1.5$.

Figures \ref{Fig22}C and \ref{Fig22}D depict the outputs of Eq.
\ref{eq:2}, with Figure \ref{Fig22}D incorporating pink noise to
simulate the internal variability of the climate system. The climate
simulation exhibits qualitative similarity to the proxy temperature
records depicted in Figures \ref{Fig10} and \ref{Fig12}, and with
those reported in \citet{Esper2024}.

A quasi-millennial cycle is evident, with amplitudes smaller during
the first millennium compared to the second. This asymmetry, observed
in many paleoclimatic reconstructions, is attributable to the Bray-Hallstatt
cycle (\textasciitilde 2318 years), which is characterized by grand
solar minima occurring around $\sim$370 and $\sim$1530 AD, respectively.
The maxima of the quasi-millennial cycle occur around 95, 1175, and
2060 AD, producing warm and cold periods such as the Roman Warm Period
(RWP), the Little Ice Age (LIA), and the Current Warm Period (CWP),
as shown in the TSI model depicted in Figure \ref{Fig12}A and in
the climate record plotted in Figure \ref{Fig12}B and \ref{Fig12}C.
The model suggests that approximately 50\% of the warming between
1850--1900 and 2011--2020 could be attributed to solar activity,
which is consistent with other studies \citep[e.g.:][]{Scafetta2006b,Scafetta2009,Scafetta2007}.

This numerical experiment highlights the importance of utilizing solar
records with a substantial multi-secular variability and, in addition,
including mechanisms that render the climate hypersensitive to solar
activity changes when this is represented by TSI changes alone. The
proposed toy-model suggests that TSI alone is unlikely to represent
the primary solar forcing agent of the climate system and suggests
the involvement of alternative mechanisms, which are still poorly
understood, such as a direct solar modulation of the cloud/albedo
system. Consequently, the actual ECS values may be relatively low,
also ranging approximately from 1.0°C to 1.5°C.

The interpretation of the climate system derived from the above empirical
evidences significantly differs from that derived from the CMIP6 GCMs
and promoted by the \citet{IPCC2001,IPCC2007,IPCC2013,IPCC2023},
and suggest that climate science is still unsettled in many key issues.

\section{Conclusion}

The findings outlined above underscore significant uncertainties in
climate modeling, climate data, solar records, and solar-climate interactions,
leaving unresolved the key question of whether observed warming is
primarily driven by anthropogenic factors, natural processes, or their
interplay. Empirical methodologies, such as those utilized by \citet{Scafetta2023a,Scafetta2024}
and \citet{Connolly2023}, highlight this ongoing ambiguity.

Concerns are mounting regarding the limitations of the CMIP GCMs employed
by the IPCC in its assessment reports from \citeyear{IPCC2007}, \citeyear{IPCC2013},
and \citeyear{IPCC2021}. These models appear unable to accurately
replicate natural climate variability across different timescales,
highlighting critical unresolved issues in fundamental climate dynamics.
Also the magnitude of solar variability across temporal scales requires
further investigation, particularly given the strong correlations
identified between solar proxy records and climate patterns throughout
the Holocene. \citet{Schmutz2021} argued that such strong correlations
challenge the validity of the low-variability TSI models, such as
those proposed by \citet{Matthes2017}, \citealp{Kopp2016} and \citet{Wu2018}.
Since these models serve as solar forcing inputs for the CMIP6 GCMs,
their choice needs to be reconsidered.

Additionally, the climate system's apparent oversensitivity to solar
activity, as observed by various researchers \citep[and many others]{Soon2000,Shaviv2008,Scafetta2009},
complicates the physical issue. Empirical estimates indicate that
the climate's sensitivity to solar activity fluctuations may be four
to six times greater than its sensitivity to radiative forcings \citep{Scafetta2023a},
raising important questions about the physical mechanisms underlying
the solar-climate interactions. The climate system's oversensitivity
to changes in TSI implies that radiative solar forcing alone may not
be the primary mechanism driving climate dynamics. Alternative processes,
including solar-based corpuscular forcing mechanisms, warrant further
exploration. Such mechanisms may significantly influence cloud formation
via cosmic ray flux \citep[cf.][]{Svensmark2019} or other interplanetary
particle interactions with the Earth's atmosphere \citep{Scafetta2016,Scafetta2020}.

Cloud-climate interactions introduce additional complexities to attribution
analyses, emphasizing the need for precise modeling of cloud processes.
Several key observations highlight their critical role:
\begin{enumerate}
\item \emph{cloud feedback uncertainty} --- variability among climate models
in representing cloud feedback significantly contributes to the wide
range of ECS estimates derived from the GCMs. The \citep{IPCC2021}
acknowledges ``\emph{medium}'' to ``\emph{low}'' confidence in
GCM representations of cloud dynamics;
\item \emph{impact of cloudiness on heat transfer} --- even minor changes
in cloud cover can drastically affect atmospheric heat transfer for
a much greater extent than significant shifts in CO\textsubscript{2}
concentrations. For example, on heavily overcast days, surface irradiance
can drop from 1,000 W/m\textsuperscript{2} to 200 W/m\textsuperscript{2}
\citep{vanWijngaarden2025};
\item \emph{trends in global cloud cover} --- satellite observations from
1983 to 2010 reveal a decline in global cloud cover from approximately
68\% to 65\%, corresponding with a \textasciitilde 0.4°C increase
in global surface temperatures over the same period (Figure \ref{Fig19}A);
\item \emph{regional correlations with cloud dynamics} --- \citet{L=0000FCdecke2024}
identified strong correlations between central European sunshine durations
and the Atlantic Multidecadal Oscillation (AMO) over the period from
1880 to 2020;
\item \emph{flood frequencies and solar proxy records} --- Holocene flood
frequency records, likely influenced by variations in cloud cover,
exhibit significant correlations with solar proxies (Figure \ref{Fig17}).
\end{enumerate}
The high equilibrium climate sensitivity (ECS) estimates of the GCMs
primarily stem from assumptions about a strong water vapor positive
feedback mechanism, initially hypothesized by \citet{Manabe1975}.
This hypothesis suggests that warming leads to increased evaporation,
further amplifying warming. However, uncertainties surrounding water
vapor and cloud feedbacks have been acknowledged since the 1960s \citep{Moller1963}
and are still unresolved (Figure \ref{Fig5}). Recent analyses indicate
that numerical climate models may overestimate evaporation rates \citep{Ma2025,Simpson2024}.
Additionally, top-of-atmosphere (TOA) radiative flux measurements
suggest that long-wave feedbacks --- such as those associated with
cirrus cloud formations --- may be negative rather than positive
\citep{Lindzen2011}. \citet{Schildknecht2020} also argues that a
proper accounting of cloud-cover dynamics brings ECS estimates closer
to or below 1°C.

A number of empirical analyses suggest a moderate climate sensitivity
to radiative forcing, with ECS estimates varying depending on the
data and assumptions considered. Realistic ECS values could be approximated
at least at $2.1\pm0.7$ °C (17--83\% range) by combining several
lines of climate sensitivity evidence \citep[e.g.:][]{Lewis2023,Scafetta2022a,Scafetta2023b}.
However, significantly lower ECS estimates --- even as low as $1.1\pm0.4$
°C (17--83\% range) --- are also possible if solar variability plays
a dominant role through mechanisms alternative to the TSI forcing
\citep[cf.][]{Scafetta2013}. Such low ECS estimates indicate the
existence of balanced positive and negative feedback mechanisms, and
have been supported by several researchers \citep{Rasool1971,Lindzen2011,Harde2014,Monckton2015,Bates2016,Knutti2017,Schildknecht2020,Stefani2021}.

In any case, it is also possible that feedback mechanisms could exhibit
scale-dependent behavior. Over short and decadal timescales, they
may tend to show slightly negative responses, primarily due to rapid
cloud-related adjustments. Over longer timescales, however, feedback
responses may become increasingly positive as surface albedo effects
linked to glacial melting and vegetation expansion gain influence.
Consequently, ECS values around $1.1\pm0.4$ °C or slightly larger
may be realistic when climate change is assessed over interannual
to millennial timescales.

Biases in global surface temperature records, particularly over land
areas, raise concerns about the reported global and regional warming
trends. The urban heat island effect, exacerbated by increasing urbanization,
likely contributes to these biases \citep{Scafetta2021a,Soon2023}.
Temperature homogenization algorithms, designed to correct non-climatic
warm biases in the meteorological records, appear to blend them into
surrounding rural and suburban stations rather than fully removing
urban-induced thermal anomalies \citep{Daleo2016,ONeill2022,Katata2023}.
Consequently, land surface temperature records --- particularly those
from the Northern Hemisphere --- tend to show exaggerated warming
trends compared to alternative datasets (Figure \ref{Fig13}), including
the satellite-based lower troposphere temperature measurements and
sea surface temperature records (Figure \ref{Fig14}).

The \citet{IPCC2023}, unequivocally attributes the observed global
surface warming from 1850--1900 to 2011--2020 almost entirely ($\sim$100\%)
to anthropogenic drivers; such assessment supports the climate alarmist
frameworks that dominate European policymaking that yielded to policies
such as the EU Green Deal and Net-Zero by 2050. However, the CMIP
GCM prediction that without anthropogenic drivers the climate would
have remained nearly stable from 1850--1900 to 2011--2020 cannot
be validated because of missing data. Moreover, the IPCC assessments
are based on surface temperature datasets and theoretical simulations
derived from GCMs that appear flawed in many ways and that incorporate
equally problematic low-variability TSI forcing functions. These models
contradict each other regarding the ECS value (Figure \ref{Fig5})and
appear to conflict with a substantial body of empirical evidence spanning
the entire Holocene. Notably, they fail to accurately reconstruct
the secular-scale warm periods of the past including the MWP, RWP
and HTM (Figures \ref{Fig10}, \ref{Fig11}), which should be a prerequisite
for trusting the GCM-based climate attributions for the period from
1850--1900 to 2011--2020 depicted in Figure \ref{Fig2}, underestimate
the warming from the 1900s to the 1940s (Figure \ref{Fig8}) and overestimate
the global surface warming of the last 40 years (Figure \ref{Fig6}).

In contrast, semi-empirical and empirical modeling approaches, which
try to account for natural variability, emphasize a significant role
of solar variability in climate change. The IPCC tends to ignore the
efforts of empirical modelings of the climate system. Yet, holistic
empirical models are necessary to properly guiding the research, in
particular when the reductionist ones are demonstrated to poorly reconstruct
natural climate variability in key patterns. The discussed empirical
models project moderate warming trajectories for the 21\textsuperscript{st}
century, diverging from the CMIP6 GCM projections, and would not support
the necessity of Net-Zero by 2050 policies to satisfy the climatic
targets of the \citet{ParisAgreement} because the global surface
warming may remain approximately below 2°C throughout the 21\textsuperscript{st}
century also under realistic and moderate development scenarios, such
as the SSP2-4.5 (Figure \ref{Fig21}). More specifically, the proposed
empirical evidence could suggest that the observed global surface
warming likely results from multiple contributing factors such as:
\begin{itemize}
\item \emph{warm bias} --- around 20\% of reported global warming may stem
from non-climatic warm biases in the global surface temperature records;
\item \emph{solar contribution} --- natural solar-induced variability,
operating through multiple mechanisms, could account for up to 50\%
of the reported warming;
\item \emph{anthropogenic contribution} --- human activities are estimated
to contribute roughly 30\% to the reported warming.
\end{itemize}
The above empirical assessment is compatible with a low climate sensitivity
to anthropogenic emissions, where ECS estimates are as low as $1.1\pm0.4$
°C. If correct, climate risks may be significantly lower than those
projected by the CMIP6 GCMs, supporting adaptive strategies and cost-effective
mitigation technologies instead of expensive Net-Zero policies \citep{Scafetta2024}.

Climate science remains far from settled, yet trillions of dollars
continue to be allocated toward policies aimed at mitigating extreme
hypothetical warming scenarios based on potentially flawed GCM outputs.
Historically, atmospheric CO\textsubscript{2} levels have been 10
to 20 times higher than current concentrations during approximately
95\% of Earth's history since complex life emerged 600 million years
ago \citep{Davis2017}. Notably, CO\textsubscript{2} concentrations
often lag temperature changes across different timescales, suggesting
temperature fluctuations may drive CO\textsubscript{2} variations
rather than vice versa \citep{Shakun2012,Koutsoyiannis2024}.

Since 1900, atmospheric CO\textsubscript{2} concentrations have increased
from 295 ppm to 425 ppm, primarily due to human emissions and, to
a lesser extent, rising sea surface temperatures. While higher CO\textsubscript{2}
levels can negatively impact marine biodiversity by lowering the pH
of the sea surface microlayer \citep{Davis2023}, they also contribute
to planetary greening through fertilization effects, enhancing crop
yields \citep{Zhu2016,Piao2019,McKitrick2025}.

As CO\textsubscript{2} is a greenhouse gas, climate hazard assessments
related to its future emissions depend largely on the Equilibrium
Climate Sensitivity (ECS) and Transient Climate Response (TCR) values
of the climate system, as well as on the projected emission scenarios.
Notably, extreme concentration pathways --- such as SSP3-7.0 and
SSP5-8.5 --- are increasingly regarded as ``\emph{unlikely}'' and
``\emph{very unlikely}'', respectively, and emphasizing climate
model simulation based on such scenarios could be misleading \citep{Ritchie2017,Hausfather2020,Pielke2022,Burgess2023}.
This fact should be taken into account when developing climate related
policies for not harming the economy with unnecessary mitigation policies,
which should be based on realistic risks. In this regards, the meta
economic analysis by \citet{Tol2023} appears particularly compelling
when comparing the costs and benefits of the Paris climate targets.

Advancing climate science requires directly confronting uncertainties
in detection, attribution, and modeling. Further research on unresolved
issues is critical for improving climate risk assessment and developing
more effective strategies for addressing future environmental challenges.

\subsection*{Data Availability (accessed on April 10, 2025)}
\begin{itemize}
\item Many climatic records can be downloaded from the cited references
and from the KNMI Climate Explorer. \\
Available from: \href{https://climexp.knmi.nl/}{https://climexp.knmi.nl/}
\item CMIP6 GCM simulations. Available from \href{https://climexp.knmi.nl/selectfield_cmip6_knmi23.cgi}{https://climexp.knmi.nl/selectfield\_cmip6\_knmi23.cgi}
\item HadSST4. Available from: \href{https://www.metoffice.gov.uk/hadobs/hadsst4/}{https://www.metoffice.gov.uk/hadobs/hadsst4/}
\item NOAA/CIRES/DOE 20th Century Reanalysis (V3). \\
Available from: \href{https://www.psl.noaa.gov/data/gridded/data.20thC_ReanV3.html}{https://www.psl.noaa.gov/data/gridded/data.20thC\_ReanV3.html}
\item AMO unsmoothed from the Kaplan SST V2. \\
Available from: \href{https://psl.noaa.gov/data/correlation/amon.us.long.data}{https://psl.noaa.gov/data/correlation/amon.us.long.data}
\item UAH MSU lt v 6.1\\
Available from: \href{https://www.nsstc.uah.edu/data/msu/v6.1/tlt/uahncdc_lt_6.1.txt}{https://www.nsstc.uah.edu/data/msu/v6.1/tlt/uahncdc\_lt\_6.1.txt}
\item NOAA-STAR v. 5 \\
Available from: \\
\href{https://www.star.nesdis.noaa.gov/data/mscat/MSU_AMSU_v5.0/Monthly_Atmospheric_Layer_Mean_Temperature/Global_Mean_Anomaly_Time_Series/}{https://www.star.nesdis.noaa.gov/data/mscat/MSU\_AMSU\_v5.0/Monthly\_Atmospheric\_Layer\_Mean\_ Temperature/Global\_Mean\_Anomaly\_Time\_Series/}
\item ISCCP Global Cloud Cover\\
Available from: \href{https://isccp.giss.nasa.gov/pub/data/D2BASICS/B8glbp.dat}{https://isccp.giss.nasa.gov/pub/data/D2BASICS/B8glbp.dat}
\end{itemize}

\subsection*{Declaration}
\begin{itemize}
\item The author declares that he has no known competing financial interests
or personal relationships that could have appeared to influence the
work reported in this paper.
\item The author is a Guest Editor of this journal and was not involved
in the editorial review or the decision to publish this article.
\end{itemize}

\end{document}